\documentclass[twocolumn,10pt,letter]{IEEEtran}
\pdfoutput=1
\usepackage[table]{xcolor}
\usepackage{amsmath}
\usepackage{amsthm}
\usepackage{amssymb}
\usepackage{arcs}
\usepackage{color}
\usepackage{geometry}
\usepackage{graphicx}
\usepackage{latexsym}
\usepackage{psfrag}
\usepackage{tikz}
\usetikzlibrary{trees}
\usepackage{xcolor}
\usepackage{algorithm}
\usepackage{caption}
\usepackage[noend]{algpseudocode}
\usepackage[tight]{subfigure}
\usepackage{float}
\usepackage{pifont}
\usetikzlibrary{arrows,shapes}
\usepackage{multirow}
\usepackage{float}
\usepackage{dblfloatfix}
\usepackage{cite}
\renewcommand{\baselinestretch}{1.0}
\newcommand{\kETAL}    {{\em et~al.}}

\newcommand{\cmark}{\ding{51}}%
\newcommand{\xmark}{\ding{55}}%

\IEEEoverridecommandlockouts
\makeatletter
\def\ps@headings{%
\def\@oddhead{\mbox{}\scriptsize\rightmark \hfil \thepage}%
\def\@evenhead{\scriptsize\thepage \hfil \leftmark\mbox{}}%
\def\@oddfoot{}%
\def\@evenfoot{}}
\makeatother
\usepackage{amssymb}
\usepackage{amsmath}
\usepackage{graphicx}
\usepackage{latexsym}
\usepackage{psfrag}

\def\BibTeX{{\rm B\kern-.05em{\sc i\kern-.025em b}\kern-.08em
    T\kern-.1667em\lower.7ex\hbox{E}\kern-.125emX}}
\makeatletter
\@addtoreset{equation}{section}
\makeatother
\renewcommand{\thesection}{\arabic{section}}
\renewcommand{\thesubsection}{\thesection.\Alph{subsection}}
\renewcommand{\theequation}{\thesection.\arabic{equation}}
\theoremstyle{definition}

\begin{document}
\author{ Reza Tourani, Travis Mick, Satyajayant Misra, Gaurav Panwar \\*
Department of Computer Science, New Mexico State University \\
\{rtourani, tmick, misra, gpanwar\}@cs.nmsu.edu\\
~\thanks{This work was supported in part by the U.S. NSF grants:1345232 and 1248109 and the U.S. DoD/ARO 
grant: W911NF-07-2-0027.}
}
%
%
\title{Security, Privacy, and Access Control in Information-Centric Networking: A Survey}
\maketitle
\pagestyle{empty}
\thispagestyle{empty}
\begin{abstract}

Information-Centric Networking (ICN) replaces the widely used host-centric networking paradigm in 
communication networks (e.g., Internet, mobile ad hoc networks) with an information-centric paradigm, 
which prioritizes the delivery of named content, oblivious of the contents' origin. 
Content and client security, provenance, and identity privacy are intrinsic in the ICN paradigm versus 
the current host centric paradigm where they have been instrumented as an after-thought. 
By design, the ICN paradigm inherently supports many security and privacy features, such as provenance 
and identity privacy, which are still not effectively available in the host-centric paradigm. 
However, given its nascency, the ICN paradigm has several open security and privacy concerns. 
In this article, we survey the existing literature in security and privacy in ICN and present open questions. 
More specifically, we explore three broad areas: security threats, privacy risks, and access control 
enforcement mechanisms.

We present the underlying principle of the existing works, discuss the drawbacks of the proposed approaches, 
and explore potential future research directions. 
In security, we review attack scenarios, such as denial of service, cache pollution, and content poisoning. 
In privacy, we discuss user privacy and anonymity, name and signature privacy, and content privacy. 
ICN's feature of ubiquitous caching introduces a major challenge for access control enforcement that requires 
special attention. 
We review existing access control mechanisms including encryption-based, attribute-based, session-based, and 
proxy re-encryption-based access control schemes. We conclude the survey with lessons learned and scope for 
future work.

{\em Keywords}--Information-centric networking, security, privacy, access control, architecture, DoS, content poisoning.
\end{abstract}

\section{Introduction}
\label{sec01}
%
According to the Cisco Visual Networking Index forecast, video traffic (including
VoD, P2P, Internet, and TV) will comprise $90\%$ of all Internet traffic by 
2019\footnote{Cisco Visual Networking Index:Forecast and Methodology}.
The majority of this traffic is currently served to end users with the help of
content delivery networks (CDNs), with servers that reside close to the network 
edge.
This has helped reduce core network traffic and improve delivery latency. 
Despite the scalability that CDNs have so far provided, the current host-centric
paradigm will not continue to scale with the proliferation of mobile devices and the 
Internet of Things (IoTs) coupled with the rapidly increasing volume of video traffic.
In the IoT domain, every node can be a provider.
This results in several many-to-many communications, which increases the size of routing 
tables and requires maintenance of per node multicast trees, thus undermining scalability.
Not only have these trends been putting pressure on Internet Service Providers
(ISPs) and content providers, but they have also motivated the research
community to explore designs for a more scalable Internet, with a primary
objective of efficient content delivery.
One of the products of this endeavor is the Information-Centric Networking (ICN)
paradigm~\cite{JacSmeTho09, TarAinVis09, GhoSheKop11}.

ICN shifts the networking paradigm from the current host-centric paradigm, where all 
requests for content are made to a host identified by its IP address(es), to a content-centric 
paradigm, which decouples named content objects from the hosts where they are located. 
As a result, named content can be stored anywhere in the network, and each 
content object can be uniquely addressed and requested. 
Several ICN architectures such as Named-data networking/content-centric networking 
(NDN/CCN)~\cite{JacSmeTho09}, Publish-Subscribe Internet Routing Paradigm (PSIRP)
~\cite{TarAinVis09}, Data Oriented Network Architecture (DONA)~\cite{KopChaChu07}, 
and Network of Information (NetInf)~\cite{AhlDamMar08} have been proposed.
Though they differ in their details, they share several fundamental properties:
unique name for content, name-based routing, pervasive caching, and assurance of
content integrity.
ICN enhances several facets of user experience as well as security, privacy, 
and access controls.
However, it also gives rise to new security challenges.  
%
%

In this article, we explore ICN security, privacy, and access control concerns 
in-depth, and present a comprehensive study of the proposed mechanisms in the 
state of the art.
%
We categorize this survey into three major domains, namely
security, privacy, and access control.
In the security section, we address {\em denial of service} (DoS and distributed 
DoS or DDoS) attacks and vulnerabilities unique to ICN, including {\em cache 
pollution}, {\em content poisoning}, and {\em naming attacks}.
Despite many similarities between a classical DoS attack and the DoS attack in
ICN, the latter is novel in that it abuses ICN's stateful forwarding plane.
The attack aims to overload a router's state tables, namely the pending interest
table (PIT). 
The cache pollution attack targets a router's content locality with the
intention of altering its set of cached content resulting in an increase
in the frequency of content retransmission, and reduced network goodput.
%
%
%
%

In the privacy section, we study the privacy risks in ICN under four classes: 
{\em client privacy}, {\em content privacy}, {\em cache privacy}, and 
{\em name and signature privacy}~\cite{ChaDecKaa13}.
We explore the implications of each of these risk classes and elaborate on
relevant proposed solutions.
Due to ICN's support for pervasive caching, content objects can be replicated 
throughout the network.
Though this moves content close to the edge and helps reduce network load and 
content retrieval latency, it comes at a cost---publishers lose control over 
these cached copies and cannot arbitrate access.
Thus, there is need for efficient access control, which allows reuse of cached 
content and prevents unauthorized accesses. 

Access control mechanisms based on {\em content encryption}, {\em clients' identities}, 
{\em content attributes}, or {\em authorized sessions} have been proposed in the literature.
We review these proposed mechanisms and highlight their benefits and drawbacks
in detail in the access control section.
In the three domains, we present a summary of the state of the art and also discuss 
open research challenges and potential directions to explore. 
We conclude the survey with a summary of lessons learned. 

Before we dive into the discussion, we briefly review some
representative ICN architectures in Subsection~\ref{sec01-01}.  
Following that we identify previous surveys in ICN covering different ICN
architectures, naming and routing, DoS attacks, mobility, and potential research
directions in Subsection~\ref{sec01-02}.
%
\subsection{Overview of the Proposed Information-Centric Networking Architectures}
\label{sec01-01}
%
%
%

In this subsection, we review some representative ICN architectures including 
DONA~\cite{KopChaChu07}, CCN~\cite{JacSmeTho09, CCNx}, NDN~\cite{NDN}, 
PSIRP/PURSUIT~\cite{TarAinVis09, PSIRP, PURSUIT}, NetInf~\cite{AhlDamMar08}, 
and MobilityFirst~\cite{SesNagNel11, MobilityFirst}.
%
%
We refer interested readers to two surveys~\cite{AhlDanImb12, XylVerSir14} for
more details on other ICN architectures, such as SAIL~\cite{SAIL},
4WARD~\cite{4WARD}, COMET~\cite{Sal10, COMET}, CONVERGENCE~\cite{CONVERGENCE},
and CONET~\cite{DetBelSal2011}.
In this survey, we will focus on research relevant to three architectures in particular, 
namely CCN~\cite{JacSmeTho09, CCNx}, NDN~\cite{NDN}, and PSIRP/PURSUIT~\cite{TarAinVis09, PSIRP, PURSUIT}.
These three have received the most attention from the community in the past and 
continue to be favored as architectures of choice.
%
%

The {\it Data Oriented Network Architecture} (DONA)~\cite{KopChaChu07} was
proposed by Koponen~\kETAL~at UC Berkeley in $2007$.
DONA uses a flat self-certifying naming scheme.
Each name consists of two parts; the first is the cryptographic hash of the
publisher's public key, and the second is an object identifier, which is
assigned by the publisher and is unique in the publisher's domain.
To achieve self-certification, the authors suggested that publishers use a
cryptographic hash of the object as the object identifier.
A subscriber can then easily verify the integrity of an object simply by hashing
it and comparing the result to the object's name.
DONA's resolution service is composed of a hierarchically interconnected network
of resolution handler (RH) entities, which are tasked with publication and
retrieval of objects. 

To publish an object, the owner sends a {\it REGISTER} message including the
object name to its local RH.
The local RH, keeps a pointer to the publisher and propagates this message to its
parent and peer RHs, who then store a mapping between the local RH's address and
the object name.
A subscriber interested in the object sends a {\it FIND} message with the object
name to its own local RH.
The local RH propagates this request to its parent RH.
The propagation continues until a match is found somewhere in the hierarchy.

After finding a match, the request is forwarded towards the identified publisher.
The authors proposed two methods of object delivery from a publisher to a requester.
In the first method, the publisher sends the object using the underlying IP
network.
The second method takes advantage of path symmetry: the request
message records the path it takes through the network.
After reaching the publisher, the object traverses the reverse path from the
publisher to the requester.
Exploiting this routing model, RHs on the path can aggregate the request
messages for an object and form a multicast tree for more efficient object 
dissemination/delivery.

{\it Content-centric Networking} (CCN)~\cite{JacSmeTho09, CCNx} was proposed by 
researchers at Palo Alto Research Center in $2009$.
In 2010, {\it Named Data Networking} (NDN)~\cite{NDN}, which follows the same
design principles, was selected by the US National Science Foundation (NSF) as
one of four projects to be funded under NSF's Future Internet Architecture
program.
Both CCN and NDN share the same fundamentals, such as a hierarchical naming
scheme, content caching, and named content routing 
(NDN was CCN before it branched out).
The hierarchical naming allows the provider's domain name to be used in making
routing decisions.
In the client-driven CCN/NDN, a client sends an interest packet into the network
to request a content by its name.

Routers, equipped with a content store (CS), a pending interest table
(PIT), and a forwarding information base (FIB), receive the interest and 
perform a CS lookup on the content name.
If the content is not available in the CS, the router performs a PIT lookup to
check whether there is an existing entry for the requested content.
If the PIT lookup is successful, the router adds the incoming interest's
interface to the PIT entry (interest aggregation) and drops the interest.
If no PIT match is found, the router creates a new PIT entry for the interest
and forwards the interest using information from the FIB.

An interest can be satisfied either by an intermediate forwarding router which
has cached the corresponding content chunk, or the content provider.
In both cases, the content takes the interest's reverse-path back to the requester.
Upon receipt of a content chunk, a router forwards the chunk along the interfaces 
on which it had received the corresponding interests for the chunk.
The router may cache a copy of the content in its CS in addition to
forwarding it through the designated faces.
%
%
%
%
%
%
%
%
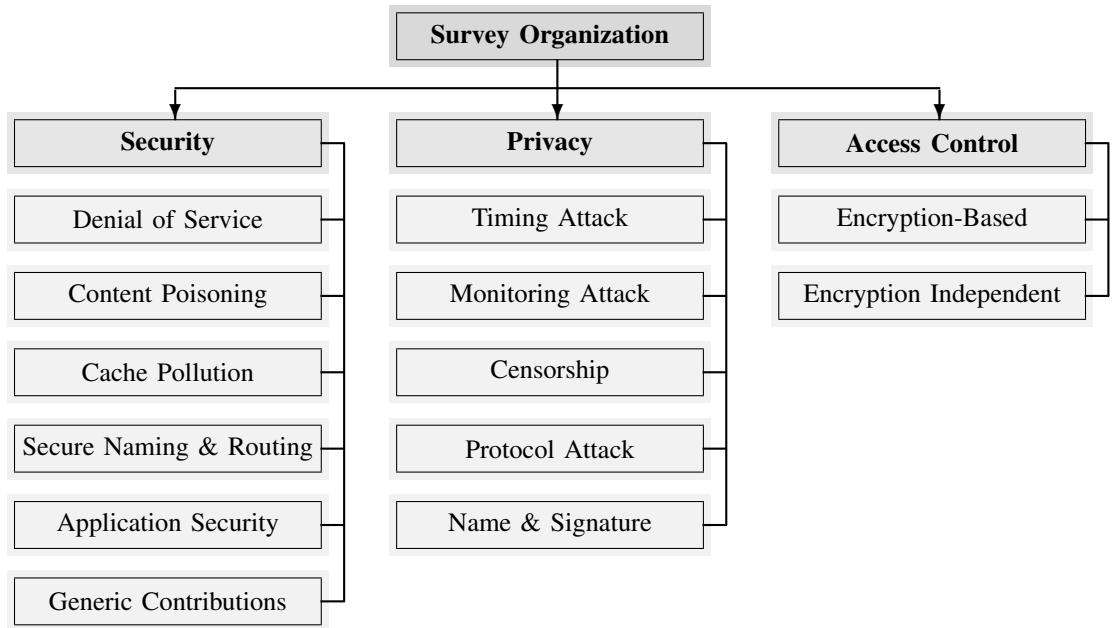
\begin{figure*}[!ht]
\setlength{\unitlength}{0.08in}
\centering
\begin{picture}(72,42)
\put(0,3){\colorbox{gray!10}{\framebox(20,3){Generic Contributions}}} 
\put(0,8){\colorbox{gray!10}{\framebox(20,3){Application Security}}} 
\put(25,8){\colorbox{gray!10}{\framebox(20,3){Name \& Signature}}} 
%
\put(0,13){\colorbox{gray!10}{\framebox(20,3){Secure Naming \& Routing}}} 
\put(25,13){\colorbox{gray!10}{\framebox(20,3){Protocol Attack}}} 
%
\put(0,18){\colorbox{gray!10}{\framebox(20,3){Cache Pollution}}} 
\put(25,18){\colorbox{gray!10}{\framebox(20,3){Censorship}}} 
%
\put(0,23){\colorbox{gray!10}{\framebox(20,3){Content Poisoning}}} 
\put(25,23){\colorbox{gray!10}{\framebox(20,3){Monitoring Attack}}} 
\put(50,23){\colorbox{gray!10}{\framebox(20,3){Encryption Independent}}} 
\put(0,28){\colorbox{gray!10}{\framebox(20,3){Denial of Service}}} 
\put(25,28){\colorbox{gray!10}{\framebox(20,3){Timing Attack}}} 
\put(50,28){\colorbox{gray!10}{\framebox(20,3){Encryption-Based}}} 
\put(0,33){\colorbox{gray!20}{\framebox(20,3){\bf Security}}} 
\put(25,33){\colorbox{gray!20}{\framebox(20,3){\bf Privacy}}} 
\put(50,33){\colorbox{gray!20}{\framebox(20,3){\bf Access Control}}} 
\put(25,40){\colorbox{gray!30}{\framebox(20,3){\bf Survey Organization}}} 
%
\thicklines
\put(10.95,38.1){\line(1,0){50.1}}
\put(36,40){\vector(0,-1){4}}
\put(11,38){\vector(0,-1){2}}
\put(61,38){\vector(0,-1){2}}
\put(45.5,9.5){\line(1,0){1.6}}
\put(45.5,14.5){\line(1,0){1.6}}
\put(45.5,19.5){\line(1,0){1.6}}
\put(45.5,24.5){\line(1,0){1.6}}
\put(45.5,29.5){\line(1,0){1.6}}
\put(45.5,34.5){\line(1,0){1.6}}
\put(20.5,4.5){\line(1,0){1.6}}
\put(20.5,9.5){\line(1,0){1.6}}
\put(20.5,14.5){\line(1,0){1.6}}
\put(20.5,19.5){\line(1,0){1.6}}
\put(20.5,24.5){\line(1,0){1.6}}
\put(20.5,29.5){\line(1,0){1.6}}
\put(20.5,34.5){\line(1,0){1.6}}
%
\put(70.5,24.5){\line(1,0){1.6}}
\put(70.5,29.5){\line(1,0){1.6}}
\put(70.5,34.5){\line(1,0){1.6}}
\put(22.05,4.5){\line(0,1){30}}
\put(47.05,9.5){\line(0,1){25}}
\put(72.05,24.5){\line(0,1){10}}
\end{picture}
\vspace{-0.5cm}
\caption{The organization of the survey.}
\label{fig01-02}
\end{figure*}

The {\it Publish Subscribe Internet Technology} (PURSUIT)~\cite{PURSUIT} project
and its predecessor {\it Publish Subscribe Internet Routing Paradigm}
(PSIRP)~\cite{TarAinVis09, PSIRP}, were funded by FP7 (European Union's research
and innovation program) to produce a publish-subscribe protocol stack.
A PURSUIT network is composed of three core entities, namely Rendezvous Nodes
(RNs) which form the REndezvous NEtwork (RENE), the topology manager, and
forwarders.
Similar to DONA, PURSUIT uses a flat naming scheme composed of a scope ID, which
groups related information objects, and a rendezvous ID, which ensures that each
object's identifier is unique in its group.
A publisher advertises its content by sending a {\it PUBLISH} message to its
{\it local RN} (the RN in the publisher's vicinity), which routes the message to the RN 
designated to store the content name defined by the scope ({\it designated RN}). 
The local RN makes this decision using a distributed hash table (DHT).
A subscriber interested in the content object sends a {\it SUBSCRIBE} message to
its local RN, which will also be routed to the designated RN using the DHT.

Upon receipt of a {\it SUBSCRIBE} message by the designated RN, the topology
manager is instructed to generate a delivery path between the publisher and the
subscriber.
The topology manager then provides the publisher with a path through the
forwarders.
In PURSUIT, network links are each assigned a unique string identifier, which
the topology manager uses to create a routing Bloom filter for each flow.
The generated Bloom filter is then added to each packet's header, and is used by
the intermediate forwarders for content delivery.

{\it Network of Information} (NetInf)~\cite{AhlDamMar08} was initially conceived
in the FP7 project 4WARD~\cite{4WARD}.
NetInf employs a flat naming scheme with a binding between names and their
locators, which point to the content's location.
As several nodes can cache copies of the data, an object may be bound to more
than one locator.
Two models of content retrieval are offered by NetInf: name resolution and
name-based routing.
In the name resolution approach, a publisher publishes its data objects to the
network by registering its name/locator binding with the name resolution service
(NRS).
An interested client resolves the named data object into a set of locators and
subsequently submits a request for the object, which will be delivered by the
routing forwarders to the best available cache.

The routing forwarders, after obtaining the data, deliver it back to the
requester.
In the name-based routing model, a client directly sends out a {\it GET} message
with the name of the data object.
This message is forwarded to an available storage node using name-based routing,
and the data object, once found, is forwarded back to the client.

{\em MobilityFirst}~\cite{SesNagNel11, MobilityFirst} was funded by the NSF's
future Internet Architecture program in $2010$.
The main focus of this architecture is to scale in the face of device mobility, hence it includes
detailed mechanisms for handling mobility, wireless links, multicast,
multi-homing, security, and in-network caching.
Each network entity (including devices, information objects, and services) is
assigned a globally unique identifier (GUID), which can be translated into one or 
more network addresses.
To advertise a content, a publisher requests a GUID from the naming service and
registers this name with a global name resolution service (GNRS).

The registered GUID is mapped, by a hash function, to a set of GNRS servers,
which are connected through regular routing.
A subscriber can then obtain the content name from a Name Certification Service
(NCS) or use a search engine to resolve a human-readable name into the
corresponding GUID.
A subscriber submits a {\it GET} message, containing both the GUID of the desired
object and its own GUID, to its local router.
Since routers require the network address, the request will be forwarded
to the GNRS to map the GUID into actual addresses.
The result of this query is a set of partial or complete routes, or a set of
addresses.
%
%
%
%
%
%
\begin{figure*}[!ht]
\begin{center}
\begin{tikzpicture}[level distance=2cm,
  level 1/.style={sibling distance=3.25cm}]
  \node {Security}
    child {node [align=center] {Denial of \\ Service \\ ~\cite{AfaMahMoi13,GasTsuUzu13,ComConGas13,DaiWanFan13,WanZhoQin14,NguCogDoy15,WanCheZho12,VirMarSis13,WanZhoQin13,WanCheZho14,LiBi14}}
    }
    child {node [align=center] {Content \\ Poisoning \\ ~\cite{GasTsuUzu13,KimNamBi15,GhaTsuUzu14,GhaTsuUzu14elements,GhaTsuUzu14Network-layer}}
    }
    child {node [align=center] {Cache \\ Pollution \\ ~\cite{ParWidLee12,XieWidWan12,ConGasTeo13,KarGue15anfis,MauRasGer15}}
    }
    child {node [align=center] {Secure Naming, \\ Routing \& Forwarding \\ ~\cite{WonNik10,DanGolOhl10,ZhaChaXio11,HamSerFad12,RemCatSac09,AlzReeVas12,AlzVasRee13securing,AlzVasRee14,AlzVasRee13,YiAfaMoi13,AfaYiWan15}}
    }
    child {node [align=center] {Application \\ Security \\ ~\cite{FotMarPol10,GoeChoFra13semantic,GoeChoFra13,KarGue15fuzzy,BurGasNat13,BurGasNat14,ViePol13,SalRen12,AmbConGas14,AsaNamKaw15,WonVerMag08,SeeKutSch14,SeeGilKut14,YuAfaCla15}}
    }
    child {node [align=center] {Other General \\ Contributions \\ ~\cite{FotMarPol10Towards,FotMarPol11,MarBarFie12,WahSchVah12,GhaNarOra14,LooAia15}}
    };
\end{tikzpicture}
\end{center}
\caption{ICN security sub-categories and the state-of-the-art.}
\label{fig02-01}
\end{figure*}
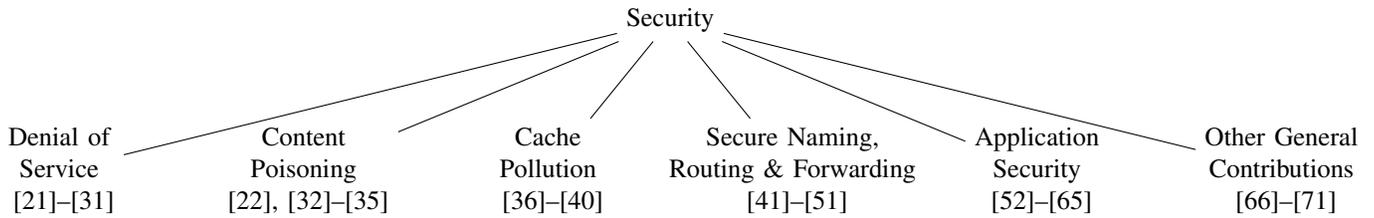

Upon receiving this information, the requesting router attaches the
destination network address to the {\it GET} message and forwards 
it into the network.
Any router on the forwarding path may contact the GNRS for an updated destination 
address or route; routes may change due to events, such as provider's mobility, 
congested link, and link failure.
The publisher, upon receiving the {\it GET} message, sends the requested object
back to the source GUID following the same procedure.
MobilityFirst provides a combination of IP routing and name-based routing by name 
resolution and data routing processes.
On-path caching is employed to satisfy subsequent requests for previously served GUIDs.
This is in contrast to off-path caching, which causes an update in the GNRS service, 
where the new caching node's network address is added to the GUID's record.
%
%
\subsection{Review of Existing ICN Surveys and Overview Literature}
\label{sec01-02}
%
%
Ahlgren~\kETAL~\cite{AhlDanImb12} reviewed the different proposed
information-centric architectures.
In addition to describing the architectures in detail, the authors also
presented their open challenges.
Following this survey, Xylomenos~\kETAL~\cite{XylVerSir14} surveyed the
proposed ICN architectures, comparing their similarities and differences, 
and discussing their weaknesses.
Tyson~\kETAL~focused on mobility in information-centric networks
in~\cite{TysSasRim12}.
Several benefits of node mobility were discussed by the authors, as well as
mobility-related challenges such as provider mobility and cached content
discovery.
Zhang, Li and Lin~\cite{ZhaLiLin13} and Zhang~\kETAL~\cite{ZhaLuoZha2015} explored
proposed caching approaches in information-centric networking. 
In~\cite{BarChoAhm12}, Bari~\kETAL~reviewed the state-of-the-art in naming and
routing for information-centric networks and explored the requirements for ideal
content naming and routing.
Future research directions in information-centric networking were discussed by
Pan~\kETAL~\cite{PanPauJai11}.

Aamir and Zaidi~\cite{AamZai14} surveyed denial-of-service attacks in
information-centric networks and identified interest flooding, request piling,
content poisoning, signature key retrieval, and cache pollution as DDoS vectors.
AbdAllah~\kETAL~\cite{AbdHasZul15} recently discussed security attacks in ICN.
The authors classified attacks into four categories: routing, naming, caching,
and miscellaneous.
The paper focused on discussing the ways an attacker can orchestrate these
attacks as well as the applicability of current IP-based solutions to 
information-centric networks.

In other overview work, Marias~\kETAL~\cite{MarBarFie12} identified security and 
privacy concerns in a future Internet architecture.
They reviewed physical layer security, network coding security, and network infrastructure 
security literature and identified authentication and identity management as core
building blocks of a secure network, and discussed implementation challenges. 
%
However, the authors did not elaborate on the attacks that are inherent to ICN,  
such as cache pollution, content poisoning, DoS/flooding, and the timing attack.
Furthermore, a review of existing access control mechanisms for ICN has been
neglected.
Wahlisch~\kETAL~\cite{WahSchVah12} discussed the threats and security problems
that arise due to stateful data planes in ICN.
The authors categorized these attacks into three classes: resource exhaustion,
state decorrelation, and path and name infiltration.
Despite presenting a thorough attack classification, this paper did not discuss
any mitigation to the aforementioned attacks.

In~\cite{FotMarPol10Towards}, Fotiou~\kETAL~discussed the security requirements and
threats in pub/sub networks including client privacy, access control, content integrity,
confidentiality, and availability, and subscriber and publisher authentication, and user 
subscription anonymity.
However, they did not propose any solutions. 
Loo~\kETAL~\cite{LooAia15} studied the security challenges faced by the NetInf 
architecture from the perspectives of both applications and infrastructure.
The authors divided their concerns into eight categories: access
control, authentication, non-repudiation, data confidentiality, data integrity,
communication security, availability, and privacy.
However, the descriptions of the problems and proposed solutions are at a high level 
and lack details or scope of future challenges. 

{\bf Novel Contributions of this Survey:} All the existing surveys have either 
not dealt with security, privacy, and access control or have 
looked at them to a very limited extent. 
The work of AbdAllah~\kETAL~\cite{AbdHasZul15} is the first survey dealing with  
security in ICNs, but it is not comprehensive. 
The survey deals more with the generic security concerns, without covering the 
ICN-specific body of the work in depth.
Also, access control in ICNs has not be considered in any survey.
{\em To the best of our knowledge, we are the first to present a comprehensive
survey of the state-of-the-art in security, privacy, and access control in
the context of ICN.}
We present each of these three aspects independently, surveying the state of the 
art, lessons learned, and the shortcomings of proposed approaches.
We also discuss existing challenges and propose potential directions and solutions.

The rest of the paper is organized as depicted in the Fig.~\ref{fig01-02}.
As depicted in the figure, we classify the state of the art in security and privacy 
in terms of attacks and corresponding proposed mitigations. 
As for access control, we divide the state of the art in terms of the mechanism used in the 
proposed solutions, which either address authentication and/or authorization.
In Section~\ref{sec02}, we review the security issues of different ICN
architectures, their proposed solutions, and existing open problems.
Different privacy issues, proposed solutions, and open challenges are presented
in Section~\ref{sec03}.
Access control enforcement mechanisms, their drawbacks, and existing open challenges 
are presented in Section~\ref{sec04}.
In Section~\ref{sec05}, we summarize the state of the art and present a comprehensive 
discussion of future research directions. 
%

\section{Security in ICN}
\label{sec02}
%
%
In this section, we review vulnerabilities in ICN and discuss the state-of-the-art
solutions, then conclude this section with open problems and potential solutions
to be explored. 
This section is divided into subsections based upon the particular types of
attacks.
In Fig.~\ref{fig02-01}, we show our categorization of the state of the art 
in security research. 
We divide the literature in the state of the art into six categories based on the 
particular attack and its mitigation approaches: 
DoS; content poisoning; cache pollution; secure naming, forwarding, and routing; 
application security; and other general contributions (i.e., contributions that 
cannot be grouped into one of the above specific subcategory). 
In the following subsections, we discuss each of these subcategories in detail in the 
order they appear here. 
%
%
%
%
%
\begin{figure*}[!h]
\begin{center}
\begin{tikzpicture}[level distance=1.5cm,
  level 1/.style={sibling distance=5cm},
  level 2/.style={sibling distance=4cm}]
  \node {Denial of Service}
    child {node [align=center] {Rate Limiting}
      child {node [align=center] {Per Face Information \\ ~\cite{AfaMahMoi13,GasTsuUzu13,ComConGas13}}}
      child {node [align=center] {PIT Size Monitoring \\ ~\cite{DaiWanFan13}}}
    }
    child {node [align=center] {Statistical Modeling \\ ~\cite{WanZhoQin14,NguCogDoy15}}
    }
    child {node [align=center] {Other Countermeasures}
      child {node [align=center] {PIT Modification \\ ~\cite{WanCheZho12,WanZhoQin13,WanCheZho14}}}
      child {node [align=center] {Client's Proof-of-Work \\ ~\cite{LiBi14}}}
    };
\end{tikzpicture}
\end{center}
\caption{Denial of Service countermeasure sub-classes and the state-of-the-art.}
\label{fig02-01-DoS}
\end{figure*}
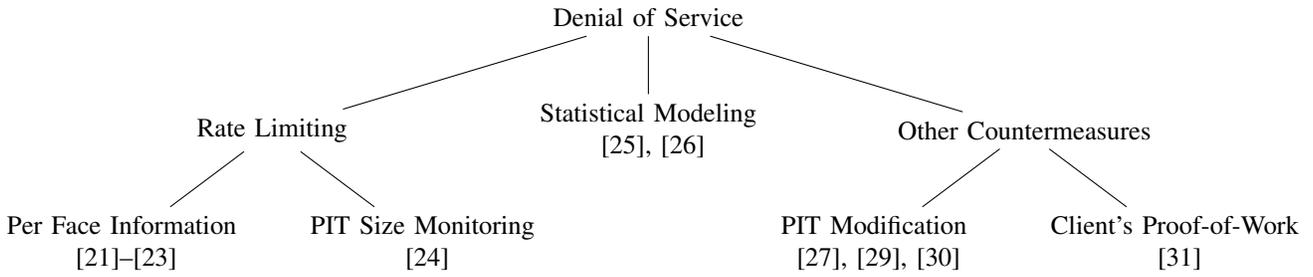
%
\subsection{Denial of Service (DoS) Attack}
\label{sec02-01}
%
DoS attacks aim to overwhelm the network services by inundating them with requests; 
e.g., server(s) inundated with requests for service (content, domain name queries, etc.)~\cite{SpeLee04,MirPriRei02,MirPriRei03}.
In ICN, DoS attacks may target either the intermediate routers or the content providers.
The most basic type of attack, interest flooding, involves an attacker sending
interests for a variety of content objects that are unlikely to exist in
the targeted routers' caches.
This attack applies to pull-based (consumer-driven) architectures such as CCN/NDN, DONA, and NetInf, 
where the intermediate entities are the attack targets (e.g., PIT in CCN/NDN, RH 
in DONA, and NRS in NetInf).
\begin{figure}[!b]
\centering
\includegraphics[height=1.4in]{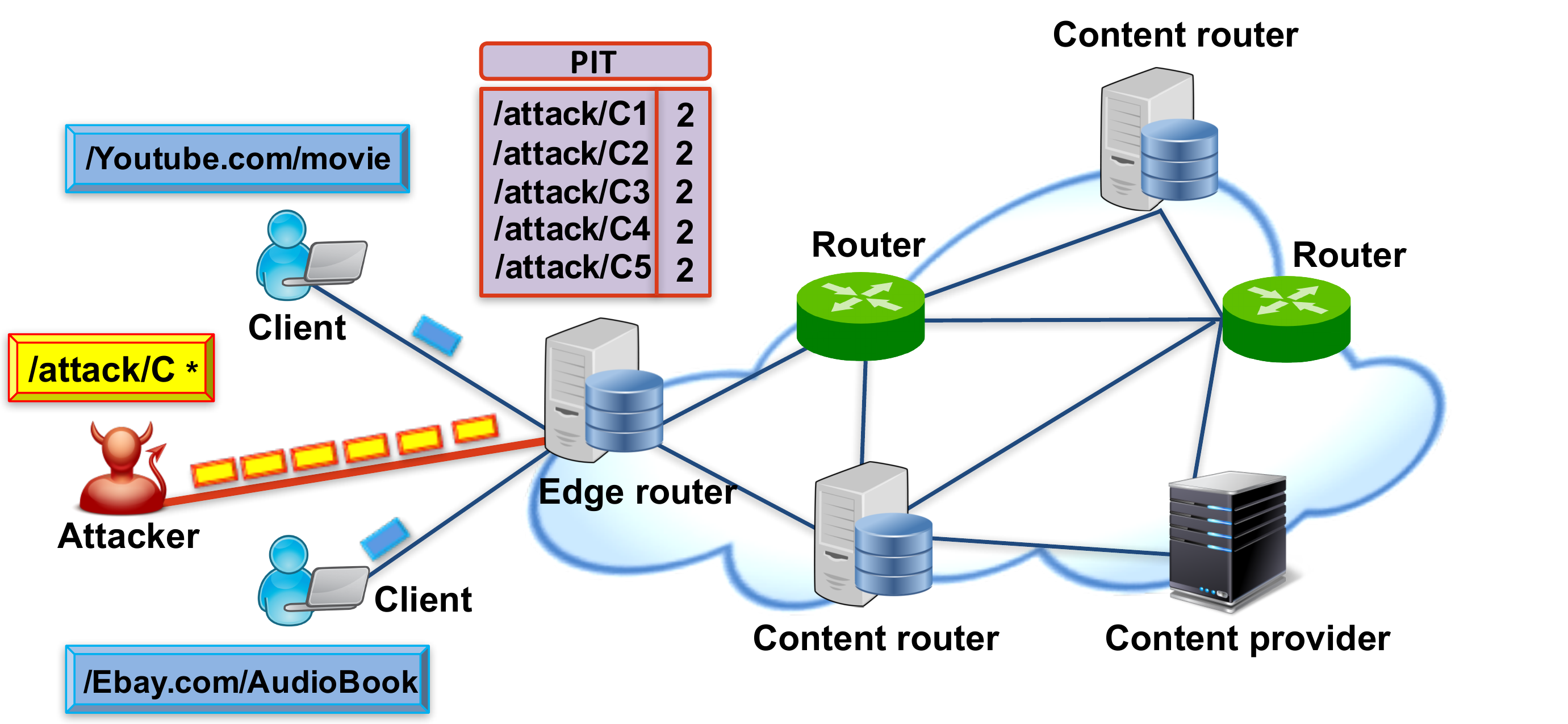}
\caption{Denial of Service (DoS) attack scenario: The attacker fills-up the edge router's PIT 
with a disproportionate number of requests.}
\label{fig02-02}
\end{figure}

The attack scenario in CCN/NDN is depicted in Fig.~\ref{fig02-02}, which shows clients and 
an attacker connected to an edge router, which can cache content. 
The network is composed of a content provider at one end (on the right) and the routing 
core consisting of routers without content cache and the routers with content cache. 
In this scenario, the edge router connected to the attacker as well as legitimate clients has 
its PIT filled up disproportionately by the attacker's interests. 
The interest name {\em /attack/C*} refers to some undefined content name that may not exist, 
is inaccurate, or is a request for dynamic content to be created on-the-fly. 

This attack is more severe when the attacker requests fake content
objects (i.e., names with a valid prefix and an invalid suffix) or dynamic objects, which 
need to be generated by the provider on demand.
Requests for fake objects will result in the provider dropping the interest; 
while the PIT entries on the targeted router(s) (e.g., routers on the path) will only get  
purged on expiration (expiration time can be large for interests).
On the other hand, dynamic content requests will have to be served by the provider. 
However, these requests/replies burden the forwarding routers as well as they may not be aggregated 
(most dynamic content is not popular), and may also cause DoS at the provider.
%

Fig.~\ref{fig02-01-DoS} illustrates the DoS countermeasures categorization.
We categorize the research in DoS mitigation into three broad categories: {\em rate limiting} approaches 
in which a router mitigates DoS attacks by throttling interests it receives from its downstream neighbors; 
{\em statistical modeling} approaches, where a node detects DoS by using statistical information on PIT 
occupancy. 
%
The last category includes several approaches that include using stateless forwarding and client's 
proof-of-work.
%
\subsubsection{{\it {\bfseries Rate Limiting-Based Countermeasures}}}
\label{sec02-01-01}
%
A large body of literature exists on rate limiting-based DoS mitigation approaches in which 
a router detects a DoS attack by monitoring the timeout rates of interests on its faces and/or 
size of its PIT occupied with interests. 
When attack is detected a router limits the interest arrival rate on its suspicious faces.
We sub-categorize the rate limiting approaches further into per-face information monitoring 
and PIT size monitoring approaches.
%
\paragraph{\underline {Per-Face Monitoring Approaches}}
%
In general, in the per-face monitoring approaches, a router stores information, such as the 
number of timed-out interests and the ratio of incoming interests to outgoing content.
Using the collected information, the router detects an ongoing attack and mitigates it by 
rate limiting the faces through which it receives malicious interests.

Afanasayev~\kETAL~\cite{AfaMahMoi13} proposed three approaches to coping with
interest flooding attacks in NDN.
Their vanilla approach is a slight modification of the well-known Token Bucket
algorithm, in which each router limits the number of pending interests for each 
interface proportional to its uplink capacity (bandwidth-delay product).
This technique cannot differentiate between an attacker and a legitimate user's interests. 
Hence an attacker can commandeer the entire uplink capacity with its interests, 
hence reducing the satisfaction rate of legitimate clients' interests.

The authors augmented this vanilla approach by introducing a concept of
per-interface fairness, where the outgoing link capacity is shared fairly
among traffic from all incoming interfaces (each incoming interface has its own queue).
This prevents traffic from a minority of incoming interfaces from consuming 
the entire link capacity.  
%
%
An interface with a high interest arrival rate is subjected to packet queuing for fairness. 
This approach improves fairness, but there is still no distinction between an attacker and 
a legitimate client. 
%
%

The last proposal differentiates interest timeout events from interest
satisfaction events.
Each router gives the interfaces with higher satisfaction rates a greater share of 
the outgoing link capacity.
However, this approach can unduly penalize interfaces that have interests that follow a larger path length. 
The greater the path length, the larger the probability of congestion and interest drops, which reduces 
the satisfaction rate of the corresponding interface. 
Also, with more routers along the path the probability of rate limiting of a flow increases. 
To address this drawback, the authors suggested that routers explicitly announce
their interest satisfaction ratio limits to their downstream neighbors, who can 
accordingly adjust their own acceptance thresholds.
This algorithm, despite being more effective, still applies penalties at the granularity of 
interface, not flow. 
Legitimate users' flows will still suffer.

Gasti~\kETAL~\cite{GasTsuUzu13} also explored DDoS attack scenarios in
NDN, focusing primarily on interest flooding.
The authors divided interest flooding scenarios into classes depending
on whether the attackers request $(1)$ existing or static, $(2)$ dynamically
generated, or $(3)$ non-existent content objects.
The attack target for Types $(1)$ and $(3)$ is only the network-core infrastructure,
while the Type $(2)$ attack targets both content providers and
the network-core.
The authors noted that malicious requests for existing or static content has
limited effect due to content caching at intermediate routers.

In contrast, requesting dynamically generated content not only consumes
intermediate routers' resources (such as PIT space and bandwidth), but also
keeps the providers busy.
It was noted that non-existent content is the type most likely to be
used in attacks against infrastructure.
To mitigate the attack, the authors suggested that routers keep track of the number 
of pending interests per outgoing face, as well as the number of unsatisfied interests 
per incoming face and/or per-name prefix.
Rate limiting is applied when these counters exceed a predefined
threshold.
We note that the per-name prefix based rate limiting is a better approach than per-interface 
rate limiting. 

Compagno~\kETAL~\cite{ComConGas13} designed Poseidon, a collaborative mechanism
for interest flooding mitigation.
Poseidon involves two phases: detection and reaction.
Detection is performed individually at the router which monitors two values over a time window:
ratio of incoming interests to outgoing content, and the amount of PIT state
consumed by each interface.
%
%
When a pre-set threshold is reached the router invokes the collaborative mitigation mode.  
The router rate limits its interfaces with abnormal interest arrival rates and sends 
attack notification to its downstream routers.
This helps downstream routers to detect the attack at an earlier stage.

The authors noted that rate-limiting was more effective at reducing the attacked
router's PIT size than the notification mechanism, however notification improved 
the satisfaction rate of requests.
This mechanism also does not address the differentiation between the attacker and 
the legitimate user. 
Legitimate clients collocated on the same interface with an attacker can be adversely affected.
%
%
\paragraph{\underline {Approaches that Monitor PIT Size}}
%
PIT size growth rate can be used to detect DoS attacks as well. 
In most of the proposed approaches, a router constantly monitors the size of its PIT.
If the PIT size reaches a threshold, the router enters the mitigation phase.

Dai~\kETAL~\cite{DaiWanFan13} proposed an approach inspired by the IP-traceback 
approach for mitigating interest flooding.
The scheme allows an attack to be ``traced back'' to the attacker.
The {\em interest traceback} procedure is triggered when a router's PIT size 
exceeds a predefined threshold.
On trigger, the router generates a spoofed data packet for the
longest-unsatisfied interest in the PIT.
The spoofed data will be forwarded to the attacker, causing its edge router
to be notified of the malicious behavior; in response, the edge router can
rate-limit the attacker's interface.

Similar to other rate-limiting approaches, this mechanism may also have a
negative impact on legitimate clients.
A legitimate client that mistakenly requests a non-existent (or yet-to-be-created) 
content, will be unfairly penalized.
Additionally, since rate limiting only occurs at the edge router, this scheme
may be ineffective if an edge router is compromised or is non-cooperative with its peers.
%
%
\subsubsection{{\it {\bfseries Statistical Modeling-Based Countermeasures}}}
\label{sec02-01-02}
%
The statistical modeling-based approaches rely on statistical information of a router's  
PIT and interfaces to identify an abnormal traffic pattern.
For instance, Wang~\kETAL~\cite{WanZhoQin14} proposed an interest flooding detection and 
mitigation mechanism based on fuzzy logic and routers cooperation.
In the detection part, the core routers monitor their {\it PIT Occupancy Rate} 
(POR) and {\it PIT Expiration Rate} (PER), which represent the rate addition of new entries into a 
PIT and the rate of PIT entry expiration, respectively.
The collected real-time POR and PER values are used through fuzzy inference 
rules to identify if they are normal or abnormal.

If either value is abnormal, the router triggers a mitigation mechanism. 
The router identifies the targeted prefix and the interface
on which the most interests for that prefix have arrived; applies 
rate-limiting to that interface; and notifies its downstream neighbor on the 
interface of the targeted prefix for more rate control. 
Simulation results show the schemes' effectiveness in reducing PIT
memory consumption and increasing legitimate interest satisfaction.
However, the assumption that the attackers only target a specific name
prefix makes mitigation only effective in dismantling attacks against specific
publishers not against the network infrastructure itself.
Moreover, a distributed DDoS attack is still feasible. 
%

Nguyen~\kETAL~proposed an interest flooding detector based on statistical
hypothesis testing theory~\cite{NguCogDoy15}.
The scheme is based upon the fact that when under attack, the interest rate on
an interface is greater than that during normal conditions.
Meanwhile, the data rate under both hypotheses remains the same; therefore, the
data hit-ratio in attack scenarios is lower than that in normal conditions.
Unlike other solutions, this scheme takes the desired false alarm probability 
as a parameter and calculates the detection threshold accordingly.
However, the evaluation uses only a simple binary tree graph with eight 
clients and one attacker. The effectiveness of the scheme for larger networks 
or during distributed attacks is difficult to analyze.
\begin{table*}[!ht]
\centering
\caption{Classification of DoS/DDoS Mitigation Approaches and Their Salient Features}
\label{table:02_01}
\begin{tabular}{|l c c c c c|}
 \hline
 {\bf Mechanism}&{\bf Target}&{\bf Content Type}&{\bf Mitigation Approach}&{\bf Router's Functionality}&{\bf Scope} \\
 \hline
 \hline
 {\bf Rate Limiting} & & & & & \\
 \hline
 \multirow{2}{*}{Afanasayev~\kETAL~\cite{AfaMahMoi13}} & \multirow{2}{*}{Router} & \multirow{2}{*}{Non-Existent} & Rate Limiting \& Per-face Fairness & PIT Extension & Individual Routers \\
 & & & Per-face Statistic \& Priority & Storing Statistics & Router Collaboration\\
 \multirow{2}{*}{Gasti~\kETAL~\cite{GasTsuUzu13}} & Provider & Dynamic & \multirow{2}{*}{Rate Limiting \& Per-face Statistics} & \multirow{2}{*}{Storing Statistics} & \multirow{2}{*}{Individual Routers} \\
 & Router & Existing \& Non-Existent & & &\\
 Compagno~\kETAL~\cite{ComConGas13} & Router & Non-Existent & Rate Limiting \& Per-face Statistics & Storing Statistics & Router Collaboration \\
 Dai~\kETAL~\cite{DaiWanFan13} & Router & Non-Existent & Rate Limiting \& PIT Size Monitoring & Not Applicable & Router Collaboration \\
 \hline
 \hline
 {\bf Statistical Modeling} & & & & & \\
 \hline
 Wang~\kETAL~\cite{WanZhoQin14} & Router & Non-Existent & Fuzzy Logic-based Detection & Storing Statistics & Router Collaboration \\
 Nguyen~\kETAL~\cite{NguCogDoy15} & Router & Non-Existent & Statistical Hypotheses Testing Theory & Storing Statistics & Individual Routers \\
 \hline
 \hline
 {\bf Other Countermeasures} & & & & & \\
 \hline
 Wang~\kETAL~\cite{WanCheZho12} & Provider & Existing & Caching Period Increase & Not Applicable & Individual Routers \\
 Wang~\kETAL~\cite{WanZhoQin13} & Router & Non-Existent & Decoupling Malicious Interest from PIT & Additional Queue & Individual Routers \\ 
 Wang~\kETAL~\cite{WanCheZho14} & Router & Existing & Bigger PIT and Cache & Not Applicable & Individual Routers \\
 Li~\kETAL~\cite{LiBi14} & Provider & Dynamic & Client's Proof-of-Work per Interest & Not Applicable & Not Applicable \\ 
 \hline
\end{tabular}
\end{table*}
%
\subsubsection{{\it {\bfseries Other Countermeasures}}}
\label{sec02-01-03}
%
This category of DoS mitigation includes approaches that change routers' 
structures, such as PIT and content store, or inherently reduce the clients' request rates 
by requesting proof-of-work.
%
\paragraph{\underline {Approaches that Modify Router's PIT or Cache}}
%
The approaches in this category focus on DoS attacks targeting the routers' PITs.
The solutions proposed include augmenting the routers with bigger PIT, longer caching period, 
and removing suspicious interests from routers' PITs.
For instance, Wang~\kETAL~\cite{WanCheZho12} investigated the effect of content caching on
DoS attacks, focusing on CCN in particular.
They compared the DoS attacks targeting content providers in IP-based
and content-centric networks, and proposed a queuing theory based model 
for DoS attacks modeling. 
This model considers the caching period of content objects as well as queuing
delay at repositories.
The authors concluded that DoS attacks in CCN (also applies to NDN) have 
limited effectiveness in comparison to DoS attacks on IP networks due 
to satisfaction on interests at intermediate routers.
%
Due to this phenomenon, interest flooding can be localized significantly by increasing 
routers cache sizes and the timeout period of content in caches.

Despite the correctness of the authors' models, the authors use several unrealistic 
assumptions.
%
The authors assumed that an attacker only requests content objects that are
available at the content provider(s) and may be cached.
However, this is not a complete attack scenario; an attacker 
can request either non-existent content or dynamically-generated content 
(which may be unpopular and hence useless when cached).
Also, the analysis provided does not account for cache replacement policies,
which would affect the content caching period.
Furthermore, intermediate routers would be more vulnerable targets to DoS than
content providers.
However, the impact of DoS on routers was not discussed.

Virgilio~\kETAL~\cite{VirMarSis13} analyzed the security of the existing
PIT architectures under DDoS attack.
The authors compared three proposed PIT architectures: $(1)$ SimplePIT, which
stores the entire URL, $(2)$ HashPIT, where only a hash of the URL is stored, 
and $(3)$ DiPIT (distributed PIT), where each interface uses a Bloom
filter to determine which content objects should be forwarded.
The authors concluded that all three proposed PIT architectures are vulnerable
to DDoS attack, and they all perform the same under normal traffic conditions.
While SimplePIT and HashPIT suffer from memory growth in the face of DoS,
DiPIT does not consume extra memory.
The Bloom filter's inherent false positive rate has the potential to cause data 
to be forwarded unnecessarily, and therefore waste bandwidth.
Although this paper showed the effects of DDoS on different PIT architectures
through simulation, the authors did not propose any viable solution.

Wang~\kETAL~\cite{WanZhoQin13} proposed a mechanism which copes with interest
flooding by decoupling malicious interests from the PIT.
The mechanism requires that each router monitors the number of expired interests
for each name-prefix, then adds a prefix to the malicious list (m-list) if this
count exceeds a chosen threshold. 
To prevent legitimate name-prefixes from staying in the m-list, each m-list
entry is assigned an expiry time, after which the prefix is removed from the
m-list.
However, an m-list entry's expiry timer is reset if a new interest arrives for
the same prefix.

The authors overcome the extra load on the PIT table size by putting information 
in the interest.
Although this helps routers keep the sizes of their PITs manageable, they
will still be responsible for forwarding the malicious interests; thus network
congestion and starvation of legitimate clients are still possible.
This mechanism also puts additional processing burden on the routers and 
increases packet overhead. 

Wang~\kETAL~\cite{WanCheZho14} modeled the interest flooding attack in NDN by 
considering factors, such as routers' PIT sizes, round trip times, PIT
entries' TTLs, content popularity distribution, and both malicious and
legitimate interest rates.
The authors derived a DoS probability distribution, which evaluates the 
probability that a legitimate interest will be dropped due to starvation. 
Simulation results confirmed the validity of the model.
The authors suggested that the effectiveness of DoS could be reduced by using
bigger PITs, bigger content stores, and shorter TTLs for PIT entries.
Nonetheless, these suggestions do not actually address the problem: an
attacker could easily increase its request rate proportionally.
%
%
\paragraph{\underline {Approaches that Require Client's Proof-of-Work}}
%
Proof-of-work approaches, reduce the request rate from clients (because of the 
delay in obtaining the proof) and serve as a barrier which only serious clients 
will overcome to use the network. 
In the ICN literature, there has been one such work. 
Li and Bi~\cite{LiBi14} proposed a DoS countermeasure for dynamic content requests using 
proof-of-work.
As opposed to static content, which is signed once when it is generated, a dynamic
content object is generated and signed upon interest arrival.
A high rate of dynamic content requests can thus overload the content provider 
with signature computation, causing DoS.
To deter potential attackers the authors proposed a proof-of-work mechanism where 
the client requests a meta-puzzle from the content provider.
Upon receiving the meta-puzzle, the client generates the actual puzzle and
solves it (similarly to how blocks are mined in Bitcoin).
The puzzle solution and the current timestamp form a part of the interest, 
which is verified by the provider. 
%

\subsubsection{{\it {\bfseries Summary and Future Directions in DoS Mitigation}}} 
\label{sec02-01-04}
In Table~\ref{table:02_01}, we summarize all the proposed DoS mitigation
mechanisms in terms of the entity implementing the mechanism, whether the attack
model involves existent, dynamic, or non-existent content requests, the nature of the 
mitigation approach, the extra functionality needed in the routers, and the 
level of collaboration required between routers.
DoS attacks, in general, either target the routers~\cite{AfaMahMoi13,ComConGas13,DaiWanFan13,WanZhoQin13,WanCheZho14,NguCogDoy15} 
and/or the content providers~\cite{WanCheZho12,GasTsuUzu13,LiBi14}.
An attacker tries to exhaust either the routers' PITs or content providers' resources by requesting dynamic or 
non-existent content with a high rate, which causes unbounded service delays for legitimate clients.

The majority of the proposed solutions~\cite{AfaMahMoi13,ComConGas13,DaiWanFan13,GasTsuUzu13}, especially against the interest 
flooding based DoS attacks, are variants of a rate limiting mechanism on the suspicious interfaces or name prefixes.
The major drawback of the rate limiting based solutions is that they may penalize legitimate clients also. 
No scheme performs per-flow based rate-limiting, which has the highest fairness. 
The closest is the approach by Gasti~\kETAL~\cite{GasTsuUzu13} where prefix based rate-limiting was proposed. 
There is need for more fine-grained rate-limiting to better distinguish malicious from benign requests. 

Other proposed mechanisms including per-interest client's proof-of-work~\cite{LiBi14}, fuzzy logic-based detection~\cite{WanCheZho14}, 
statistical hypotheses testing theory~\cite{NguCogDoy15}, and increasing the caching time~\cite{WanCheZho12} have also been proposed 
to solve the problem. 
However, these mechanisms either require storage of per content statistics at the routers or are not computationally scalable, 
especially in the real time.
A better mechanism may be one that removes the suspicious requests from the PIT~\cite{WanZhoQin13}, similar to the 
publish-subscribe Bloom filter based self-routing~\cite{PSIRP,PURSUIT}.
This mechanism can be augmented by adopting a self-routing approach for the suspicious interests and using the available 
stateful routing for the legitimate interests.

Another potential direction is employing a software-defined networking (SDN) approach in which a network controller with an 
overall aggregated view of the network detects and mitigates the DoS attack in its early stages.
It can be achieved by the collaboration of routers at different levels of the network hierarchy, specifically for filtering 
the communication flows that share malicious name prefixes.
Exploiting a more sophisticated interest aggregation method, which aggregates the malicious interests with same prefix 
(regardless of their suffixes) into one PIT entry, can also slow down the PIT exhaustion.
We also believe some of the current IP-based detection and defense mechanisms~\cite{ZarJosTip13} might be relevant for 
ICN DoS mitigation.
This is a significant area of interest.

An attacker can orchestrate a DoS attack in publish/subscribe networks by manipulating the z-filter 
in a content packet.
This causes each intermediate router to forward the packet to all of its interfaces, creating congestion 
in the network.
However, DoS attack in publish/subscribe networks has not received much attention from the community, except the work proposed by 
Alzahrani~\kETAL~\cite{AlzReeVas12, AlzVasRee13securing}.
We believe that DoS in publish/subscribe networks is a legitimate security concern, which requires more in depth analysis and solutions.

All the proposed mechanisms try to address interest flooding in CCN and NDN architectures.
However, the rate limiting and proof of work approaches can be applied to other architectures, where the attacker targets 
the intermediate entities such as DONA's resolution handler and NetInf's name resolution server.
%
%

%
%
%
%
%
\subsection{Content Poisoning Attack}
\label{sec02-02}
%
The objective of the {\it content poisoning attack} is to fill routers' caches with
invalid content.
To mount this attack, an attacker must control one or more intermediate routers to 
be able to inject its own content into the network.
The injected content has a valid name corresponding to an interest, but
a fake payload or an invalid signature.
This attack is applicable to all ICN architectures, however, it is less effective 
in architectures using self-certifying names. 
With self-certifying names the digest of the packet's content is the name of the packet. 
Thus it is easier to verify the correctness of a content chunk by comparing the hash of the 
chunk against the digest and drop packets whose hash does not match. 
%
%
%
\begin{figure}[!ht]
\centering
\includegraphics[height=1.4in]{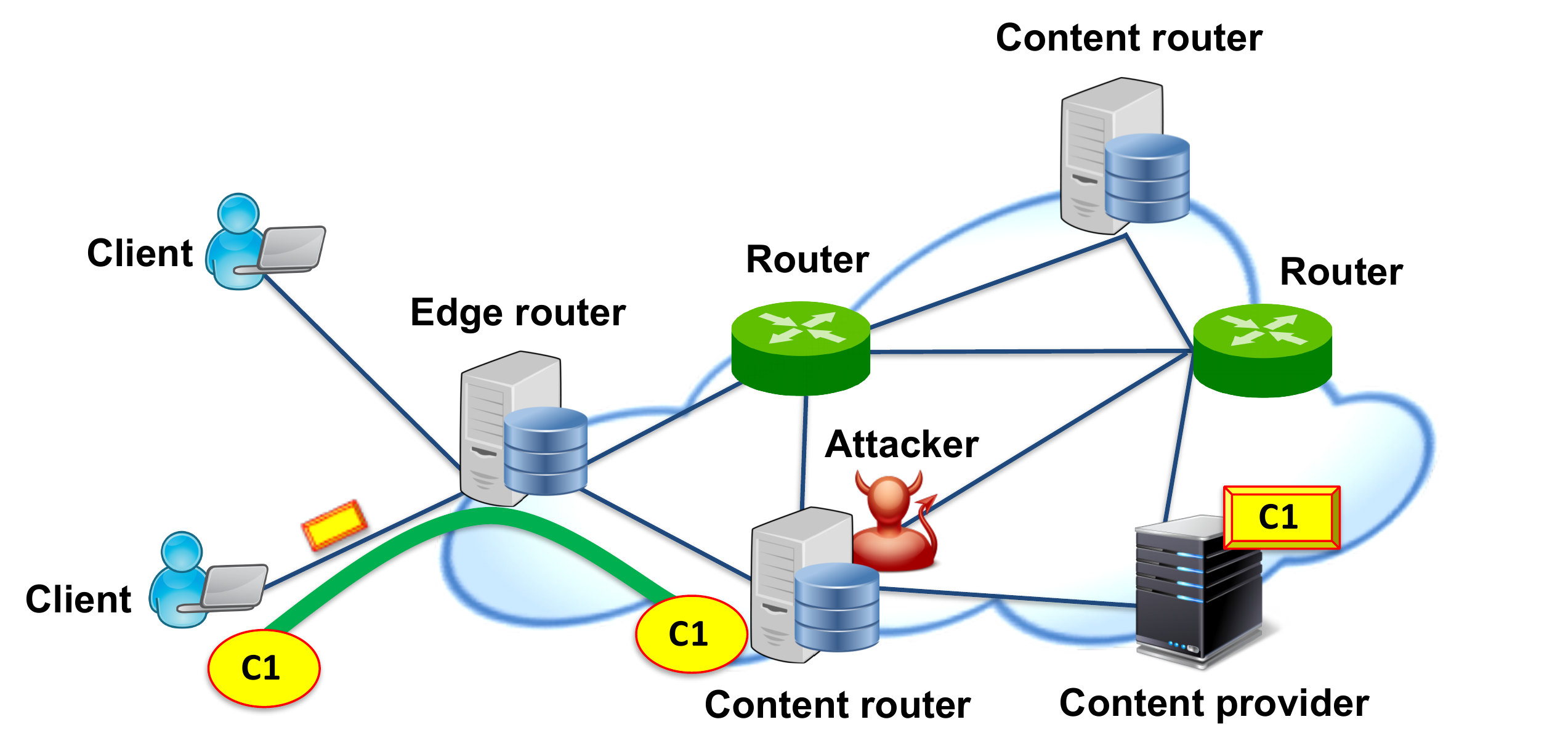}
\caption{Content poisoning attack scenario.}
\label{fig02-03}
\end{figure}

We illustrate The poisoning attack in Fig.~\ref{fig02-03}. 
The attacker is one of the routers on the path between the client and provide 
returning an invalid content (oval C1) instead of the genuine content
(double-border rectangle C1) corresponding to the requested name.
This attack can have potentially devastating consequences: unless the content are validated an attacker 
can fill the network with poisoned content objects, while useful content find no place in the caches.

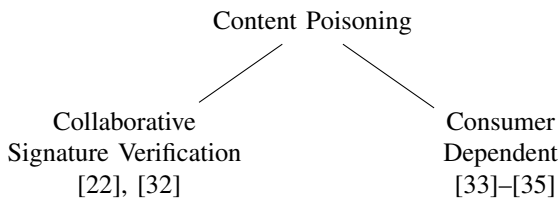
\begin{figure}[!ht]
\begin{center}
\begin{tikzpicture}[level distance=1.75cm,
  level 1/.style={sibling distance=5cm}]
  \node {Content Poisoning}
    child {node [align=center] {Collaborative \\ Signature Verification \\ ~\cite{GasTsuUzu13,KimNamBi15}}
    }
    child {node [align=center] { Consumer \\ Dependent \\ ~\cite{GhaTsuUzu14,GhaTsuUzu14elements,GhaTsuUzu14Network-layer}}
    };
\end{tikzpicture}
\end{center}
\caption{Content poisoning countermeasure sub-classes and the state-of-the-art.}
\label{fig02-02-Poisoning}
\end{figure}

Fig.~\ref{fig02-02-Poisoning} illustrates our categorization of content poisoning countermeasures.
The first category, {\it collaborative signature verification}, refers to those mechanisms in which 
routers cooperate with each other to verify the content signature.
The {\it consumer dependent} category includes those approaches that either rely on using additional 
fields in request and data packets or clients' feedback.
We start with the first category.
%
%
%
%
\subsubsection{{\it {\bfseries Collaborative Signature Verification Countermeasures}}}
\label{sec02-02-01}
%
This category refers to the approaches that propose router verification of signatures of packets they forward.
To distribute and reduce the load of signature verification, the routers flag the verified chunks 
to signal their peers that the packet has been validated, and/or verify the signature of the chunks upon cache hit 
(only verify popular content).

Gasti~\kETAL~\cite{GasTsuUzu13} were the first to discuss the content/cache
poisoning attacks.
As their first countermeasure, the authors suggested the use of a
``self-certifying interest/data packet'' (SCID) to help routers validate
received content chunks.
Before sending an interest, a client is required to obtain the desired chunk's
hash, name, and signature from the content provider. 
This information is attached to the interest.  
On obtaining a content chunk, a router can check its validity
by comparing its hash to the hash from the interest information it has.
This method is less computationally intensive than traditional RSA signature
verification, however it requires the client to obtain the hashes for each data 
chunk/packet beforehand and for the routers to store them until verification.
This increases content retrieval latency and router storage overhed, thus limiting 
scalability.

As an improvement, the authors proposed cached content signature
verification by routers.
In the basic version, each router randomly selects and verifies content chunks, 
dropping those whose signatures cannot be validated. 
To prevent redundant verification, routers collaboratively select a range of
content chunks to verify. 
The scope of this collaboration can vary from a neighborhood to an organization.
To reduce collaboration overhead, the authors also suggested client feedback
based decision-making in which a client may inform its edge router about each
content chunk's validity.
However, this type of feedback can also be used by 
malicious clients to mislead routers by reporting legitimate content
objects as fake, or vice-versa.
\begin{table*}[!ht]
\centering
\caption{Content Poisoning Countermeasures are Classified to Collaborative Verification and Consumer Dependent Classes}
\label{table:02_02}
\begin{tabular}{|l c c|}
 \hline
 {\bf Mechanism}&{\bf Mitigation Approach}&{\bf Overhead} \\
 \hline
 \hline
 \multicolumn{2}{|l}{{\bf Collaborative Signature Verification}} & \\
 \hline
 Gasti~\kETAL~\cite{GasTsuUzu13} & Self-Certifying Interest \& Collaborative Signature Verification & Hash Value Comparison \& Random Signature Verification \\
 Kim~\kETAL~\cite{KimNamBi15} & Collaborative Signature Verification of Serving Content & Signature Verification on Cache Hit \\
 \hline
 \hline
 {\bf Consumer Dependent} & & \\
 \hline
 Ghali~\kETAL~\cite{GhaTsuUzu14} & Client Feedback, Content Ranking & Content Ranking Calculation \\
 Ghali~\kETAL~\cite{GhaTsuUzu14elements,GhaTsuUzu14Network-layer} & Interest-Key Binding \& Adding the Provider's Public key to the Content & PPKD Comparison \& Signature Verification \\
 \hline
\end{tabular}
\end{table*}
The mechanism proposed by Kim~\kETAL~\cite{KimNamBi15} was inspired by check before 
storing (CBS)~\cite{BiaDetCap13}, which probabilistically verifies content items, only storing 
validated content items in the cache.
The authors measured that generally around 10\% of the cached contents 
are requested again before their expiration from their caches.
Hence, they divided the cache into serving content, which will be requested while 
they are cached, and by-passing content, which will be dropped from the cache before 
subsequent interests.

The authors used a segmented LRU policy for cache replacement: a content is initially put 
in the by-passing content segment of the cache. 
%
The proposed countermeasure only verifies the signature of a serving content, that is a content that has a cache hit. 
At that point the content's signature is verified and it is moved to the serving content cache segment.
%
%
To avoid multiple verifications of a chunk, the verified chunk is marked in the serving content cache segment. 

The authors simulations showed that the approach resulted in a reduction in the number of poisonous 
content cached; however, the scheme has some drawbacks. 
Any chunk that is requested twice still needs to be verified, thus adding to the latency and computation. 
An attacker can enforce verification of every fake content, by requesting it twice; at scale this could lead to 
a DoS/DDoS attack. 
The authors show that with an increase in the serving content cache segment proportion the overall 
content hit rate goes down. 
But they do not mention if this reduction in hit-rate is for fake content or for usable serving 
content; this could have a significant bearing on system efficiency.   
%
%
%
\subsubsection{{\it {\bfseries Consumer Dependent Countermeasures}}}
\label{sec02-02-02}
%
In the consumer dependent countermeasures, the clients either give feedback on the 
legitimacy of the received content or include the providers' public keys 
in their request packets to enable verification.
Ghali~\kETAL~\cite{GhaTsuUzu14} proposed a content poisoning mitigation
mechanism while introducing an updated definition for fake content.
The authors defined a fake content as one with a valid signature using the wrong
key, or with a malformed signature field.
The authors discussed the applicability of existing solutions such as signature
verification by intermediate routers, which is infeasible at line speed.
On the other hand, although self-certifying names are more efficient as a 
countermeasure, issues such as efficient content hash retrieval and handling of 
dynamic content objects need solutions.
Hence, the authors proposed a ranking mechanism for cached content using exclusion-based 
feedback.

Exclusion is a selector feature in the CCN and NDN architectures~\cite{Exclusion}, 
which allows a client to exclude certain data (either by hash or name suffix) from 
matching its interest, effectively overriding a match on the requested name's prefix.
Clients can use this feature to avoid receiving data objects that are known to
be unwanted, corrupted, or forged. 
In the proposed approaches, a detector function ranks content based on three factors: 
number of exclusions, exclusion time, and exclusion-interface ratio.
The exclusion time defines the recency of a particular data name exclusion.   
A content goes down in rank if it has more exclusions, a recent exclusion, or 
if the router receives exclusion feedback for it from multiple clients on 
different interfaces.
To overcome poisoning, if a router has multiple cached contents with names that match that requested in the interest, 
then the router returns the highest ranked content.

The drawbacks of this approach are: it is highly dependent on client feedback; 
non-cooperative and/or malicious clients can undermine its effectiveness; storage of 
multiple copies of same content undermines cache efficiency.
Furthermore, the exclusion feature is not present in all ICN architectures.

Ghali~\kETAL~\cite{GhaTsuUzu14elements, GhaTsuUzu14Network-layer} noted that 
content poisoning mitigation is contingent on network-layer trust management.
According to them, cache poisoning attack in ICN is due to interest ambiguity and lack of a trust model.
The former arises from the interest packet structure, which considers the content name as 
the only compulsory field, while neglecting two other fields, the content digest and the 
publisher public key digest (PPKD). 
The latter refers to the lack of a unified trust model at the network layer.

As a solution, the authors suggested to clarify interest ambiguity by adding a binding between content 
name and the provider's public key, an Interest-Key Binding (IKB), to the interest packet.
The only modification at the content provider is the addition of the provider's public 
key to the content's KeyLocator field.
An intermediate router, upon receiving a content, matches the hash of the public 
key present in the KeyLocator field with the interest's PPKD (available in the PIT).
The content will be forwarded if these match, and will be discarded otherwise.

The client-side complexity of this approach is in obtaining the provider's public key in advance.
In order to bootstrap a trust model, the authors proposed three approaches: a pre-installed public 
key in the client's software application, a global key name service similar to DNS, and a global 
search-based service such as Google.
To reduce core routers' workload, the authors proposed that edge routers perform the IKB check 
for all content packets, while core routers randomly verify a subset of content packets.
Nevertheless, this mechanism does not scale. 
Signature verification, which is a public key infrastructure (PKI) based verification, is slow 
and cannot be performed at line speed, even if only some randonly chosen routers or only edge routers 
perform the verification.
Some other weaknesses of the mechanisms proposed by the authors include the assumption that the 
verifying router is trusted--perhaps the router is malicious, then it can verify an incorrect IKB 
to be correct~\cite{GasTsuUzu13, GhaTsuUzu14, GhaTsuUzu14elements, GhaTsuUzu14Network-layer}. 
Further, the schemes lacked detailed analysis of scalability and overhead.   
%

\subsubsection{{\it {\bfseries Summary and Future Directions in Content Poisoning Mitigation}}} 
\label{sec02-02-03}
Table~\ref{table:02_02} summarizes the basic techniques used in the proposed countermeasures 
and their overheads.
In this attack, the attacker's goal is to fill the routers' caches with fake contents, that 
are either content with valid names and invalid payloads or content with invalid signatures.
All of the proposed mechanisms require the intermediate routers to verify the data packets' 
signatures~\cite{GasTsuUzu13,KimNamBi15}, compare the content hash in interest and data 
packets~\cite{GasTsuUzu13,GhaTsuUzu14elements,GhaTsuUzu14Network-layer}, or to rank the 
contents based on the clients' feedback~\cite{GhaTsuUzu14}.
Signature verification approaches suffer from delays, which undermine scalability.
The client feedback based content ranking approach can be undermined by malicious clients.

We believe that the hash verification based approach is the more promising approach on account 
of low amortized cost to intermediate routers. 
More study need to be conducted to identify a suitable cryptographic hash function. 
Another approach is to trace the fake content back to its origin by leveraging the history of 
each interface on the route.
After successfully detection of the attack origin, a mitigation mechanism can be orchestrated. 
For instance, a router may prevent caching the content chunks that arrive from a suspicious 
interface or have the same name prefixes as the fake content.
We believe that there is still need for more efficient and scalable mitigation approaches. 
%

%
%
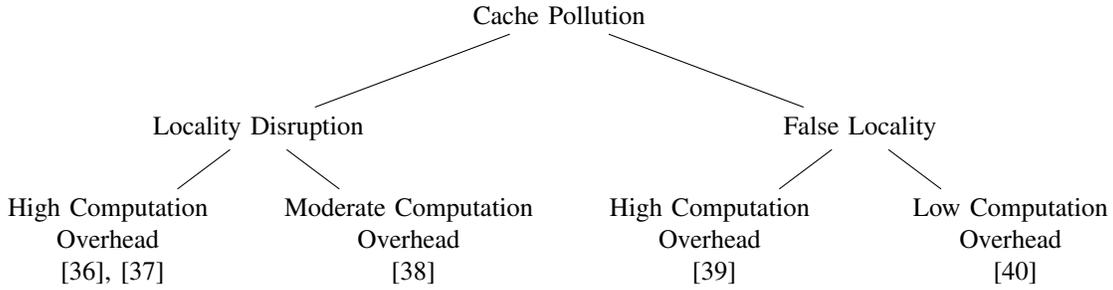
\begin{figure*}[!ht]
\begin{center}
\begin{tikzpicture}[level distance=1.5cm,
  level 1/.style={sibling distance=8cm},
  level 2/.style={sibling distance=4cm}]
  \node {Cache Pollution}
    child {node [align=center] {Locality Disruption}
      child {node [align=center] {High Computation \\ Overhead \\ ~\cite{ParWidLee12,XieWidWan12}}}
      child {node [align=center] {Moderate Computation \\ Overhead \\ ~\cite{ConGasTeo13}}}
    }
    child {node [align=center] {False Locality}
      child {node [align=center] {High Computation \\ Overhead \\ ~\cite{KarGue15anfis}}}
      child {node [align=center] {Low Computation \\ Overhead \\ ~\cite{MauRasGer15}}}
    };
\end{tikzpicture}
\end{center}
\caption{State-of-the-art in cache pollution countermeasures.}
\label{fig02-03-Pollution}
\end{figure*}
%
%
\subsection{Cache Pollution Attack}
\label{sec02-03}
%
Caching in ICN is effective, especially if the universe of on 
the Internet follows a popularity distribution (e.g., Zipf distribution), where a small number of popular 
contents are requested frequently, while the rest of the contents are requested sparingly.
The popular (frequently requested) contents can be caches in the network, thus reducing request 
latency and network load. 
However, an attacker can undermine this popularity based caching by skewing content popularity 
by requesting less popular content more frequently. 
%
%
This is the {\it cache pollution attack}. 

In this subsection, we explore two classes of cache pollution attacks: {\em locality disruption} 
and {\em false locality}.
In the locality disruption attack, an attacker continuously requests new, unpopular contents to 
disrupt cache locality by churning the cache. 
In the false locality attack, on the other hand, the attacker's aim is to change the popularity 
distribution of the local cache to favor a set of unpopular contentx by repeatedly requesting the 
unpopular contents set. 
In principle, this attack is feasible in all ICN architectures. 
However, in publish/subscribe architectures (e.g., PSIRP and PURSUIT) the attack may have minimal impact. 
The one-time subscription mechanism used in publish/subscribe architectures means a subscriber 
cannot artificially increase a content's popularity by requesting it multiple times.

Fig.~\ref{fig02-03-Pollution} illustrates the cache pollution attacks categorization: locality disruption and false locality.
%
%
The attack countermeasures are further subcategorized according to their computation overhead 
at the intermediate routers. 
%
We note that the approach proposed by Karami~\kETAL~\cite{KarGue15anfis} addresses both locality 
disruption and false locality threats.
\begin{table*}[!h]
\centering
\caption{Cache Pollution Countermeasures Classified to Locality Disruption and False Locality Classes}
\label{table:02_03}
\begin{tabular}{|l c c c c|}
 \hline
 \multirow{2}{*}{{\bf Mechanism}}&\multirow{2}{*}{{\bf Detection \& Mitigation Approaches}}&\multirow{2}{*}{{\bf Attack Type}}&\multicolumn{2}{c|}{{\bf Router's Overhead}} \\
 \cline{4-5}
 & & & {\bf Storage} & {\bf Computation} \\
 \hline
 \hline
 {\bf Locality Disruption} & & & & \\
 \hline
 Park~\kETAL~\cite{ParWidLee12} & Cached Content Matrix Ranking & Low-rate Locality Disruption & Low & High \\
 Xie~\kETAL~\cite{XieWidWan12} & Probabilistically Caching Popular Content & Locality Disruption & Moderate & High \\
 Conti~\kETAL~\cite{ConGasTeo13} & Random Content Sampling for Attack Threshold Detection & Locality Disruption & Low & Moderate \\
 \hline
 \hline
 {\bf False Locality} & & & & \\
 \hline
 Karami~\kETAL~\cite{KarGue15anfis} & Adoptive Neuro-Fuzzy Inference System Replacement Policy & Locality Disruption \& False Locality & Moderate & High \\ 
 Mauri~\kETAL~\cite{MauRasGer15} & Honeypot Installation \& Hidden Monitoring & False Locality (by Content Provider) & Moderate & Low \\
 \hline
\end{tabular}
\end{table*}
%
%
\subsubsection{{\it {\bfseries Locality Disruption Mitigation Approaches}}}
\label{sec02-03-01}
%
In the proposed approaches to mitigate locality disruption, the routers either cache the content with certain popularity
(attack prevention) or have to periodically evaluate the popularity of their cached content (attack detection).
We subcategorized these prevention and mitigation mechanisms based on their computation overhead on the routers into high 
and moderate subcategories.
%
%
\paragraph{\underline {Approaches with High Computation Overhead}}
%
Several proposed locality disruption mitigation approaches require complex and iterative procedures per content 
caching decision at intermediate routers, thus incurring high computation overhead.
%
%
For instance, Park~\kETAL~\cite{ParWidLee12} proposed a cache pollution detection scheme based on randomness check.
The iterative scheme takes advantage of matrix ranking and sequential analysis for detecting 
a low-rate pollution attack: an attacker requesting chunks at a low rate to bypass any 
rate filters.
The detection scheme starts with the routers mapping their cached content onto an $n\times n$ binary matrix 
$M$, where $n \simeq [\sqrt S_c]$ and $S_c$ is the average number of cached contents. 
The authors employ two cryptographic hash functions for mapping a content name to location in the matrix and 
evaluate its rank $M$. 
The ranking process is iterated $k$ times, and the attack alarm is triggered if the matrix-rank reaches a 
pre-defined threshold.
Due to its focus on low-rate attacks the scheme does not consider popular contents, which are removed 
from consideration.

The authors showed the effectiveness of their scheme in detecting low-rate locality-disruption attacks.
However, this scheme is not applicable to the harder to detect false locality attack.
Furthermore, the proposed approach is computationally heavy for the caching routers.

Xie~\kETAL~\cite{XieWidWan12} proposed {\it CacheShield}, a mechanism providing robustness against 
the locality disruption attack.
It is composed of two main components: a probabilistic shielding function, and a vector of content 
names and their corresponding request frequencies.
When a router receives a request for a content chunk, if the chunk is in its CS, it replies with 
the content.
Otherwise, the router forwards the interest towards the provider.
When a chunk arrives at the router, the shielding function defined as, $1/(1+e^{\frac{p-t}{q}})$, 
where $p$ and $q$ are pre-defined system-wide constants and $t$ denotes the $t^{th}$ request for 
the given chunk, is used to calculate the probability of placing the content in the CS. 

If the chunk is not placed in the CS, then the router either adds the chunk's name with a frequency 
of one in the vector of content names, if it does not exist; 
if the name exists, then the frequency is incremented by one. 
A chunk is placed in the CS when the request frequency of the exceeds 
a pre-defined threshold.
This approach suffers from the fact that the shield function's parameters $p$ and $q$ are constants and 
can be easily deduced (if not known), and hence an attacker can easily calculate the value of $t$. 
Then the attacker has to just ensure that it requests the unpopular contents more than $t$ times.  
Additionally, the portion of the CS used to store the name vector adds to the storage overhead. 
%
%
\paragraph{\underline {Approaches with Moderate Computation Overhead}}
%
There are other proposed approaches that use only a subset of the content at a router to perform attack 
detection, hence do not suffer from high overhead. 
For instance, to overcome the shortcomings of CacheShield, Conti~\kETAL~\cite{ConGasTeo13} proposed a machine-learning approach.
They evaluated the impact of cache pollution attacks on different cache replacement policies and network 
topologies.
They proposed a detection algorithm, which operates as a sub-routine of the caching policy.
The algorithm is composed of a learning step and an attack-testing step.
It starts by checking the membership of an arrived content in a sample set chosen from the universe of contents.
If the content belongs to the sample set, the learning step will be triggered with the goal of identifying 
an attack threshold (defined as $\tau$) for evaluating the contents. 

The value of $\tau$ is used by the attack test sub-routine in the testing step.
The attack test sub-routine compares the calculated $\tau$ with another value $\delta_m$, which is a 
function with parameters, such as content request frequency and the size of the measurement interval, 
of all contents in the sample set. 
If $\delta_m$ is greater than $\tau$, then the mechanism detects an attack. 
The drawback of this approach is that it only detects the attack, but does not identify the attack 
interests, or content chunks. 
Further, the assumption that the adversary's content requests can only follow a uniform distribution is 
simplistic and may not reflect the reality.  
%
%
%
\subsubsection{{\it {\bfseries False Locality Mitigation Approaches}}}
\label{sec02-03-02}
The false locality attack can be orchestrated by malicious consumers and/or producers.
A malicious consumers' goal is to alter the content popularity in the local caches, while malicious producers' 
intent is to store its content in the routers' caches.
As with the cache pollution attack, we subcategorize the proposed countermeasures into high and low 
computational overhead.
%
\paragraph{\underline {Approach with High Computation Overhead}}
%
%
Karami~\kETAL~\cite{KarGue15anfis} proposed an Adaptive Neuro-Fuzzy Inference System (ANFIS) based cache 
replacement policy resilient to cache pollution.
The policy has three stages: input-output data pattern extraction, accuracy 
verification of the constructed ANFIS structure, and integration of the structure as a cache 
replacement policy.
In the first stage, an ANFIS structure is constructed according to the properties of the cached content.
Variables such as a content's time duration in cache, request frequency, and standard deviation of the 
request frequency, are all fed into a nonlinear system.
%
The system returns a goodness value between $0$ and $1$ per content ($0$ indicates false-locality, $0.5$ 
indicates locality-disruption, and $1$ indicates a valid content).

The system iteratively evaluates the goodness of the cached contents that have been cached beyond a predefined time period. 
%
The system selects the contents with goodness values less than a goodness threshold, ranks them,  
and applies cache replacement over the content with low goodness values.
The authors showed the advantages of their proposed mechanism over CacheShield in terms of hit damage-ratio (proportion 
of hits that cannot occur due to the attack), percentage of honest consumers receiving valid contents, and communication overhead.
However, this mechanism needs to store historical and statistical information for each cached content--a significant 
memory overhead.
Additionally, the iterative computation of statistics undermines scalability.
%
%
\paragraph{\underline {Approaches with Low Computation Overhead}}
%

Mauri~\kETAL~\cite{MauRasGer15} discussed a cache pollution scenario in NDN, where a malicious provider 
intends to malign the routers' cache to preferentially store its own content for lower latency.
The authors assumed that the provider used colluding terminal nodes (bots or zombies) to request its content(s). 
This results in a disproportionately larger portion of the attacker's content catalog to move down to the network edge, 
thus improving its delivery latency.
The authors proposed a mitigation mechanism for this attack that used a honeypot installed close to potential zombies, 
which monitors and reports the malicious interests to the upstream routers.
A router gathers these interests into a blacklist; the interests in this blacklist are routed to the provider using 
the standard NDN routing protocol, not the CS or nearest replica.
The proposed solution incurs low computation overhead on the routers, however, it requires additional infrastructure.
\begin{table*}[!ht]
\centering
\caption{Secure Naming Approaches are Classified According to their Underlying Cryptographic Schemes}
\label{table:02_04}
\begin{tabular}{|l c c c|}
 \hline
 {\bf Mechanism}&{\bf Crypto}&{\bf Provenance}&{\bf Drawbacks} \\
 \hline
 \hline
 {\bf RSA Crypto} & & & \\
 \hline
 Wong~\kETAL~\cite{WonNik10} & RSA & Pub. Key Digest & PKG Requirement for Private key Generation \\
 Dannewitz~\kETAL~\cite{DanGolOhl10} & RSA & Pub. Key Digest & Lack of Evaluation \& Scalability Issue \\
 \hline
 \hline
 {\bf IBC Crypto} & & & \\
 \hline
 Zhang~\kETAL~\cite{ZhaChaXio11} & IBC & IBC Signature & Scalability Issue \& Public key Length \\
 Hamdane~\kETAL~\cite{HamSerFad12} & HIBC & IBC Signature & Signature Verification Overhead \\
 \hline
\end{tabular}
\end{table*}
%

\subsubsection{{\it {\bfseries Summary and Future Directions in Cache Pollution Mitigation}}}
\label{sec02-03-03}
In Table~\ref{table:02_03}, we summarize the proposed cache pollution solutions based on their detection 
and mitigation approaches, and the nature of the attack. 
We also present the storage and computation overheads for each solution at the routers.
Cache pollution is divided into false locality and locality disruption attacks.
The objective of these attacks is to degrade cache effectiveness and increase the content retrieval latency.
Some of the proposed approaches~\cite{KarGue15anfis,ParWidLee12,XieWidWan12} incur high computation 
cost at the intermediate routers, which undermines their scalability.
Other proposed mechanisms either only detect the cache pollution attack~\cite{ConGasTeo13} or 
address the less severe malicious provider attack scenario~\cite{MauRasGer15}.
All the proposed mechanisms except~\cite{MauRasGer15} can be applied to ICN architectures 
that leverage caching. 

We believe that the key aspect of a solution is in designing a robust caching mechanism, 
which not only increases the resiliency of the cache against these attacks, but also 
improves the overall network latency and users quality of experience.
One possible direction is further exploration of collaborative caching. 
Proposed collaborative caching schemes have aimed at improving cache utilization and reducing 
latency~\cite{WanBiWu13,ChoLeePar12,XuLiLin13}.
However, the positive impacts of collaborative caching mechanisms on mitigating cache pollution 
attack have not been explored.
With collaborative caching and feedback between the caches, mechanisms can be designed to contain 
or root out cache pollution attack attempts. 
For instance, a coalition of collaborative caches can exchange cache states and cached 
content popularity to reduce caching of unpopular content~\cite{MicTouMis16,SahLukYla13}.
%
%
%
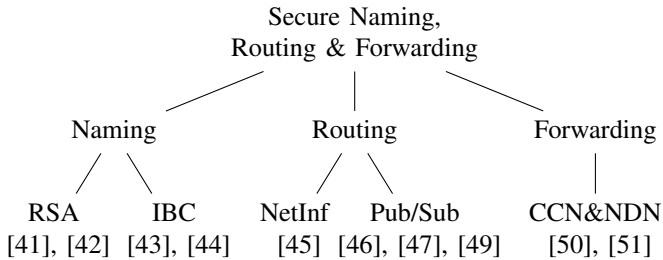
\begin{figure}[!hb]
\begin{center}
\begin{tikzpicture}[level distance=1.3cm,
  level 1/.style={sibling distance=3.2cm},
  level 2/.style={sibling distance=1.6cm}]
  \node [align=center] {Secure Naming, \\ Routing \& Forwarding}
    child {node [align=center] {Naming}
      child {node [align=center] {RSA \\ ~\cite{WonNik10,DanGolOhl10}}}
      child {node [align=center] {IBC \\ ~\cite{ZhaChaXio11,HamSerFad12}}}
    }
    child {node [align=center] {Routing}
      child {node [align=center] {NetInf \\ ~\cite{RemCatSac09}}}
      child {node [align=center] {Pub/Sub \\ ~\cite{AlzReeVas12,AlzVasRee13securing,AlzVasRee13}}}
    }
    child {node [align=center] {Forwarding}
      child {node [align=center] {CCN\&NDN \\ ~\cite{AfaYiWan15,YiAfaMoi13}}}
    };
\end{tikzpicture}
\end{center}
\caption{The state-of-the-art in secure naming, routing, and forwarding.}
\label{fig02-04-Naming}
\end{figure}
%
%
\subsection{Secure Naming, Routing, and Forwarding}
\label{sec02-04}
%
Content naming scheme (name schema) is an integral aspect of ICN.
In ICN, a verifiable binding between the content name and its provider can help 
nullify attacks such as content poisoning.
Secure naming is also essential for verifying provenance of a content (an important feature of ICN). 
Secure routing and forwarding on the other hand are essential aspects of any network architecture. 
All three architectures we are discussing (NetInf, Publish/Subscribe, and CCN/NDN) have their own nuances in  
routing and forwarding, each leveraging their core-features. 
In this subsection, we discuss the proposed security enhancements on these routing and forwarding approaches. 
Secure routing has been the focus of the NetInf and Pub/Sub approaches while secure forwarding has been the focus in 
CCN/NDN (NDN in particular).  
%
%

Fig.~\ref{fig02-04-Naming} categorizes the proposed mechanisms into {\it secure naming}, {\it routing}, 
and {\it forwarding} categories.
The approaches in the secure naming category are sub-categorized, based on their underlying 
cryptographic schemes, to {\it RSA-based} and {\it IBC-based} subcategories.
As mentioned, we sub-categorize secure routing and forwarding based on the underlying 
architectures.
%
%
\subsubsection{{\it {\bfseries Secure Naming}}}
\label{sec02-04-01}
%
All the proposed naming schemes can be easily categorized into either those that use RSA cryptography 
and those that use identity-based cryptography. 
We follow this categorization. 
%
%
\paragraph{\underline {Approaches using RSA}}
%
The approaches using RSA either use the provider's public key or its digest 
to guarantee content provenance.
Wong~\kETAL~\cite{WonNik10} proposed a secure naming scheme to establish trust between content providers and clients.
The scheme uses a metadata composed of three identifiers: authority identifier (ID), which is generated from the 
provider's public key; content identifier, which is the cryptographic hash of the content; and algorithmic 
identifier, which binds the content identifier with a set of the content fragment/chunk identifiers.
Based on the URI naming convention, the authority field is mapped to the provider's public key and the 
resource path field holds the content identifier. 
%
%
The content metadata are disseminated into a set of network nodes that function as part of a domain name 
system and also store the metadata in a DHT.
For content retrieval, a client queries the DNS to resolve the content name into a digital certificate.
By extracting the authority identifier from the certificate, the client obtains the metadata that has to be 
resolved by the DHT.
The query to the DHT returns the content and algorithmic ID, which the client uses to request the content.
This approach suffers from scalability concerns such as header overheads and the 
latency due to DNS and DHT queries, which the authors have not discussed.

In a similar vein, Dannewitz~\kETAL~\cite{DanGolOhl10} proposed a naming scheme for NetInf. 
They proposed an information object (IO) for each content as a tuple composed of the content ID, the content, and a piece of metadata.
The content ID follows a self-certifying flat structure containing type, authentication, and label fields.
The type field specifies the hashing function used for ID generation.
The authentication field is the hash value of the provider's public key; and the label field contains a 
number of identifier attributes and is unique in the provider's domain.
The IO contains the provider's complete public key and its certificate, a signature over 
the self-certified data, and the hash function used for the signature.
This scheme has several weaknesses: the IO field can be a big transmission overhead; the 
signature verification if it happens per chunk can be expensive, and if it happens after the whole content 
is downloaded can enable cache poisoning or pollution attacks. 
%
%
\paragraph{\underline {Approaches that Employ IBC Cryptographic Scheme}}
%
In this subcategory the approaches use a binding between the content name and the corresponding 
provider's public key.
Zhang~\kETAL~\cite{ZhaChaXio11} proposed a name-based mechanism for efficient trust management in 
content-centric networks.
This mechanism takes advantage of identity-based cryptography (IBC), in which either the provider's 
identity or the content name prefix is used as the public key.
A trusted private key generator (PKG) entity generates the private key corresponding to the public key.
For the content name prefix to be used as the public key a name resolution service is required to 
register the name prefix (for uniqueness).
%

Despite its advantages, use of  IBC implies that PKI is still needed to secure communication between the 
PKG and other network entities.
Additionally, the use of the content name prefix as the public key is a new approach and needs further investigation. 
Another significant drawback is the need for a trusted PKG, which is another entity that needs to be added into the system; 
which undermines usability. 

Hamdane~\kETAL~\cite{HamSerFad12} proposed a hierarchical identity-based cryptographic (HIBC) naming scheme for NDN.
This scheme ensures a binding between a content name and its publisher's public key.
The identity-based content encryption, decryption, and signature mechanisms follows~\cite{ZhaChaXio11}.
Different from the previous work, the authors proposed a hierarchical model in which a root PKG is responsible 
only for generating private keys for the domain-level PKGs.
The domain-level PKGs perform the clients' private key generation.
This scheme has the same scalability concerns as the previous scheme on account of the encryption/decryption costs. 
In fact, the overhead is higher as the size of the public key is longer and additively grows with the depth of the hierarchy. 

Table~\ref{table:02_04} summarizes the existing secure naming schemes and presents the type of cryptography used, 
the mechanism for ensuring provenance, and the nature of the encryption infrastructure.
We note that the proposed naming schemes have significant overheads. 
Reducing these overheads or at least amortizing their cost over the complete set of interests/responses 
is an open research area.  
\begin{table*}[!hb]
\centering
\caption{Secure Routing and Forwarding Approaches are Classified with Regard to the Architectures}
\label{table:02_04_01}
\begin{tabular}{|l c c c|}
 \hline
 {\bf Mechanism}&{\bf Architecture}&{\bf Objective}&{\bf Proposed Solution} \\
 \hline
 \hline
 {\bf Secure Routing} & & & \\
 \hline
 Rembarz~\kETAL~\cite{RemCatSac09} & NetInf & Secure inter-domain communication & Name Resolution Service \& Gateway redirection \\
 \hline
 \hline
 Alzahrani~\kETAL~\cite{AlzReeVas12, AlzVasRee13securing} & Pub/Sub & DoS resistant Bloom Filter-based routing & Employing temporary link identifier for z-Filter Generation \\
 Alzahrani~\kETAL~\cite{AlzVasRee13} & Pub/Sub & Malicious publisher with fake z-Filter & Publisher’s edge router validates z-Filters in data packet \\
 \hline
 Fotiou~\kETAL~\cite{FotMarPol11} & Pub/Sub & Identity-based Authentication for DoS mitigation & Each router validates signature, routing using z-Filter \\
 \hline
 \hline
 {\bf Secure Forwarding} & & & \\
 \hline
 Yi~\kETAL~\cite{YiAfaMoi13} & CCN/NDN & Prefix Hijacking \& PIT Overload & Employing NACK Packet for Unsatisfied Interest \\
 Afanasayev~\kETAL~\cite{AfaYiWan15} & CCN/NDN & Secure Namespace Mapping & Associating a Name Prefix to Globally Routable Prefixes \\
 \hline
\end{tabular}
\end{table*}
%
\subsubsection{{\it {\bfseries Secure Routing}}}
\label{sec02-04-02}
%
We categorize the proposed secure routing schemes, according to their underlying architectures, 
into secure 
routing in NetInf and Publish/Subscribe networks.
%
\paragraph{\underline {Approaches for Secure Routing in NetInf}}
%
Two approaches have been proposed to secure routing in NetInf. 
Both aim to establish secure communication between public and private domains 
(Rembarz~\kETAL~\cite{RemCatSac09}).
%
%
The first approach, {\em gateway-centric} approach, uses a gateway to route all communications 
between the public and private networks.
%
%
A publisher in the private domain publishes a content to a private name resolver, PNR, which 
resides in the private domain.
The PNR informs a public name resolver (NR) in the public domain, about the published content's 
identifier along with the gateway's location; instead of the actual publisher's location.
A public subscriber resolves the content identifier at the public NR and obtains 
the gateway address.
The subscriber successfully authenticates itself to the gateway for the gateway to resolve the 
content identifier at the PNR and delivers the content from the publisher to the subscriber.

In the second approach, the publisher in the private domain publishes its private data identifier 
to a PNR.
The PNR creates a mapping between the content identifier {\it ID} and a generated alternative 
identifier {\it ID'} that is sent to the NR.
A subscriber, in the public domain, contacts the NR to resolve {\it ID'} to its location.
The authentication happens at the PNR.
This mechanism removes the gateway, a single point of failure, in the first approach.
However, the PNR's computation and communication overhead for subscribers authentication and 
authorization (especially when the private network serves large amounts of requests) undermines 
the scalability of this approach.
%
%
\paragraph{\underline {Approaches for Secure Routing in Publish/Subscribe}}
%
The proposed approaches for secure routing in publish/subscribe (pub/sub) networks focus on designing DoS-resistant 
self-routing mechanisms and key management approaches that prevent malicious publishers 
from generating fake routes.
Alzahrani~\kETAL~\cite{AlzReeVas12,AlzVasRee13securing} proposed a DoS-resistant self-routing 
mechanism using Bloom filters.
In pub/sub networks, each network link is assigned a unique identifier (LID), which is 
represented in the form of a Bloom filter.
When a network entity requests for a path from the client to the content location (publisher or a cache), 
an entity called the topology manager (TM), resident in one 
or more routers, generates a filter (z-filter) that specifies the delivery path from a publisher 
to the subscriber by OR-ing the Bloom filters (LIDs) of the links on the delivery path.
At the intermediate routers, an AND operation between the z-filter (in the packet header) and the 
routers' LIDs on the path identifies the delivery links.

This mechanism is vulnerable against DoS attack. An attacker can collect enough z-filters 
and reuse them to overload the frequently used delivery path(s) with bogus traffic.
As a remedy, the authors suggested the use of temporal link identifiers that become stale after a pre-defined time period. 
This temporal, per-flow z-filters was designed to restrict the attacker's impact. 
The remedy introduces two drawbacks; first, the number of z-filter updates increases with 
a decrease in the time interval--a trade-off between attack mitigation and computation overhead at the TM.
Second, the size of the packet header (includes the z-filter) increases with the number of links in 
the delivery path.
The authors also investigated factors that affect the z-filter's size in~\cite{AlzVasRee14}.
%


Alzahrani~\kETAL~\cite{AlzVasRee13} proposed a key management protocol for publish-subscribe networks 
which utilized dynamic link identifiers.
Following up on~\cite{AlzReeVas12,AlzVasRee13securing} the authors proposed an 
enhancement that prevents a malicious publisher from generating fake z-filters by enabling the publisher's edge 
router to verify the TM generated z-filter. 
Fake z-filters can enable the transmission of a large number of packets aimed at overwhelming unwitting subscribers.  
%
%
The TM shares a symmetric key with the publisher's edge router and uses it to cryptographically hash the 
corresponding z-filter and it's generation timestamp, and forwards both to the publisher.
The publisher adds these information to each packet that it forwards towards the subscriber.
The proposed mechanism is vulnerable against the malicious publisher colluding with its edge router.
In addition, this mechanism requires stateful routers, which are vulnerable against DoS attacks 
(similar to CCN/NDN DoS-flooding attack). 
%

Fotiou~\kETAL~\cite{FotMarPol11} reviewed a clean-slate PSIRP networking
architecture and highlighted its security assurances.
The architecture employs self-certifying names, each composed of a rendezvous
identifier (RID) and a scope identifier (SID).
%
To preserve information security, content transmissions are encrypted and
include packet-level authentication (PLA): packet header contains the sender's signature,  
public key, and certificate.
The forwarding mechanism utilizes a z-filter generated by the
topology manager to define the information delivery path.
%
%
As already discussed, z-filters suffer from scalability and false positives. 
Apart from that, the use of per-packet cryptographic signatures in PLA 
makes line-speed operations difficult. 

%
\subsubsection{{\it {\bfseries Secure Forwarding}}}
\label{sec02-04-03}
%
The secure forwarding category includes mechanisms that either secure the forwarding 
plane or create a secure namespace mapping, which allows interest forwarding for name prefixes not in the 
routers FIB tables.

Yi~\kETAL~\cite{YiAfaMoi13} augmented the NDN forwarding plane to thwart security problems, such 
as prefix hijacking and PIT overload (cases of authenticated denial of service). 
In prefix hijacking, an attacker announces the victim's prefix and drops the interest.
The authors suggested the use of interest NACKs whenever requests are not satisfied for reasons, 
such as network congestion, non-existent content, and duplicate content. 
The interest NACK helps reduce the size of the PIT on account of the NACK removing a PIT entry.
Additionally, it mitigates the prefix hijacking vulnerability, by providing extra time for the 
router to query other faces for a content match.
However, this requires each router to store RTT information for each interest--a significant overhead for core routers. 
Additionally, with the NACK consuming an interest in the PIT, there is no scope for bogus interest aggregation; this could 
exacerbate interest based DoS attacks. 
%
%
%
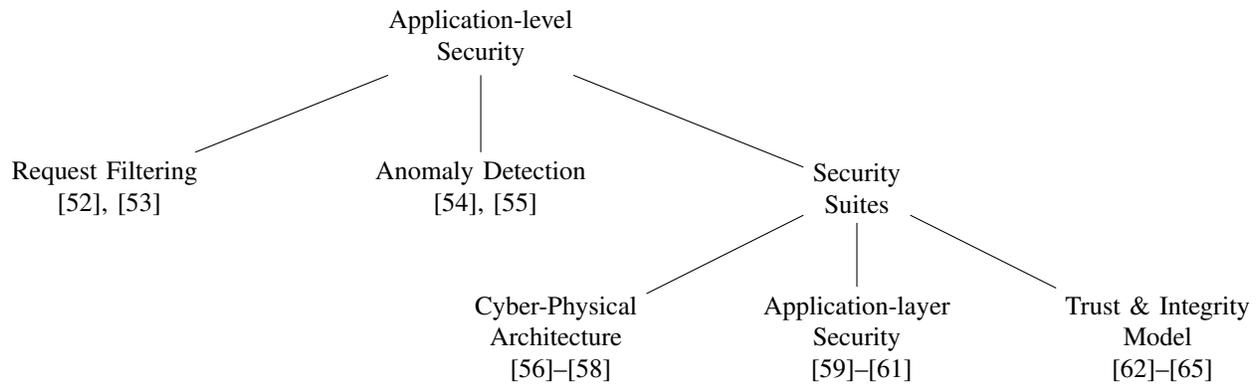
\begin{figure*}[!ht]
\begin{center}
\begin{tikzpicture}[level distance=2cm,
  level 1/.style={sibling distance=5cm},
  level 2/.style={sibling distance=4cm}]
  \node [align=center]{Application-level \\ Security}
    child {node [align=center] {Request Filtering \\ ~\cite{FotMarPol10,GoeChoFra13semantic}}
    }
    child {node [align=center] {Anomaly Detection \\ ~\cite{GoeChoFra13,KarGue15fuzzy}}
    }
    child {node [align=center] {Security \\ Suites}
      child {node [align=center] {Cyber-Physical \\ Architecture \\ ~\cite{BurGasNat13,BurGasNat14, ViePol13}}}
      child {node [align=center] {Application-layer \\ Security \\ ~\cite{SalRen12,AmbConGas14,AsaNamKaw15}}}
      child {node [align=center] {Trust \& Integrity\\ Model \\ ~\cite{WonVerMag08,SeeKutSch14,SeeGilKut14,YuAfaCla15}}}
    };
\end{tikzpicture}
\end{center}
\caption{Application-level security sub-classes and the state-of-the-art.}
\label{fig02-05-Application}
\end{figure*}

Afanasayev~\kETAL~\cite{AfaYiWan15} proposed a secure namespace mapping scheme, which allows interest forwarding for name 
prefixes that are not in the FIB---useful to handle node mobility.
The proposed mechanism is built upon two main concepts: {\em link object} and {\em link discovery}.
The link object is an association between a name prefix and a set of globally routable prefixes.
By creating and signing a link object, the content owner maps its own name prefix to those globally routable prefixes.
The authors designed an NDN based DNS service (NDNS), where the mapping between the name prefix and the 
globally routable prefixes are stored, and the service provides this mapping to a requesting entity. 

For link discovery, a client queries the NDNS iteratively for each component of the requested name prefix.
If a client sends an interest that a router cannot satisfy using its FIB, that router returns a NACK.
After the NACK reaches the client, its local forwarder discovers and validates the link object corresponds to the name prefix.
After that, the client embeds the link object in its original interest and forwards it to the network. 
Although this scheme is a good initial solution to provider mobility it suffers from overheads. 
When a provider moves, the current routable prefix, which is in the FIB of the routers, will results in 
interests being routed to the provider's former location until the FIB entries time out; 
a waste of bandwidth in high traffic scenarios.

Table~\ref{table:02_04_01} summarizes the proposed secure routing and forwarding approaches 
and presents  the architecture, the objective of the proposed mechanism, and solution to that problem.
Among the proposed mechanism, the work by Afanasayev~\kETAL~\cite{AfaYiWan15} is the most 
important as it has addressed the producer mobility; an open challenge in the ICN community.

\subsubsection{{\it {\bfseries Summary and Future Directions in Secure Naming, Routing, and Forwarding}}} 
\label{sec02-04-04}
The proposed approaches for secure naming and routing in the ICN architectures are a good first attempts to 
address the malicious attacks possible. 
A content naming scheme with a verifiable binding between the content name and its provider is 
essential to nullify attacks such as content poisoning attack and is integral to ICN.
However, in all proposed approaches~\cite{DanGolOhl10,HamSerFad12,WonNik10,ZhaChaXio11} this binding comes at the 
high cost of signature verification (complete verification of binding requires signature verification of each chunk), which 
would prevent intermediate routers from verifying signature of all arriving packets to maintain line speed. 
There is still a need for more scalable and computationally efficient approaches.
The identity based cryptographic approaches~\cite{HamSerFad12,ZhaChaXio11} require the 
client to trust a third party for private key generation; a practice that significantly 
undermines the applicability of these approaches.

A secure and efficient naming scheme is still an open challenge. 
Any such scheme should include metadata, such as the content hash and the provider's 
identity and signature for enhanced security. 
For instance, a potential secure naming approach can be signature of the manifest 
(includes chunk names and hashes) by the content provider.
This is currently an important area of research with proposals being made to the ICN 
Research Group, an Internet Research Task Force~\cite{ICNRG}.
On the other hand, secure routing and forwarding (and routing and forwarding in general) do not perform acceptably  
consumer and/or producer mobility. 
Even though this has not appeared in the literature, employing Bloom filter based routing (z-filter) in pub/sub networks 
leads to a potential routing attack.
Unless the Bloom filter is authenticated by an intermediate router, an attacker or a malicious router can easily 
modify the bits in the filters to either overload the network or disrupt content delivery. 
%
%
Developing efficient mechanisms to help routers validate the integrity and authenticity of the z-filters needs more research focus.
%

%
\subsection{Application-level Security}
\label{sec02-05}
%
We have classified the works in ICN application layer security into three major subtopics:
{\it filtering}, {\it anomaly detection}, and {\it security suites}.
Fig.~\ref{fig02-05-Application} illustrates our categorization and the sub-categorization within the 
categories followed by a mention of the corresponding state-of-the-art.
The filtering category deals with the identification and removal of unwanted content, 
such as spam, forged content, and content from untrusted publishers at the application layer.
Anomaly detection includes the detection of undesired activities,
such as flooding, misbehavior of network elements, and malicious traffic.

We have designated application-specific security measures as security suites, which combine different 
cryptographic techniques to achieve specific goal(s).
We sub-categorize the mechanisms in security suites into {\;it cyber-physical 
architecture}, {\it application layer security}, and {\it trust and integrity model}.
The cyber-physical architecture subcategory deals with the proposed ICN-based architectures 
for smart grid, smart home, and Internet of things.
The application layer security, reviews the security applications for ICN, 
such as secure email, covert channel, and information sharing.
The trust and integrity model subcategory include the proposed mechanisms that build trust in the network.
%
\subsubsection{{\it {\bfseries Request Filtering}}}
\label{sec02-05-01}
%
The state-of-the-art in request filtering either utilizes content ranking or exploits providers' 
information, such as public keys and name prefixes, to block spams and blacklisted content.
Fotiou~\kETAL~\cite{FotMarPol10} proposed an anti-spam mechanism for
publish/subscribe networks.
It is based on an {\em inform-ranking} process, with content ranked based on votes from
publishers and subscribers.
Each publisher serving a content implicitly votes for that content. 
%
After the content is published, it is voted on by subscribers.
%
%
After the votes are collected, they are used to rank the content objects and 
identify spam objects.

Simulations showed that the mechanism filters spams better than other existing schemes, 
which only consider the publisher votes for ranking.
However, this scheme's reliance on user feedback may counteract its effectiveness. 
Not only are typical users unlikely to vote on the content, but malicious users can hijack 
the voting process. 
Moreover, the voting process itself confers non-negligible communication overhead.

Goergen~\kETAL~\cite{GoeChoFra13semantic} designed a semantic firewall for
content-centric networks.
Unlike IP firewalls, which filter at flow-level granularity, the proposed firewall
can filter content based on provider and/or name.
For provider-based filtering, the firewall used provider's public key to identify 
disallowed providers and filter contents with invalid signatures.
For content-name filtering requests with blacklisted keywords in the name are filtered.
Both types of filtering can be performed on either interests or the
content chunks.

Additionally, the firewall could monitor for abnormal behavior on each of its interfaces
and filter abnormal peers (e.g., high request volume or high drop rate).
A minimalistic evaluation showed that the firewall's latency increases slowly 
with an increase in the number of filtering rules.
However, latency and scalability in the face of large number of content chunks or large content universe 
has not been analyzed. 
%
%
%
\subsubsection{{\it {\bfseries Anomaly Detection}}}
\label{sec02-05-02}
%
Most proposed anomaly detection mechanisms aim to detect abnormal 
behaviors by using classification or fuzzy logic algorithms on routers statistical information. 
Goergen~\kETAL~\cite{GoeChoFra13} proposed a mechanism for CCN to detect attack patterns 
based on the activities of the FIB, PIT, and CS.
To detect abnormal behavior, each node periodically evaluates 
per-second statistics, such as bytes sent/received, content items received, and  
interests received, accepted and dropped.
%
%
The mechanism uses a support vector machine (SVM) to classify a particular time period as either anomalous or benign. 
The results show the efficacy of this method for attack detection;
however, its ability to detect low-rate attacks is questionable.
Furthermore, the computation cost of SVM at the network elements may be prohibitive; a software-defined networking based 
approach may be a good direction to explore. 

Karami~\kETAL~\cite{KarGue15fuzzy} proposed a combined Particle Swarm Optimization (PSO) 
meta-heuristic, k-means clustering, and a fuzzy detection algorithm for CCN to classify normal/abnormal behaviors.
%
The fuzzy approach is notable for its low false-positive rate; however, at the cost of an 
increased false-negative rate. 
An attacker with sufficient resources can produce a large amount of traffic to ensure its 
malicious packets get through the system without detection.
%
%
\subsubsection{{\it {\bfseries Security Suites}}}
\label{sec02-05-03}
%
Here we discuss the several security suites proposed for ICN architectures based on our 
categorization: cyber-physical architecture, application layer security, and trust and integrity model.  
%
\paragraph{\underline {Cyber-Physical Architecture}}
%
This subcategory includes ICN inspired communication architectures 
for cyber-physical system, such as the Internet of Things (IoT) and smart grid networks.
Burke~\kETAL~\cite{BurGasNat13} presented a security framework for a CCN-based
lighting control system.
In the first variation of the protocol, control commands required a three-way handshake
and were transmitted in a signed content payload; in the second, the commands
were immediately sent as a signed interest.
The framework uses an authentication manager to manage the network's PKI, and
employs shared symmetric keys for communication. 
To reduce the burden of key storage on the embedded devices, these symmetric
keys can be generated on-demand by a pseudorandom function.
These shared symmetric keys can then be used to enforce encryption-based access
control.

The authors in~\cite{BurGasNat14} employ a similar architecture for secure
sensing in IoT.
The system uses a trusted authorization manager (AM) to generate the root keys, which are 
used to sign other keys used.
The AM associates a producer with a namespace, which is listed in the producer's
certificate.
Each sensor is also assigned an access control list, which specifies
the permissions of each application with respect to that node.
While this scheme is flexible, it suffers from a significant overhead problem---power-constrained
devices such as sensing nodes are required to perform asymmetric-key cryptography.

Vieira and Poll~\cite{ViePol13} proposed a security suite for C-DAX, an
information-centric Smart Grid communication architecture.
The proposed security suite employs content-based cryptography, in
which content topics are used as public keys, and the corresponding secret
keys are generated by a security server.
For each topic, write-access secrets and read-access secrets must be
distributed to each authorized publisher and subscriber, respectively.
While the scheme provides sufficient security and flexibility for typical
applications, its reliance on a central security server constitutes a single
point of failure. 
In a high-impact critical infrastructure such as the Smart Grid, the failure or
compromise of this service could have dire consequences.
Also, requiring cyber-physical devices to store two keys for each topic 
limits scalability.
\begin{table*}[!ht]
\centering
\caption{Categorization of Application Security approaches}
\label{table:02_05}
\begin{tabular}{|l c c|}
 \hline
 {\bf Mechanism}&{\bf Application}&{\bf Approach} \\
 \hline
 \hline
 {\bf Filtering} & & \\ 
 \hline
 Fotiou~\kETAL~\cite{FotMarPol10} & Anti-spam mechanism & Information ranking based on publishers and subscribers votes \\
 Goergen~\kETAL~\cite{GoeChoFra13semantic} & Semantic firewall & Filtering by content name, provider's public key, and anomaly detection \\
 \hline
 \hline
 {\bf Anomaly Detection} & & \\ 
 \hline
 Goergen~\kETAL~\cite{GoeChoFra13} & Traffic anomaly detection at routers & Statistical data analyses and SVM classification \\
 Karami~\kETAL~\cite{KarGue15fuzzy} & Anomaly detection mechanism & Fuzzy detection algorithm and traffic clustering \\
 \hline
 \hline
 {\bf Security Suites} & & \\ 
 \hline
 {\it Cyber-Physical Architecture} & & \\
 Burke~\kETAL~\cite{BurGasNat13} & Lighting control system & Submitting commands as signed content or signed interest \\
 Burke~\kETAL~\cite{BurGasNat14} & Secure sensing in IoT & Assigning a sensor an ACL for content publishing \\
 Vieira~\kETAL~\cite{ViePol13} & Security suite for Smart Grid & Content-based cryptography and access level distribution via security server \\
 \hline
 {\it Application-Layer Security} & & \\
 Saleem~\kETAL~\cite{SalRen12} & Secure email service & Asymmetric crypto with emails as independent objects \\
 Ambrosin~\kETAL~\cite{AmbConGas14} & Ephemeral covert channel & Time difference analysis between cache hit and cache miss \\
 Asami~\kETAL~\cite{AsaNamKaw15} & Moderator-controlled information sharing & Publisher signature followed by moderator signature for message publications \\
 \hline
 {\it Trust and Integrity Model} & & \\
 Wong~\kETAL~\cite{WonVerMag08} & Content integrity by security plane & Content signature and publisher authentication to security plane \\
 Seedorf~\kETAL~\cite{SeeKutSch14, SeeGilKut14} & Self-certifying names and RWI binding & Employing a Web-of-Trust \\
 Yu~\kETAL~\cite{YuAfaCla15} & Trusted data publication/consumption & Schematized chain-of-trust \\
 \hline
\end{tabular}
\end{table*}
%
%
\paragraph{\underline {Application layer Security}}
%
This subcategory includes secure ICN-based application layer protocols, such as secure email service, 
covert channel, and information sharing.
Saleem~\kETAL~\cite{SalRen12} proposed a distributed secure email service for
NetInf, based on asymmetric-key cryptography.
Each email message is treated as an independent object. 
A client (user) is identified by its public key, and no domain name 
service is required thus providing scalability.
However, the subscription-based nature of the service potentially leaves users
vulnerable to spam, and no mitigation for this has yet been proposed.

Ambrosin~\kETAL~\cite{AmbConGas14} identified two different ways of creating an ephemeral
covert channel in named-data networking.
The sender and receiver require tight time
synchronization and agreement on a set of unpopular contents to exploit.
To send a ``$1$'' covertly, the sender requests an unpopular object during
a time slot; to send a ``$0$,'' no request is sent.
In the first variation, the object is assumed to be cached at the edge router
if it was requested.
The receiver then requests the same content, and measures the retrieval time to
differentiate a cache hit from a cache miss, and consequently infers the bit that
was sent.
This mechanism is accurate when the sender and receiver are
collocated behind the same edge router; therefore, its applicability
is limited.
%

Asami~\kETAL~\cite{AsaNamKaw15} proposed a moderator-controlled information
sharing (MIS) model for ICN, which provides Usenet-like functionality 
while leveraging identity-based signature scheme.
Several message groups are defined, each of which is assigned a moderator.
To publish a message in a group, the publisher signs with its secret key then sends it to
the group moderator. 
The moderator can then sign the message and relay it to the group's subscribers, or reject 
the message and drop it.
To verify a signature, the subscriber only needs to know the identities of
the publisher and moderator.
This is an example of implementation of a secure legacy application in ICN.   
%
%
\paragraph{\underline {Trust and Integrity Model}}
%
This subcategory focuses on directions, such as dedicated security plane, 
self-certifying names to real-world identities binding, and trust schema creation.
Wong~\kETAL~\cite{WonVerMag08} proposed a separate security plane for 
publish/subscribe networks for assuring content integrity.
The security plane takes over the distribution of authentication
materials and associated content metadata from the data plane.
The materials distributed by the security plane would include
the content name and ID, the Merkle tree root, the
publisher's public key, and the publisher's signature.
To prevent the insertion of malicious metadata, publishers 
identify themselves to the security plane and submit to a challenge-response
authentication.
We believe that while it is convenient for data to be separated from its 
authentication materials, a separate control plane is ultimately unnecessary.
The integrity assurances can be provided by implementing simple content-signing schemes,
such as the manifest-based content authentication supported by CCN or NDN~\cite{JacSmeTho09}. 

Seedorf~\kETAL~\cite{SeeKutSch14, SeeGilKut14} proposed a mechanism for binding self-certifying names and 
real world identities (RWIs) using a Web-of-Trust (WoT).
A WoT is a directional graph, in which nodes (users) are identified by an RWI-public key digest pair. 
Edges represent trust relationships: an edge from a node $u$ to a node $v$ 
indicates that $v$'s certificate has been signed by $u$.
User $u$ trusts another user $v$ if there exists a path starting at $u$, reaching $v$ in the WoT. 
%
%
Although this mechanisms is very useful in infrastructure-less networks (e.g., disaster response networks) it may 
suffer from inefficiencies based on the size of the WoT graph, graph updates in the event of network segmentation, 
and inaccuracies based on the basic notion of a trust chain. 

Yu~\kETAL~\cite{YuAfaCla15} presented a schematized trust model for named-data networks 
to automate data authentication, signing, and access procedures for clients and providers.
The proposed model is composed of two components: a set of trust rules, and trust anchors.
Trust rules define associations between data names and the corresponding keys that are used to sign them. 
The authors define a chain of trust, which is discovered by recursively evaluating trust rules, starting from the 
{\em KeyLocator} field in the content and ending at a trusted anchor. 
Anchors are envisioned to serve as trusted entities that help bootstrap the key discovery process. 
%
%
%
%
For data authentication, the client uses the public key in the KeyLocator of the packet and according to the trust schema, 
recursively retrieves public keys to reach a trust anchor to verify the content.
%
%

The iterative discovery and key verification step may become inefficient for mobile or IoT devices that are power constrained. 
Further the trust rules may become complex quickly within a few levels, thus requiring a mechanism for automatic creation of the 
trust chain in an application. 
%
The scheme will have limited applicability until then. 
%

\subsubsection{{\it {\bfseries Summary and Future Directions in Application Security}}}
\label{sec02-05-04}
Table~\ref{table:02_05} summarizes the proposed application-level mechanisms.
The table contains the proposed approaches reference, the corresponding application, 
and the approach's information.
We note that several interesting applications have been considered in the ICN domain.

Different ICN security applications and application-level security mechanisms, such as content filtering, 
anomaly detection, and covert channel have been proposed in the literature.
Mechanisms proposed in~\cite{FotMarPol10,GoeChoFra13,GoeChoFra13semantic,KarGue15fuzzy} attempt to detect abnormal traffic at the intermediate 
routers, spam contents based on the subscribers' and publishers' votes, or performed content filtering through the firewall.
In~\cite{BurGasNat13,BurGasNat14,ViePol13}, the authors proposed ICN inspired architectures for lighting control systems, Internet of things, and the smart grid.
In~\cite{YuAfaCla15}, Yu~\kETAL~proposed a chain-of-trust based schema for content publishers and consumers to use to share content. 
The authors in~\cite{WonVerMag08} suggested the separation of data and security planes for better content integrity assurance.
Other proposed applications include ephemeral covert channel communication~\cite{AmbConGas14}, secure email service~\cite{SalRen12}, and 
moderator-controlled information sharing~\cite{AsaNamKaw15}.

We have not found an application that incorporates all the security functionalities available in ICNs (any architecture) nor did we find a 
comprehensive application-level security suite (again for any architecture). 
That should be one of the interests of future researchers in this domain. 
\begin{figure*}[t]
\setlength{\unitlength}{0.08in}
\centering
\begin{picture}(95,24)
\put(36,19){\colorbox{gray!25}{\framebox(15,4){\bf Privacy}}}
\put(0,3){\colorbox{gray!10}{\framebox(15,10)[c]{\parbox{11\unitlength}{\bf Timing Attack {\small{\\\\ ~\cite{ChaDecKaa13}~\cite{AcsConGas13}~\cite{ComConGas} \\
~\cite{MohMekZha15}~\cite{MohZhaSch13}}}}}}}
\put(18,3){\colorbox{gray!10}{\framebox(15,10)[c]{\parbox{14\unitlength}{\bf Communication Monitoring Attack \small{\\\\ ~\cite{ChaDecKaa13}~\cite{LauLaoRod12Tech}~\cite{LauLaoRod12ACM}}}}}}
\put(36,3){\colorbox{gray!10}{\framebox(15,10){\parbox{14\unitlength}{\bf Censorship and Anonymity Attack \small{\\\\ ~\cite{AriKopRag11}~\cite{ElaBenMet11}~\cite{FotTroMar14}~\cite{DibGasTsu11} \\ ~\cite{ChuKimJan14}~\cite{TaoFeiYe15}~\cite{TouMisKli15}}}}}}
\put(54,3){\colorbox{gray!10}{\framebox(15,10){\parbox{12\unitlength}{\bf Protocol Attack \small{\\\\ ~\cite{ChaDecKaa13}~\cite{LauLaoRod12Tech}~\cite{LauLaoRod12ACM}}}}}}
\put(72,3){\colorbox{gray!10}{\framebox(15,10){\parbox{14\unitlength}{\bf Naming-Signature Privacy \small{ \\\\ ~\cite{BauDavNar12}~\cite{ChaDecKaa13}~\cite{KatSaiPsa14} \\ 
~\cite{MarGom12}~\cite{MarGomGir11}~\cite{Sol12}}}}}}
%
\thicklines
\put(43.8,19){\line(0,-1){3}}
\put(7.7,16){\vector(0,-1){3}}
\put(25.7,16){\vector(0,-1){3}}
\put(43.8,16){\vector(0,-1){3}}
\put(61.7,16){\vector(0,-1){3}}
\put(79.7,16){\vector(0,-1){3}}
\put(7.64,16){\line(1,0){72.1}}
\end{picture}
\caption{Privacy Risks and their Countermeasures.}
\label{fig03-01}
\end{figure*}
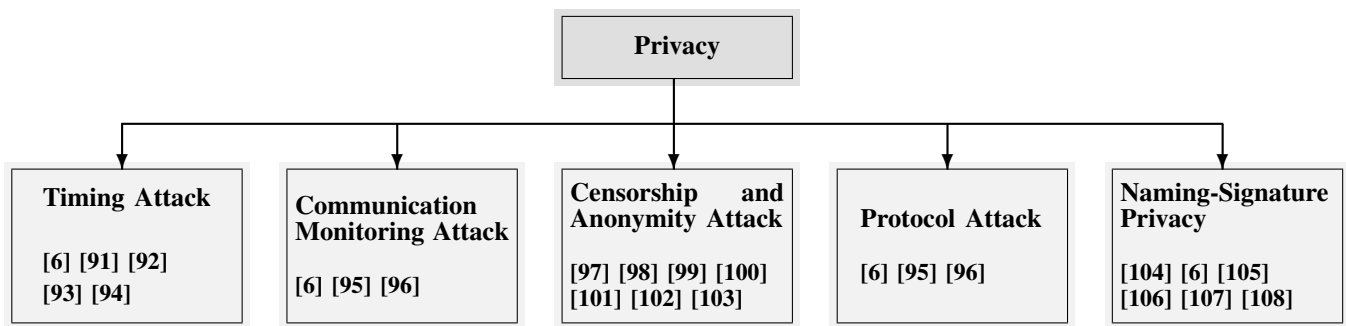
\section{Privacy in ICN}
\label{sec03}
%
In this section, we explore privacy risks in ICNs and the proposed mitigation mechanisms. 
%
%
Privacy attacks in ICN may target the routers, cached contents, content names, content signatures, 
as well as client privacy.
These privacy concerns are applicable to all architectures. 
Additionally, a few attacks are possible due to the inherent design choices
of specific architectures; we discuss them separately.
We will highlight the vulnerable design choices and discuss their advantages
and disadvantages.

Fig.~\ref{fig03-01} presents our categorization of privacy attacks in ICNs, along with the 
proposed mitigation mechanisms. 
%
We categorize privacy attacks into {\em timing attack}, {\em communication monitoring attack}, 
{\em censorship and anonymity attack}, {\em protocol attack}, and {\em naming-signature privacy}. 
In timing and communication monitoring attack (Subsections~\ref{subsec03-01},~\ref{subsec03-02}), the 
attackers probe the cached content of a router over time to identify content popularity in the 
cache or requesters’ content access behavior. 
In Subsection~\ref{subsec03-03}, we discuss the proposed approaches for anonymous communication.
The protocol attack subsection (Subsection~\ref{subsec03-04}), reviews the vulnerable design 
features of an architecture, such as longest prefix matching and the scope field.
The name of a content in ICN and its signature by design ties the content to the producer's identity, 
which raises concerns of producer (publisher) privacy. 
In Subsection~\ref{subsec03-05}, we discuss the privacy concerns from this  
exposure and review the literature on publishers’ privacy.

Before discussing the state of the art based on these categories, we mention one work that 
is general, and hence goes across several of the above categories, hence merits a standalone definition. 
%
Fotiou~\kETAL~reviewed the proposed ICN architectures and discussed the privacy requirements
and design choices for secure content naming, advertisement, lookup, and forwarding in~\cite{FotAriSar14}.
The authors classified each privacy threat as either a monitoring, decisional interference, 
or invasion attack.
The decisional interference attack either prevents a consumer from accessing certain content, 
prevents the content advertisement and forwarding of a specific provider, or allows content 
filtering based on content name.
In the invasion attack, an attacker tries to acquire sensitive information from the target.
The authors also analyzed the identified threats and ranked them according to
the DREAD model~\cite{HowLeB03}, and briefly reviewed ongoing research on privacy 
concerns in information-centric networking.
Now, we discuss the categories.
%
%
\vspace{-0.15in}
\subsection{Timing Attack}
\label{subsec03-01}
\begin{figure}[H]
\centering
\includegraphics[height=1.4in]{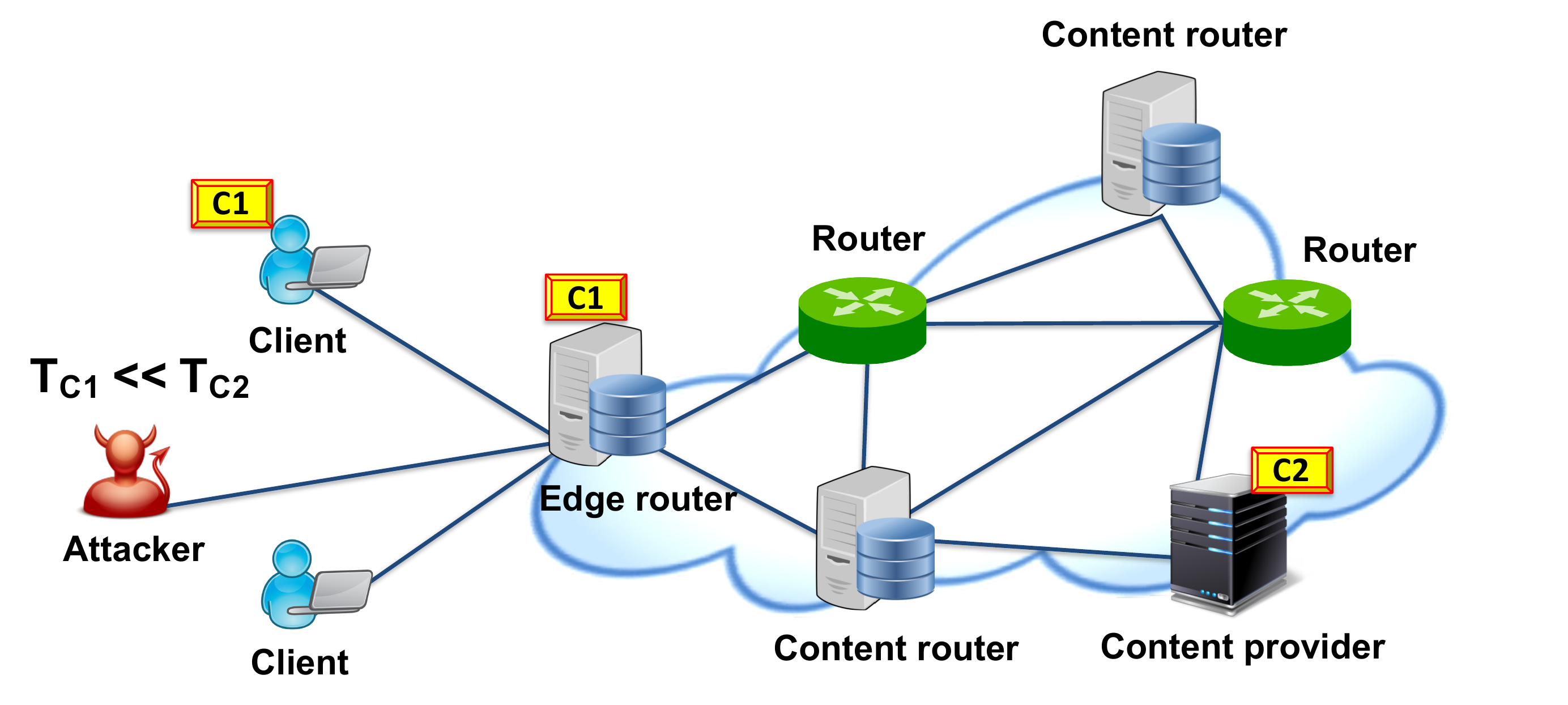}
\caption{Timing attack scenario.}
\label{fig03-02}
\end{figure}

%
Timing attack has been explored in a large body of
literature~\cite{AcsConGas13,MohZhaSch13,MohMekZha15,ChaDecKaa13,ComConGas}.
In a timing attack, an attacker probes content objects which it believes are
cached at a shared router.
The attacker leverages precise time measurements to distinguish cache hits and
cache misses, and thereby can identify which contents are cached.
A cache hit implies that the content had been requested by another client in the 
neighborhood, while a cache miss indicates that the content has not been requested 
(or has been evicted from the cache). 
An informed attacker can also ascertain whether the request is served by the 
provider or by a router somewhere along the path to the provider.  
As illustrated in Fig.~\ref{fig03-02}, a shorter latency in retrieving content
{\bf C1} in comparison to content {\bf C2} reveals the availability of {\bf C1}
in the shared edge router's cache.

We note that this attack, although feasible in all architectures employing caching, is 
less effective in the pub/sub architectures.
%
%
%
In pub/sub (specifically PSIRP/PURSUIT), when a node subscribes to a publisher's content, the latencies of 
the initial packet deliveries (already created and potentially cached packets)   
can be used to see whether the packet came from a nearby or farther cache publisher. 
The timing of subsequent (newly generated) packets do not reflect caching latencies as they are disseminated 
by the publisher and multicast into the network, and may not even be delivered from a cache. 
%
%
\begin{table*}[!t]
\centering
\caption{Summary of Timing Attack Mitigations}
\label{table:03_01}
\begin{tabular}{| l | c | c | c |}
 \cline{1-4}  
 &{\bf Acs~\kETAL~\cite{AcsConGas13}}&{\bf Mohaisen~\kETAL~\cite{MohZhaSch13, MohMekZha15}}&{\bf Chaabane~\kETAL~\cite{ChaDecKaa13}} \\  
 \hline  
 \hline  
 {\bf Approach} & \parbox{1.3in}{Delay for the first $k$ interests} & \parbox{2.0in}{Delay for the first interest from each client} & 
 \parbox{1.3in}{Delay for the first $k$ interests} \\
 {\bf Mitigating Entity} & \parbox{1.3in}{Edge routers} & \parbox{2in}{Edge routers \& access points} & \parbox{1.3in}{Edge routers} \\
 {\bf Granularity} & \parbox{1.3in}{Per content} & \parbox{2in}{Per client per content} & \parbox{1.3in}{Per content} \\
 \hline
\end{tabular}
\vspace{-0.2in}
\end{table*}
%

\subsubsection{{\it {\bfseries Timing Attack Mitigation Approaches}}}
\label{subsec03-01-01}
%
Acs~\kETAL~\cite{AcsConGas13} investigated cache privacy in CCN/NDN networks   
in the presence of timing and cache probing attackers.
They confirmed the effectiveness of these attacks in different network
topologies, and demonstrated attack feasibility even when  
the attacker and the victim are three hops away from a shared router 
(success rate of $59\%$).
They discussed two traffic classes: interactive traffic and content
distribution traffic.
For interactive content, the authors proposed the addition of a random number 
to the content name; the number is mutually agreed upon by the requester and 
the content provider. 
This prevents the attacker from successfully probing the cache for this content 
if the precise content name matching approach is employed.

However, this approach does undermine caching--cached content can no longer be reduced. 
%
As an alternative solution, the authors suggested that the requester and producer 
mark privacy-sensitive interests and content as private.
The intermediate routers do not cache these marked content, thus preventing privacy leaks.
The authors also suggested the emulation of a cache miss at a router, 
with the router applying a random delay before satisfying a content chunk request. 
But, a delay undermines user's quality of experience (QoE). 

The authors reduced the impact on QoE by using a popularity threshold.
The premise of the model is that the privacy-sensitive contents are usually unpopular, and 
that increased popularity generally results in reduction of the privacy need. 
With this addition, the router randomly delay satisfying a content for the first $k$-times it 
is requested, and deliver the content as soon as possible for the subsequent requests.  
%
%
This model reduces the latency for popular contents, but clients 
experience the extra delay for the first $k$-interests and this mechanism also requires extra 
state for maintaining the number of requests.

In~\cite{MohZhaSch13,MohMekZha15}, Mohaisen~\kETAL~took a similar approach as above and
proposed three variations of a mitigation technique for the timing attack.
In the vanilla approach, an edge router fetches content chunks from the provider
and stores the retrieval times for the corresponding first interests.
The router also tracks the interest frequency of each requested privacy-sensitive content chunk. 
Each first interest for a cached content chunk from a new client (one who has
not requested that content before) will be satisfied with a delay same as the recorded 
retrieval latency for the chunk. 
Clearly, the per-client state needed to be stored means that this approach will not scale with increasing 
number of clients. 
To reduce the storage requirements, a second approach proposed that the
edge router stores only per-interface interest retrieval time history.
Although this approach reduces state size, it also increases the potential of success of timing attack for an 
attacker on the same interface.

The last variation solved the shortcomings of the first two through cooperation
between the access points/proxies and their corresponding connected edge routers.
Here, the access point stores per-client state; and the router stores only per-face 
statistics. 
The decision to apply random delay is made by the router with the help of the
downstream access point. 
%
%
The access point flags the interest from a new client to inform the router.  
The router delays the data reply for the flagged interests. 
%
%
We believe that despite the strengths of this scheme, the use of random delays goes  
against one of the core principles of ICN--leverage caching to reduce latency. 

In~\cite{ChaDecKaa13}, Chaabane~\kETAL~also proposed applying a delay--either on all
requests for cached content, or on the first $k$-requests only--for mitigating timing attack. 
They also briefly discussed collaborative caching and random caching, to preserve cache privacy.
Collaborative caching increases the anonymous clients set by increasing the number of clients 
that share a set of routers; thus it implicitly helps to preserve privacy.
The authors provided no analysis of the caching approaches.
We believe collaborative caching is a good direction for further exploration.   
%

\subsubsection{{\it {\bfseries Summary and Future Directions in Timing Attack Mitigation}}}
\label{subsec03-01-02}
%
Table~\ref{table:03_01} summarizes the proposed solutions to the timing attack. 
We present the referenced work, the proposed solution, and the entity in the network where 
the mitigation procedure is executed. 
We have not mentioned~\cite{ComConGas} as the authors have not really presented a mitigation 
strategy. 

The majority of the proposed timing attack mitigation mechanisms~\cite{AcsConGas13,ChaDecKaa13,MohMekZha15,MohZhaSch13} 
apply an artificial delay during content forwarding, which makes them applicable to all architectures.
%
Despite the effectiveness of this approach in misleading the adversary, it
undermines the advantage of latency reduction due to caching.
Another negative impact of this approach is degradation in clients' QoE, 
especially for the popular content objects. 

One natural approach of coping with timing attack is designing an efficient collaborative caching mechanism, 
which not only increases the anonymity set of the clients but also improves system performance and 
reduces overall content retrieval latency.
Moreover, this precludes the need for artificial delays. 
Chaabane~\kETAL~\cite{ChaDecKaa13} have made an initial attempt in this direction. 
Network coding techniques can also be leveraged to design a secure and
efficient content dissemination model by coding and dispersing the chunks.

%
%
\subsection{Communication Monitoring Attack}
\label{subsec03-02}
%
In the communication monitoring attack~\cite{LauLaoRod12ACM,LauLaoRod12Tech,ChaDecKaa13},
an attacker has access to the same edge router that the victim receives content from 
(similar to timing attack). 
%
However, here an attacker targets a specific victim and tries to identify the victim's 
requested contents; this is different from timing attack where the goal is to identify 
contents popularity.
The attacker may know the victim's content consumption habits or specific characteristics, which
differentiate the victim from other clients (e.g., language, region, or
institutional affiliation).
%

\subsubsection{{\it {\bfseries Communication Monitoring Attack Mitigation Approaches}}} 
\label{subsec03-02-01}
%
Lauinger~\kETAL~\cite{LauLaoRod12Tech} proposed two types of request monitoring
attacks under the stationary content popularity model 
with a constant request rate, employing non-invasive and invasive cache probing, respectively.
The stationary popularity assumption states that the content popularity distribution
does not change over large time periods, and the interest for a content is independent of 
previous interests. 
In the non-invasive cache probing model, the authors assumed that the attacker's
requests do not change the router's cache state.
The attacker (with prior knowledge of the victim's interests) frequently probes the shared router's cache.
%
%

The unrealistic assumption in the non-invasive model that the cache probing does not change 
the content popularity leads to the proposal of the invasive cache probing attack model. 
In the invasive model, a cache miss caused by the attacker at the shared router causes 
the requested content to be cached, hence the attacker needs to differentiate cache hits 
from cache misses.
The authors also proposed a model for calculating the attacker cache-probing frequency.
%

The mitigation approaches proposed for monitoring attacks have been similar to that of the timing attacks. 
The authors in~\cite{LauLaoRod12ACM,LauLaoRod12Tech} proposed 
selective caching, in which a content will be cached only if it reaches a
specific popularity threshold.
This is congruent with the assumption that privacy risk decreases as content popularity increases.
Alternatively, a client can ensure privacy by establishing a
secure tunnel with either the content provider or a trusted proxy~\cite{DibGasTsu11,TouMisKli15}.
Another solution relies on the trustworthiness of the ISP 
to honor a client's request by not caching a content that is marked as privacy-sensitive by a provider.
However, these approaches work under the assumptions that the ISP is trustworthy
and the privacy-sensitive content are unpopularity, which may not always be valid assumptions.

Chaabane~\kETAL~explored attacks against content privacy in~\cite{ChaDecKaa13}.
The authors introduced the monitoring and censorship attacks resulting from information
exposure from caching routers.
To cope with content privacy issues, the use of secure tunneling with symmetric/asymmetric
encryption (like SSL/TLS).
However, secure tunneling undermines the utility of caching, increasing  
core network load and content retrieval latency.
As an alternative solution, the authors proposed broadcast encryption and proxy
re-encryption, which in turn suffer from significant communication and computation 
overhead.
Also, it is common knowledge that even with data encryption, monitoring of encrypted communication 
can leak information through traffic analysis. 

In~\cite{ComConGas}, Compagno~\kETAL~proposed a method to geographically localize a client 
To mount this attack, the attacker uses several distributed hosts (zombies or bots) to request 
contents that they suspect a victim(s) may request. 
The aim is to identify corresponding cache or PIT hits.
Precise time measurements and complete knowledge of the network topology and 
several other network properties are important in this attack. 
The authors noted that this attack is only effective when the victim requests unpopular 
content--a popular content is requested by many and hence monitoring a few entities is difficult. 
%
Although the study is interesting, the assumptions especially about complete 
network knowledge is strong and not practical. 
Also, the authors present no countermeasure. 
%

\subsubsection{{\it {\bfseries Summary and Future Directions in Communication Monitoring Mitigation}}}
\label{subsec03-02-02}
Solutions to this attack disable caching of sensitive content either by 
creating a secure tunnel~\cite{DibGasTsu11,TouMisKli15} or with the clients flagging the 
requests as non-cacheable for privacy~\cite{LauLaoRod12Tech,LauLaoRod12ACM}. 
These solutions are applicable for all ICN architectures.
However, we believe that undermining network's caching capabilities is not a desired 
solution--it increases communication complexity and cost.
%
%
Although we agree that secure tunneling is a viable approach, we believe an
efficient tunneling mechanism should be designed, which at least allows partial 
content caching.
Another direction to research is naming scheme randomization~\cite{AriKopRag11}, which 
would make content-name prediction difficult for attackers. 
If manifests are used (metadata to create chunk names), they can contain 
encrypted information on how to request the random chunks, which only a legitimate 
client can decrypt. 
The requirement of decryption will also serve as an attack deterrent in general. 
Strengthening the vulnerable architectural features, such as scope, exclusion, and prefix 
matching would help reduce the attack scenarios for the affected schemes. 
Of course, they come at the expense of efficiency resulting from these features. 
%
%
\subsection{Anonymity and Censorship Mitigation}
\label{subsec03-03}
%
%
\begin{figure}[!b]
\centering
\includegraphics[height=1.4in]{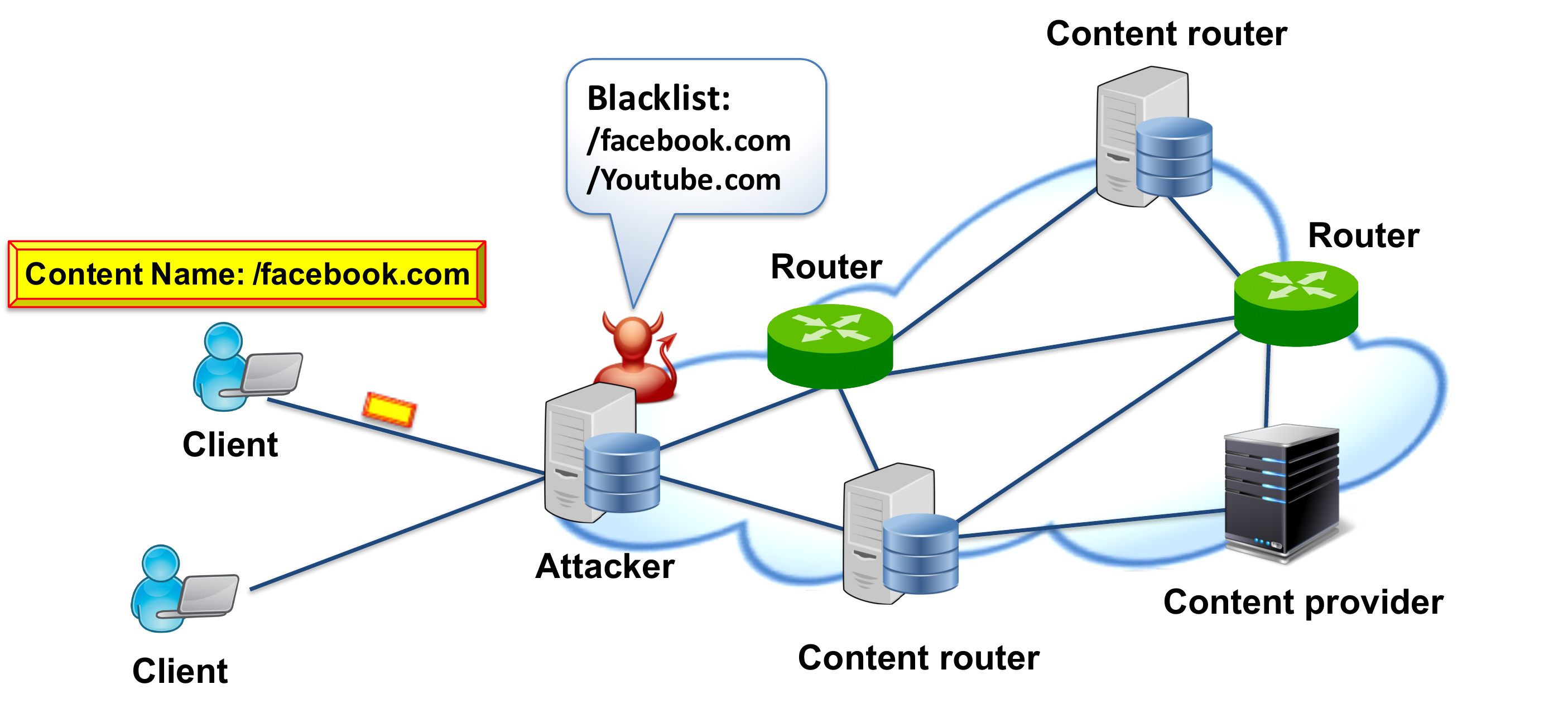}
\caption{Censorship risk due to lack of anonymity.}
\label{fig03-03}
\vspace{-0.1in}
\end{figure}
As in other networks, anonymous communication is important in ICN as well.
Lack of anonymity may reveal critical information about the clients and the requested
contents, which could be used to enable censorship.
Unlike in IP networks, in ICN the packet carries the name of the content requested. 
The name in the interest (be it a human readable name, a hashed string, or a self-certifying 
name) can be used by an intermediate router to filter and drop it. 
The name can also be used by the first-hop router or proxy to censor the clients. 
%
%
As depicted in Fig.~\ref{fig03-03}, an on-path adversary monitors the client's interest and 
compares the requested content name against its contents' blacklist for censorship.
A match results in the request being dropped--an {\em effective censorship} mechanism. 

\begin{figure}[!t]
\begin{center}
\begin{tikzpicture}[level distance=1.5cm,
  level 1/.style={sibling distance=4.7cm},
  level 2/.style={sibling distance=2.2cm}]
  \node {Anonymous Communication \& Censorship Mitigation}
    child {node [align=center] {Non-proxy-based}
      child {node [align=center] {Steganography \\ ~\cite{AriKopRag11}}}
      child {node [align=center] {Encryption \\ ~\cite{ElaBenMet11,FotTroMar14}}}
    }
    child {node [align=center] {Proxy-based}
      child {node [align=center] {Encryption \\ ~\cite{DibGasTsu11,ChuKimJan14}}}
      child {node [align=center] {Coding \\ ~\cite{TaoFeiYe15,TouMisKli15}}}
    };
\end{tikzpicture}
\end{center}
\caption{Anonymous communication and censorship mitigation approaches are categorized into whether 
they use a proxy or not.}
\label{fig03-03-Censorship}
\end{figure}
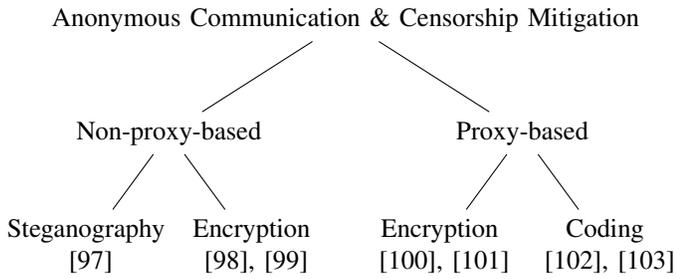

The exposure of the content name, and the semantic binding between the name and the content 
itself, raise new privacy and censorship concerns.
Several anti-censorship mechanisms have been proposed in the literature~\cite{AriKopRag11, 
DibGasTsu11, ChuKimJan14, FotTroMar14, TouMisKli15, TaoFeiYe15}.  
%
As it is illustrated in Fig.~\ref{fig03-03-Censorship}, we categorized the proposed mitigation 
mechanisms into {\it non-proxy-based} and {\it proxy-based} categories.
The non-proxy-based mechanisms employ steganographic and/or encryption to provide privacy. 
In the proxy-based category, consumers interact with a proxy that is responsible for the client and 
name privacy (by creating encrypted proxy-client tunnels). 
%
%
%
%
%
\subsubsection{{\it {\bfseries Non Proxy-based Mechanisms}}}
\label{subsec03-03-01}
%
The anti-censorship mechanisms we discuss under non-proxy-based category either 
employ steganographic techniques to obfuscate content names or use ephemeral identities 
and homomorphic cryptography to enhance clients privacy.
Thus, we categorize these approaches into {\it steganographic} and {\it encryption} subcategories.
%
%
\paragraph{\underline {Mitigation Employing Steganography}}
%
In schemes that employ steganography, the objective is to obfuscate the content chunks' names, thus 
increasing the computational complexity of deciphering the chunks' names for the attackers, who are unaware of 
the name generation schemes.
Arianfar~\kETAL~were one of the first to study this problem~\cite{AriKopRag11}. 
They proposed a name obfuscation scheme in which the content provider uses a 
secret cover file--a random file of the same size as the content. 
The provider splits the content and the cover into same sized blocks and runs an 
exclusive-or operation on all combinations of $k$~$(\geq 2)$ blocks of the content 
and the cover to create the corresponding encoded content chunks that are then 
published into the network. 
The name of an encoded chunk is the hash of the hashes of 
the names of the corresponding content and cover blocks respectively.  

Utilizing a secure back channel, the provider sends each verified requesting client 
the necessary metadata, such as the content hash, the content's length in blocks, the 
corresponding cover blocks, the names, and the name generation algorithm. 
Using this meta-data the client generates the chunk names, requests them from the 
network, and deciphers them. 
Although the chunks and their names are publicly available, an adversary cannot decipher 
the content without the metadata; and it is computationally expensive to break the scheme 
to decipher the chunk names. 

The size overhead of the scheme is significant. 
The cover file represents a $100\%$ overhead, and must be transmitted via a secure 
back channel for each client--not scalable. 
In fact, if a secure back channel exists, that can be used to send the file itself. 
%
%
\paragraph{\underline {Mitigations Employing Encryption}}
%
Approaches employing encryption either exploit temporary identities or leverage homomorphic 
cryptography to prevent client identity-based interest filtering. 
Elabidi~\kETAL~\cite{ElaBenMet11} proposed a privacy protection scheme, which enforces 
identity expiration.
The system is composed of identity providers, trust verification providers, and digital identity 
protection authorities in addition to the standard network elements.
The scheme provides users with ephemeral identifiers (by identity providers), which they communicate 
to the service providers. 
The service providers authenticate the users through a trust verification provider.
The trust verification provider informs the digital identity protection
authority when an ephemeral identity is used after its expiration.
Though this design provides the useful ``forgetfulness'' property for the identities, a 
malicious service provider could disable access or filter requests from users by corrupting the 
ephemeral identities and preventing access for clients. 
Other issues with this scheme include need for several entities and the requirement of user authentication 
by a third-party service, which raise concerns of overhead and availability.
\begin{table*}[!ht]
\centering
\caption{Summary of the Proposed Mechanisms for Anonymous Communication and Censorship Mitigation}
\label{table:03_02}
\begin{tabular}{|l c c c|}
 \hline
 {\bf Mechanism}&{\bf Approach}&{\bf Infrastructure}&{\bf Computation Complexity} \\
 \hline
 \hline
 {\bf Non Proxy-Based Approaches} & & & \\
 \hline
 {\it Steganography} & & & \\
 Arianfar~\kETAL~\cite{AriKopRag11} & Encoding interest by mixing content and cover file & Not Applicable& High (cover \& exclusive-or)\\
 \hline
 {\it Encryption} & & & \\
 Elabidi~\kETAL~\cite{ElaBenMet11} & Ephemeral identities for users & Requires three new entities & High (several interactions) \\
 Fotiou~\kETAL~\cite{FotTroMar14} & Hierarchical DNS based brokering model & Brokering Network & High (homomorphic cryptography)\\
 \hline
 \hline
 {\bf Proxy-Based Approaches} & & & \\
 \hline
 {\it Encryption} & & & \\
 DiBenedetto~\kETAL~\cite{DibGasTsu11} & TOR based model -- 2 layers of encryption & Two Proxies & Moderate (symmetric key)\\
 Chung~\kETAL~\cite{ChuKimJan14} & TOR based model -- 2 layers of encryption & Two Proxies & Moderate (symmetric key)\\ 
 \hline
 {\it Coding} & & & \\
 Tao~\kETAL~\cite{TaoFeiYe15} & Random linear network encoded interest & One Proxy & Moderate (RLNC + PKI)\\
 Tourani~\kETAL~\cite{TouMisKli15} & Huffman encoded interest & One Proxy & Low (Huffman coding)\\
 \hline
\end{tabular}
\end{table*}

Fotiou~\kETAL~\cite{FotTroMar14} proposed a mechanism to preserve content lookup
privacy by leveraging homomorphic cryptography~\cite{VanGenHal10}. 
The scheme involves cooperation between providers, clients, and a hierarchical brokering
system--a tree of brokering nodes.  
A provider publishes its content identifier to the brokering system, which
disseminates the identifier-provider pair to the leaf brokering nodes.
To locate a content, a client submits an encrypted query to the root broker node.
By employing homomorphic cryptography, the query can be resolved by the brokering
system without decryption.
When the content is found, the client will be sent an encrypted response
containing a pointer to the desired content provider.

In this scheme, a query includes a vector of sub-queries corresponding to the nodes in 
the brokering system. 
Each broker using its part in the sub-query to forward the query to its children recursively until 
the content is identified. 
%
A big pitfall of the mechanism is it requires $2^{h-1}$ decryption operations to locate a content 
at level $h$ in the tree-hierarchy. 
In addition, considering the number of messages transmitted per query, the
system scales poorly in the face of an increasing number of clients and
contents.
%
%
\subsubsection{{\it {\bfseries Proxy-based Mechanisms}}}
\label{subsec03-03-02}
%
In proxy-based approaches, a client needs to interact and share a secret with a proxy 
(a network of proxies).
The proxy is responsible for decrypting/decoding clients' requests, retrieving the 
requested content, and returning the encrypted/encoded content to the clients.
The approaches are similar in spirit to the popular Tor (The onion routing protocol--the 
popular anti-censorship tool for IP networks).
Based on how the layered-encryption is performed, we categorize the proposed proxy-based 
approaches into {\it encryption-based} and {\it coding-based}.
%
\paragraph{\underline {Encryption based Mitigation}}
%
%
%

ANDaNA~\cite{DibGasTsu11}, a tunneling-based anti-censorship protocol, uses two
proxies--one proxy adjacent to the requester, and another proxy closer to the destination--to 
create a tunnel with two layers of encryption.   
By using ANDaNA, a client decouples its identity from its request.
The first proxy is only aware of the client's identity (but not the content name),
while the second proxy can only identify the requested content (not the client's
identity). 
The interest travels unencrypted between the second proxy and the provider. 
The authors proposed an {\em asymmetric} version of the protocol where the two-layers of encryption 
are performed using the proxies public keys, with the packets decrypted by the proxies using 
their private keys. 
%
The content on its way back is encrypted using symmetric keys shared by each proxy with 
the client.   

Due to the high cost of the PKI operations, the authors proposed a {\em symmetric} key based 
session-key model to replace PKI operation. 
Despite ANDaNA's usefulness as an anti-censorship tool, it induces
significant delays in content delivery (ref. results in~\cite{DibGasTsu11}) in
comparison to Tor.
These delays are caused, in part, by the process of setting up the secure channel.

In~\cite{ChuKimJan14}, Chung~\kETAL~took a similar approach to ANDaNA and Tor.
In this approach, the client encrypts the interest packet with two symmetric keys that will be shared 
with two Anonymous Routers (ARs).
The interest's encryption order follows the onion routing model.
Different from conventional onion routing, an identifier (a hash of the content
name) is embedded in the encrypted interest to enable cache utilization 
(i.e., CS-lookup) and interest aggregation (PIT lookup) at the first AR.
The provider transmits the content to the closest (second) AR in plaintext.
The content response on the way back may be cached on the second AR, which encrypts the content  
and forwards it to the first AR.  
The first AR decrypts the content for caching before re-encrypting
it and forwarding it towards the client.
Similar to ANDaNA, this scheme suffers from the same high cost of multiple per-packet 
encryptions/decryptions. 
%
%
\paragraph{\underline {Coding based Mitigation}}
%

Unlike encryption-based anti-censorship approaches, the mechanisms in coding-based category 
employ coding techniques, such as random linear network coding and Huffman coding to protect 
clients privacy. 
In these mechanisms, a client only needs to interact with a single proxy, which performs interest 
and content encryption/decryption.
Tao~\kETAL~\cite{TaoFeiYe15} proposed a mechanism leverages ICN's inherent content chunking 
in conjunction with random linear network coding (RLNC).
To request a content chunk, the client splits the interest into small chunks and
encrypts a linear combination of the chunks with the public key of an intermediate trusted proxy.
The proxy, after receiving enough interest chunks, reconstructs the original
interest packet and sends it toward the content provider.
The content provider follows the same approach as the client, splitting the
content into small chunks and forwarding a linear combination of them 
towards the proxy.
The two major concerns of this proposed scheme are a lack of cache utilization
and the high cost of many asymmetric-key cryptographic operations. 

Tourani~\kETAL~\cite{TouMisKli15} addressed the ICN censorship problem, by proposing 
a client anonymity framework that leverages the prefix-free coding technique.
In their proposed design, each client shares a unique Huffman coding table with
an anonymizer, which may be collocated with the content provider or an
intermediate trusted router.
The client encodes the content chunk's name postfix (part of the name after the domain name) 
using its Huffman coding table, leaving the domain name in plaintext, to be used for routing. 
The authors also proposed ways to encode the whole name (when the domain is also censored) 
with the help of network entity, named the anonymizer. 

When an encoded interest reaches the anonymizer, the name is decoded
and the interest with the unencrypted name is forwarded to the content provider.
The content provider sends the content chunk in plaintext to the anonymizer (caching can 
be leveraged on the path), which then encrypts the content name and forwards it to the client. 
%
The routers between the anonymizer and the provider can identify the content, but cannot 
identify the requester, while the routers from the anonymizer to the client cannot identify the 
name, thus preserving client privacy.
%
The paper did not have a trade-off analysis between cache utilization and 
privacy preservation, and did not discuss the scope of potential differential 
cryptanalysis attacks. 
However, it is one of the approaches with the least overhead/latency.
%

\subsubsection{{\it {\bfseries Summary and Future Directions in Anonymity and Censorship Mitigation}}}
\label{subsec03-03-03}
Table~\ref{table:03_02} summarizes the existing anonymous communication mechanisms and 
presents their infrastructure requirements and computation complexities.
Note that the technique proposed by Tourani~\kETAL~has the lowest computation complexity 
and infrastructure cost.
Some of the existing anti-censorship solutions~\cite{DibGasTsu11,ChuKimJan14,TaoFeiYe15,TouMisKli15}, 
have achieved anonymous communication through secure tunneling, where the content is encrypted 
between the providers/proxy and clients.
Other approaches include a name obfuscation scheme~\cite{AriKopRag11} and a hierarchical 
brokering network~\cite{FotAriSar14} for anonymous content retrieval.
Expensive cryptographic operations~\cite{AriKopRag11,DibGasTsu11,ChuKimJan14}, requirement 
for a secure back channel~\cite{AriKopRag11}, and undermining of in-network 
caching~\cite{DibGasTsu11,TaoFeiYe15,TouMisKli15} are the main pitfalls of these mechanisms.
Except the work by Fotiou~\cite{FotAriSar14} that targets architectures with brokering network 
(e.g., PSIRP and PURSUIT), other proposed solutions (e.g., tunneling, name obfuscation, and network 
coding based mechanisms) are applicable to all ICN architectures.

There are some potential directions for future research on cache utilization optimization and
reduction in the cost of cryptographic operations.
Applying cryptographic operations on a subset of content chunks to reduce cost
has not yet been explored.
Exploiting low-complexity network coding techniques~\cite{TaoFeiYe15,TouMisKli15} instead of traditional
cryptography would be a good idea to expand the applicability of tunneling schemes. 
This is especially important given that the majority of devices in the future will be resource-constrained devices 
(e.g., mobile devices, Internet of Things, etc.).
\begin{figure}[!b]
\centering
\includegraphics[height=1.4in]{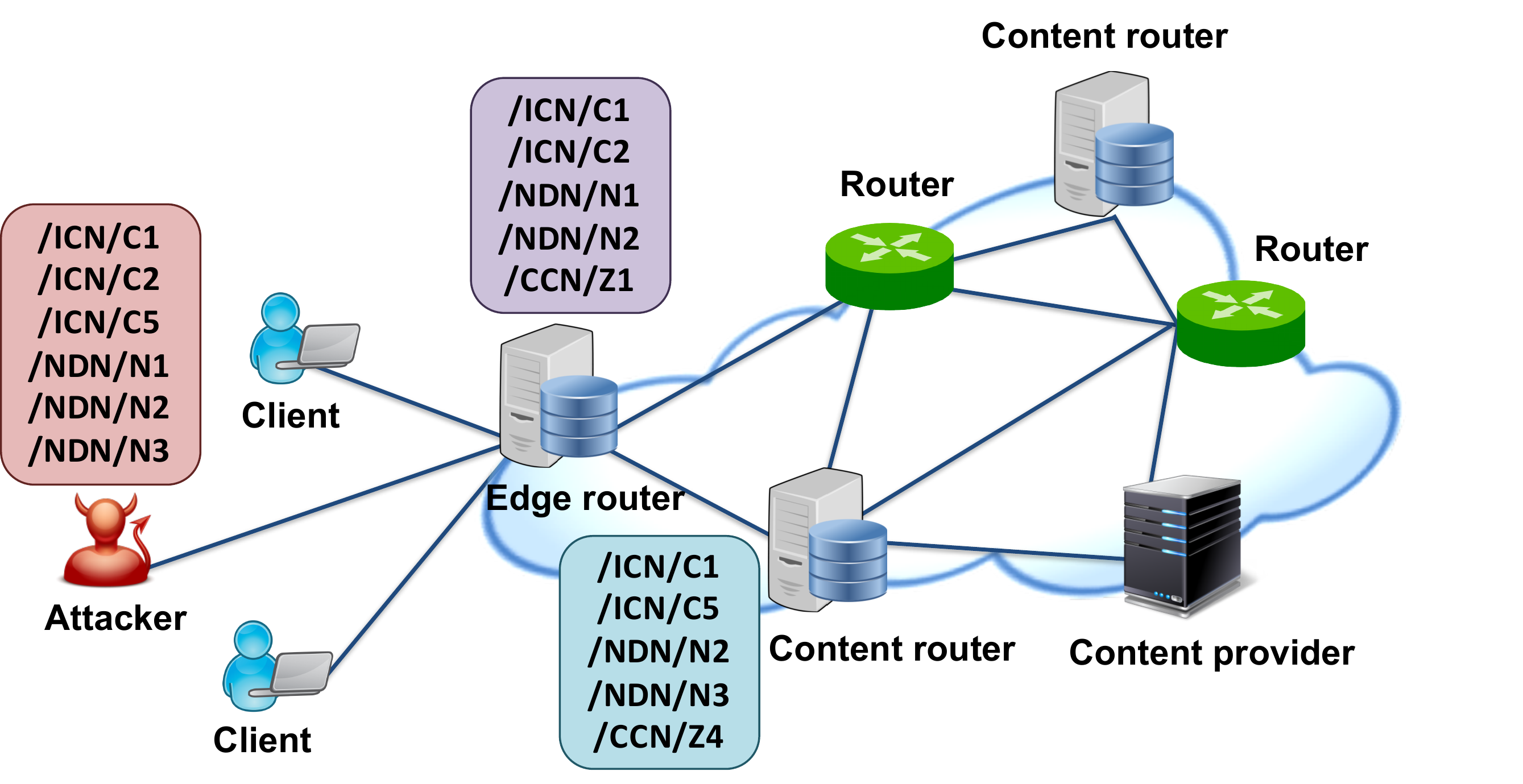}
\caption{Protocol attack scenario.}
\label{fig03-04}
\vspace{-0.1in}
\end{figure}
%
%
\subsection{Discovery and Protocol Attacks}
\label{subsec03-04}
%
Discovery and protocol attacks are a result of intrinsic design features of
CCN and NDN architectures (only applicable to these architectures).
Some examples of these features are the interest packet scope field
and the name-based matching used in NDN.
Fig.~\ref{fig03-04} illustrates a discovery attack, in which an attacker
probes all caches in a two-hop locality for content with prefixes  {\it /abc} and
{\it /XYZ}.
In this subsection, we review two of the articles that addressed the pitfalls of
these design features.

The authors in~\cite{LauLaoRod12ACM,LauLaoRod12Tech} introduced
an object-discovery attack, which abuses NDN's~\cite{CCNx} prefix matching 
and exclusion pattern features.
The attacker employs the prefix matching feature to probe for all cached content
objects under a particular name prefix starting at the root of the namespace, say {\it /www.google.com/}, 
and iteratively exploring it by using interests with exclusions and forcing intermediate 
routers to walk through the namespace.  
With the exclusion feature an attacker can discover the whole namespace (quickly for small 
namespaces) and also the names of cached content (additional monitoring attacks). 

Chaabane~\kETAL~\cite{ChaDecKaa13} also defined two protocol attacks 
based on prefix matching and scoping respectively.
The prefix matching attack works as described by Lauinger~\kETAL~\cite{LauLaoRod12Tech}. 
In the scoping attack, an attacker probes all the available content objects
in nearby caches by leveraging the scope field in the interest packet.
By carefully selecting the scope, the attacker can identify the content
available in individual routers, thereby breaching the privacy of other clients.
However, no solution has been proposed for these attacks except for the
removal of the enabling features.
%

\subsubsection{{\it {\bfseries Summary and Future Directions in Discovery and Protocol Attacks}}} 
\label{subsec03-03-04}
The use of prefix matching, exclusions, and the scope field are examples of features that can be attacked 
in some ICN architectures to probe for popular content objects and explore the content namespace. 
Prefix-matching feature is useful for legitimate clients with limited knowledge
of their desired content name (e.g., when only a prefix of the content name is known).
The scope field can also be employed by a legitimate client who would like to obtain
a content only in the case that it is available in a nearby cache.
Therefore, these features should not be completely eliminated from ICN, but
instead should be redesigned with these threats in consideration.
Potential solutions may be the use of rate-limiting requests for a specific namespace, 
similar to what is done by DNS servers today. 
We believe that there is a need for a comprehensive analysis, both analytical and experimental, of these 
features to identify their trade-offs. 
%

%
%
%
%
Developing mechanisms to help routers to validate the integrity and authenticity of the z-filters needs research focus.
%

\subsection{Name and Signature Privacy}
\label{subsec03-05}
%
Unlike the current Internet, several ICN architectures require the content
to be explicitly requested by name.
In ICN, names either follow a hierarchical human-readable or a self-certifying flat-name model.
We refer readers to a survey on ICN naming and routing~\cite{BarChoAhm12} for
more details.
In the human-readable naming convention, the content name exposes information
about the content and the provider due to the inherent semantic binding.
\begin{figure}[!ht]
\begin{center}
\begin{tikzpicture}[level distance=1.5cm,
  level 1/.style={sibling distance=5cm},
  level 2/.style={sibling distance=2.5cm}]
  \node [align=center] {Name \& Signature \\ Privacy}
    child {node [align=center] {Name Obfuscation \\ ~\cite{BauDavNar12,ChaDecKaa13,KatSaiPsa14}}
    }
    child {node [align=center] {Overlay Network \\ ~\cite{MarGomGir11,MarGom12,Sol12}}
    };
\end{tikzpicture}
\end{center}
\caption{Naming and signature privacy sub-classes and the state-of-the-art in censorship.}
\label{fig03-05-Name}
\end{figure}
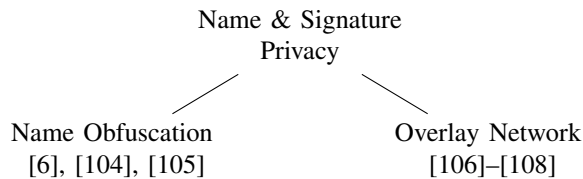
\begin{table*}[!ht]
\centering
\caption{Summary of the Proposed Approaches to Augment Name and Signature Privacy}
\label{table:03_03}
\begin{tabular}{|l c c c|}
 \hline
 {\bf Mechanism}&{\bf Approach}&{\bf Advantage}&{\bf Drawback} \\
 \hline
 \hline
 {\bf Name Obfuscation} & & & \\
 \hline
 Baugher~\kETAL~\cite{BauDavNar12} & Cryptographic Content Hash-based Naming & Easy Authentication-Provenance & Not Suitable for Dynamic Content \\
 Chaabane~\kETAL~\cite{ChaDecKaa13} & Bloom Filter-based Naming \& Group Signature & Increased Publisher Privacy & Bloom Filter Size \& False Routing \\
 Katsaros~\kETAL~\cite{KatSaiPsa14} & Employing Ephemeral Names & Increased Publisher Privacy & Undermine Caching\\
 \hline
 \hline
 {\bf Overlay Network} & & & \\
 \hline
 Martinez~\kETAL~\cite{MarGomGir11,MarGom12} & Digital Identity in an Overlay Network & Privacy for Real Identities & Additional Infrastructure \\
 Sollins~\cite{Sol12} & Overlay Network with Identifier Resolution Service & Privacy for Real Identities & Lack of Compatibility Analysis \\
 \hline
\end{tabular}
\end{table*}
%

Fig.~\ref{fig03-05-Name} illustrates our categorization of the literature in name 
and signature privacy starting with two broad categories: {\it name obfuscation} and {\it overlay 
network}.
The proposed approaches in name obfuscation try to enhance name privacy 
by switching from the human-readable naming to machine-readable naming convention.
In the overlay category the approaches use an overlay network in 
conjunction with a name resolution service to securely map the real identities to digital identities.
%
%
%
\subsubsection{\it {\bfseries Name Obfuscation}}
\label{subsec03-05_01}
%
The proposed name obfuscation use machine-readable naming 
schemes, which are generated by content digest, Bloom filter, and the use of ephemeral names.
Cryptographic hash based naming was motivated by Baugher~\kETAL~\cite{BauDavNar12}.
The main advantage of such self-verifying names (names are cryptographic hash of the content) 
is the low cost of content authentication.
In these schemes, a client obtains a content's (or chunk's) self-verifying name from
a catalog that maps contents from their human-readable names to their hashes.
The client stores the hashed name for future use and submits a request for the content 
corresponding to the hashed name into the network.
It accepts the retrieved content if its cryptographic hash matches the self-verifying 
name from the catalog.
This mechanism can also be used to preserve the privacy of the provider. 

The authors noted that hash-based naming is only useful for read-only, cacheable data objects.
Additionally, the use of the catalog to obtain self-verifying names requires
the establishment of trust between clients and the catalog publisher, which requires creation 
of trusted infrastructure in the network a potential overhead. 

Chaabane~\kETAL~in~\cite{ChaDecKaa13} discussed the privacy concerns emanating from the 
%
semantic correlation between the human-readable names and the content/provider identity, 
including potential leaks from digital signatures. 
%
%
They suggested the use of one Bloom filter for each name in the hierarchy to represent names 
without correlating with the content.
To protect publisher privacy, they proposed different schemes such as
confirmer signature, group signature, ring signature, and ephemeral identity.
All of these solutions, except ephemeral identity, achieve signature privacy by
increasing the cardinality of the anonymity-set of signers.
Under ephemeral identity, frequently changing temporary identities used by a publisher 
prevent an attacker from identifying the publisher based on its signature.
However, the probabilistic nature of Bloom filter and potential for false-positives 
may cause false routing and incorrect interest to content-chunk mapping. 
Furthermore, the size of the Bloom filters could be large and the lookup latency 
will increase with increasing levels in the name hierarchy. 

Katsaros~\kETAL~also investigated ephemeral names for content to improve publisher 
privacy~\cite{KatSaiPsa14}.
Despite the benefits of using ephemeral names for content providers, temporary
naming undermines the network's caching capability.
Contents with ephemeral names will expire and will be purged from
the caches, hence they will not be available to meet clients' requests; this is 
especially true for popular content.
%
%
\subsubsection{\it {\bfseries Overlay Network}}
\label{subsec03-05-02}
%
This category of secure naming leverages an overlay network in which entities are associated 
with identities that are only known in that domain.
The overlay network uses a name resolution service to map the entities to their identities.
Martinez~\kETAL~\cite{MarGomGir11,MarGom12} proposed such as scheme for privacy and untraceability. 
Each network entity (users, machines, services, hardware) is associated with a
digital identity and a domain.
Each domain is equipped with a Domain Trust Entity (DTE), which manages entity-identifier 
associations and identifier authentication. 
%
The DTEs form an interconnected infrastructure, which facilitates identity-based communication. 
For two entities to establish a communication channel, the first entity authenticates itself 
to the DTE infrastructure and submits a query seeking the other.
The DTE infrastructure processes the query and returns the identifier of the other entity.
The identifiers are used to establish a secure tunnel through the DTEs.
Although this overlay network preserves the entities' identities, the network's security 
can be undermined by compromised DTEs, which themselves form additional network infrastructure.

Sollins~\cite{Sol12}~discussed the design issues with names in ICN and proposed
an overlay naming system for content identification.
The naming system uses the scope of the ID space (local,
global), the ID syntax (size, structure, character set), and the ID structure 
(flat, hierarchical, composite).
In addition, identifier-object mapping requires the existence of a
naming authority to enforce ID lifetime and uniqueness and a name resolution system.
%
%
The author designed a Pervasive Persistent Object ID
(PPOID), based on the principles of layering and modularity.
With PPOID, a human-readable identifier is mapped into an ID space, which 
resolves to an ICN identifier.
Simple and expressive user-friendly identifiers at the top layer are mapped onto 
machine-readable identifiers for real-time resolution and delivery.
%
%
However, the author did not discuss the applicability of this naming
system to the existing popular ICN architectures and challenges. 
%

\subsubsection{\it {\bfseries Summary and Future Directions in Name and Signature Privacy}}
\label{subsec03-05-03}
Table~\ref{table:03_03} summarizes the proposed mechanisms for preserving name and signature privacy.
We present the referenced work, their approaches to augment the naming and signature privacy and their advantages  
along with their drawbacks.
%
The proposed approaches for name and signature privacy include overlay-based network~\cite{MarGomGir11,MarGom12,Sol12}, self-verifying 
names~\cite{BauDavNar12}, and hierarchical Bloom filter based naming~\cite{ChaDecKaa13}.
The drawbacks of the overlay-based models is their dependency on trusted entities and additional latency for resolving content names. 
The proposed hierarchical Bloom filter naming approach~\cite{ChaDecKaa13}, suffers from false positives.
The self-verifying naming approach~\cite{BauDavNar12} is only applicable to read-only content, 
not to dynamic contents, which are generated upon request.
For dynamic content no catalog can be generated ahead of time.
%
%

We believe an efficient approach in this context could be for the provider and the user to cloak their identities by using 
several certificates to map to several identities and using the identities at random. 
This is similar to the $k$-anonymity mechanism used to create an anonymity-set for an identity. 
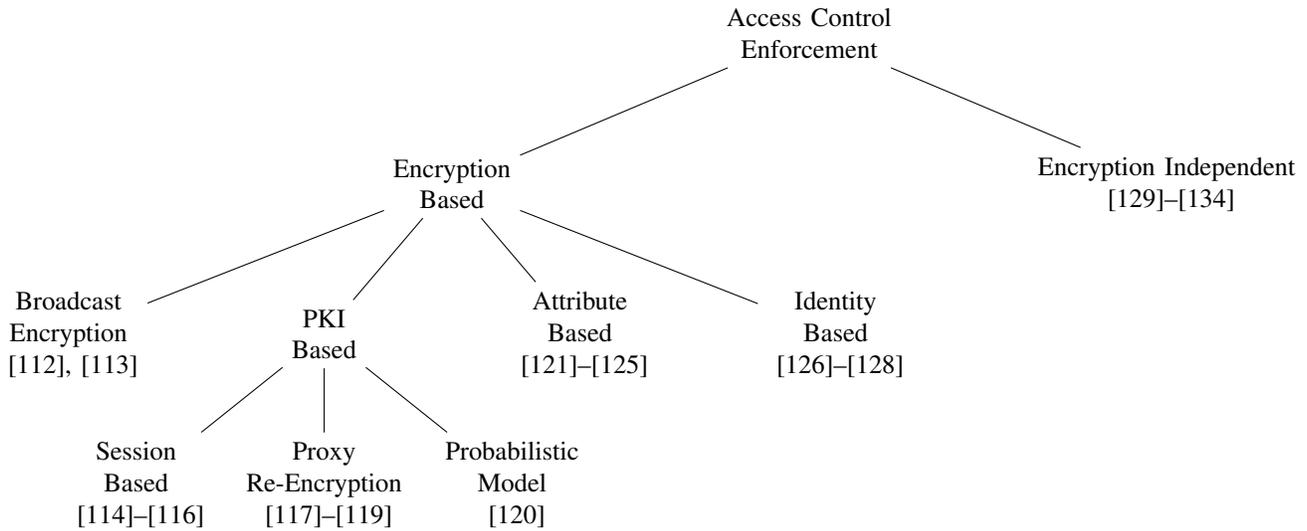
\begin{figure*}[!ht]
\begin{center}
\begin{tikzpicture}[level distance=2.0cm,
  level 1/.style={sibling distance=9.5cm},
  level 2/.style={sibling distance=3.4cm},
  level 3/.style={sibling distance=2.5cm}]
  \node [align=center] {Access Control \\ Enforcement}
    child {node [align=center] {Encryption \\ Based}
       child {node [align=center] {Broadcast \\ Encryption \\ ~\cite{MisTouMaj13,MisTouNat16}}}
       child {node [align=center] {PKI \\ Based}
          child {node [align=center] {Session \\ Based \\ ~\cite{RenAhmAbi09,RenAhmAbi10,WanXuFen14}}}
          child {node [align=center] {Proxy \\ Re-Encryption \\ ~\cite{WooUzu14,ManMarPar15,ZheWanRav15}}}
          child {node [align=center] {Probabilistic \\ Model \\ ~\cite{CheLeiXu14}}}
       }  
       child {node [align=center] {Attribute \\ Based \\ ~\cite{IonZhaSch13,LiVerHua14,LiWanHua14,DasZor15,RayKazLak15}}}
       child {node [align=center] {Identity \\ Based \\ ~\cite{HamSerFat13,HamFat15,AiaLoo15}}}
    }
    child {node [align=center] {Encryption Independent \\ ~\cite{KurUzuWoo15,FotMarPol12,Sin12,TanZhoZou14,GhaSchTsu15,LiZhaZhe15}}
    };
\end{tikzpicture}
\end{center}
\caption{A classification of existing access control enforcement mechanisms.}
\label{fig04-01}
\end{figure*}
\section{Access Control in ICN}
\label{sec04}
%
%
In this section, we explore the proposed access control (AC) enforcement mechanisms
for ICNs.
The unique characteristics of ICN, such as name based routing and in-network caching make 
AC management more important. 
By design most ICN architectures are requesting host agnostic. 
Thus, once content is disseminated in the network it can be cached and disseminated by 
network routers to satisfy requests without the routers checking if the 
requesting entity can access the content. 
This in turn could lead to content providers losing control over who accesses their content.   
Researchers in the domain have recently started exploring this problem. 
%
%
%

As depicted in Fig.~\ref{fig04-01}, we categorize the-state-of-the-art in ICN access 
control based on whether they use a particular encryption technique or are independent 
of the underlying encryption used as {\it encryption-based} and {\it encryption independent} 
categories.
The encryption-based category is further subdivided, based on the type of encryption  
into {\it broadcast encryption}, {\it PKI}, {\it attribute-based}, 
and {\it identity-based} subcategories.
The encryption independent category presents approaches that present AC frameworks that can 
use any encryption algorithm for performing AC. 
%
%
%
We discuss these categories in more details in what follows. 
%
\subsection{Encryption-Based Access Control}
\label{subsec04-01}
%
%
All proposed encryption-based approaches are conceptually similar--the content providers 
encrypt their content before disseminating them into the network.
Clients need to authenticate themselves and obtain the content decryption keys to 
be able to decrypt and consume the content.
%
%
%
\subsubsection{\it {\bfseries Broadcast Encryption Access Control}}
\label{subsec04-01-01}
%
Broadcast encryption allows a content provider to encrypt its 
content using a single key for all clients; the clients use their individual keys to 
decrypt the content.
It also allow efficient revocation of the clients (without content re-encryption).
A secure content delivery framework, which waives the necessity of an online
authentication service was proposed by Misra~\kETAL~\cite{MisTouMaj13,MisTouNat16}.
The framework uses the $(n,t)$-Shamir's secret sharing based broadcast
encryption to enforce AC.
The framework's strength is that it needs no additional authorization entity nor incurs extra 
computational overhead at the routers.
%

For secure content delivery, the provider encrypts the content with a symmetric
content encryption key and disseminates it into the network.
In addition, the provider generates and disseminates a small amount of keying material 
(called enabling block, EB, and containing $t$-key shares) into the network. 
Only authorized clients can use the EB and their individual keys to decrypt 
the content encryption key and decrypt the content after that. 
The EB is requested by the client along with the content, and is cacheable. 
%

%
%
%
Client revocation is achieved by updating the EB by the replacement of one of the 
key shares with the revoked client's share, which disables the revoked client from 
decrypting the symmetric key.
%
%
In this mechanism, the EB is an overhead (minor for large contents, but significant for small ones). 
The EB update on client revocation also consumes network bandwidth. 
\begin{figure}[!ht]
\begin{center}
\begin{tikzpicture}[level distance=2.0cm,
  level 1/.style={sibling distance=3.5cm},
  level 2/.style={sibling distance=3.4cm},
  level 3/.style={sibling distance=2.5cm}]
  \node [align=center] {PKI \\ Based}
    child {node [align=center] {Session \\ Based \\ ~\cite{RenAhmAbi09,RenAhmAbi10,WanXuFen14}}
    }
    child {node [align=center] {Proxy \\ Re-Encryption \\ ~\cite{WooUzu14,ManMarPar15,ZheWanRav15}}
    }
    child {node [align=center] {Probabilistic \\ Model \\ ~\cite{CheLeiXu14}}
    };
\end{tikzpicture}
\end{center}
\caption{A classification of the existing PKI-based access control enforcement mechanisms.}
\label{fig04-02}
\end{figure}
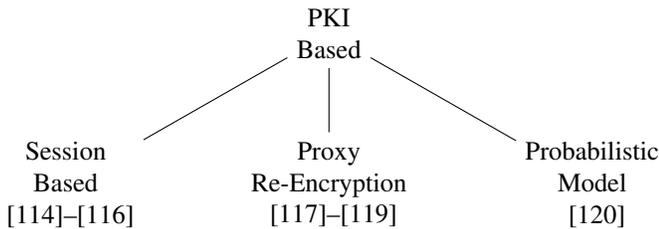
%
%
\subsubsection{\it {\bfseries PKI-Based Access Control}}
\label{subsec04-01-02}
%
As shown in Fig~\ref{fig04-02}, we categorize the PKI-based mechanisms into 
{\it session-based}, {\it proxy re-encryption}, and {\it probabilistic} subcategories.
%
%
\paragraph{\underline {Session-Based Access Control}}
%
The state-of-the-art in session-based AC suggests establishment of a secure 
session between a client-provider pair after client authentication and authorization.
Within a secure session, the client can request content from the provider.
Renault~\kETAL~\cite{RenAhmAbi09, RenAhmAbi10} proposed a session-based access
control mechanism for NetInf.
This mechanism requires a security controller, collocated with
each content storage node, to check the access rights of clients.
A client and the security controller establish a secure channel and exchange public keys using 
the Diffie-Hellman key exchange protocol, thus requiring no additional infrastructure.
%
%

The client requests a content using the content ID and its
own public key (the public key may be omitted for publicly available content).
On receiving a client's request, the security controller performs challenge-response 
with the client to verify the client's identity.
Upon verification, the controller checks whether the client is authorized to access
the content before forwarding the data; revocation can happen at this point.
The interactions take place in a secure session; the
session ends if either party explicitly requests its termination.

The main drawbacks of this scheme are: the cache between the client and the controller is effectively 
unusable and the need for the secure tunnel between the controller and the client for the duration 
of communication. 
The authors discussed the security of this mechanism against several well-known attacks, however 
they did not explore the potential for DoS/DDoS attack. 
A client can open one/more idle connections with the controller and exhaust the resources. 
Also, this connection-oriented set-up is antithetical to the connectionless ICN 
paradigm. 
%
%

Wang~\kETAL~\cite{WanXuFen14} designed a current IP-like session-based AC mechanism. 
%
The authors illustrated their design using the example of an online social
network (OSN).
%
A user registers in the OSN (content provider) by sharing a symmetric key and
its credentials with the OSN service.
Upon registration, the OSN provides a unique ID for the user.
The client logs in to interact with the OSN. 
%
It generates a new symmetric key and sends it to the OSN along with the login information.
The OSN then assigns a session ID to the client and stores a tuple consisting
of the session ID, the client ID, and the new key.

To upload content, a client needs to be authorized first. 
After authorization, the client encrypts the content with the previously shared 
symmetric key, then forwards it to the OSN along with its desired AC policy.
The OSN decrypts the content and re-encrypts the content with a newly generated symmetric key. 
Other clients request the content using its public name (obtained from a search in the OSN or 
a search engine).
The OSN, authorizes the client and its access to the content and returns the content's secure network 
addressable name, the symmetric key to decrypt the content, and the required metadata encrypted with the 
requester's session key. 
The requesting authorized client decrypts the message and requests the content by the secure name.
To prevent the public name-secure name correlation and access by revoked clients, the OSN changes the secure 
name at regular intervals. 
%
%

This scheme undermines the potency of in-network caching as renaming a popular content effectively 
invalidates it in the cache. 
It also results in a content existing under several names in the network, which violates the 
ICN's principle of content name immutability.
%
%
Also, content access overhead is high given that the process has to 
be repeated for each content.  
%
%
\paragraph{\underline {Proxy Re-Encryption-Based Access Control}}
%
In proxy re-encryption-based AC, a piece of information is re-encrypted 
by an intermediate proxy (a third party or an intermediate router) for each client.
%
%
Wood~\kETAL~\cite{WooUzu14} proposed a flexible mechanism for secure
end-to-end communication, leveraging a combination of proxy re-encryption and
identity-based encryption.
The content provider encrypts content using a symmetric key before dissemination.
A client may obtain a content from either a cache or the content provider.
Upon receiving the encrypted content, the client requests the symmetric key
from the content provider.
The provider validates the client's legitimacy and access level and sends the symmetric key to 
a validated client, encrypted with the client's identity.
The client extracts the received key and decrypts the content.

The proposed scheme reduces the cost of cryptography as only the symmetric key is 
encrypted individually for each client, the content is not.
However, contact with the content provider is required with each request,
even if the content can be retrieved from a cache.
This undermines content availability in the case of the provider's unavailability.
%

Mangili~\kETAL~\cite{ManMarPar15} proposed a framework for AC and 
track-ability in which content is broken into partitions and further into fragments 
allowing two layers of encryption by providers.
A provider encrypts the fragments into a chunk using a symmetric key 
that will be stored in the encrypted chunk.
In the second-layer of encryption, used for confidentiality and collusion prevention, 
a key-chain is generated using the ``key-regression'' key derivation algorithm~\cite{FuKamKoh06}.
An authenticated consumer regenerates the second-layer key by using a secret 
obtained from the provider.
To prevent collusion, the provider encrypts the first-layer encrypted chunks with
different second-layer keys (per user or group of users keys), which will be generates 
only for authorized clients.

On client revocation, the provider generates a new second-layer key and publishes 
the re-encrypted data.
The framework requires caching routers to regularly query the provider 
for newly encrypted chunks to replace the old ones.
Despite leveraging in-network caching, clients are required to perform per content 
authentication at the providers; requiring always online providers.
Furthermore, legitimate clients may end up with fragmented sets of chunks with 
each fragment of chunks encrypted with a different key.
This would require a client to download all the corresponding keys and identify 
which key decrypts which fragment.

Zheng~\kETAL~\cite{ZheWanRav15} proposed an AC mechanism which 
requires edge routers to perform content encryption.
The process starts with the publisher encrypting the content with its public key
and a random key $k_1$.
Upon a client's request for a content, the edge router selects a random key
$k_2$, and re-encrypts the encrypted content (as in proxy re-encryption).
The random key $k_2$ is encrypted by the publisher's public key and signed
by the edge router, and is attached to the content to be sent to the client.
To decrypt the content, the client sends the encrypted $k_2$,
the content name, and its identity to the publisher.
The publisher validates the client's identity and access level and upon validation 
uses its private key, along with $k_1$ and $k_2$ to generate the content decryption key $k$ for the client.
Upon receiving $k$, the client may decrypt the content.
Due to the randomness of the $k_2$ generated with each request, the
decryption key $k$ will be different for each client.

The performance analysis in the paper shows that the edge router's re-encryption
operation takes about $10$~seconds for a small content ($256$MB).
The need to use edge routers' resources for encryption undermines the scalability
of this solution, especially since the majority of the future Internet traffic is
expected to consist of large multimedia content.
%
%
\paragraph{\underline {Probabilistic Access Control}}
%
In the probabilistic AC, the network is equipped with Bloom filters 
for storing the authorized clients' public keys.
The intermediate routers use these Bloom filters to block unauthorized requests, 
which helps reduce clients' authentication cost.
Chen~\kETAL~proposed a probabilistic structure for encryption-based AC in~\cite{CheLeiXu14}.
Publishers and clients are equipped with public-private key-pairs, and each client
initially subscribes to a publisher by sending an interest.
The publisher stores a record for each registered client, noting
the client's credentials.
For efficiency the authors suggested PKI-bootstrapped symmetric key exchange 
between the publisher and the client.  
The content requested by the client is delivered encrypted. 
After receiving the content, the client authenticates itself to the publisher to securely obtain 
the symmetric decryption key.

The authorized clients' public keys are put into a Bloom filter, which is transmitted to 
network routers to allow them to filter invalid requests.
The interest of a client whose public key is not indexable in the content's Bloom filter is dropped. 
Although this procedure reduces network load, the recommended client revocation incurs costly content 
re-encryption and distribution.
The approach has two other drawbacks: 
Bloom filter's suffer from false positives--an unauthorized client's request can be satisfied with a small 
probability.
The size of the filters could rise rapidly with increasing number of clients. 
Second, is the need for authentication of the client at the publisher to obtain the symmetric key. 
This requires an {\em always-online} publisher (or another entity) to verify client credentials, which 
is difficult to guarantee. 
\subsubsection{\it {\bfseries Attribute-Based Access Control}}
\label{subsec04-01-03}
%
In attribute-based AC, a content is encrypted with a set of its attributes.
Each client is assigned a key, generated from the client's set of attributes.
The client can consume the content if she can use her attributes to decrypt the content-access policy, 
which is either embedded in the encrypted content or the decryption key.
Ion~\kETAL~\cite{IonZhaSch13} proposed an attributed-based encryption (ABE) mechanism
for AC enforcement that used either the key-policy or the ciphertext-policy 
based encryption models.

In the key-policy model, the content is encrypted with a key derived from the
content attributes, and the access policy is embedded in the decryption key.
A key authority grants different decryption keys to clients, based on their
attributes and access policies.
In the ciphertext-policy model, the AC policy contains the required
client attributes and is attached to the encrypted content.
The key authority issues a key for each client, in this case derived from the
client's attributes. 
Attribute and identity based encryption mechanisms require elaborate revocation 
procedures.
The authors did not describe the process of client revocation, and 
did not analyze the performance and efficiency of revocation in the scheme.

Li~\kETAL~\cite{LiVerHua14, LiWanHua14} used attribute-based encryption for access 
control enforcement in ICN.
In the proposed scheme, a trusted third party defines and manages the subject 
and object attributes by creating attribute ontology for each (ontology in this context is the 
universe of all attributes).
As the cached contents are available to all users, to 
prevent unauthorized access, the authors proposed a naming scheme, 
which preserves the privacy of the AC policy.
%
%
To publish a content, the publisher generates a random symmetric key with which
it encrypts the content.
The encrypted content, along with its corresponding metadata, is 
disseminated into the network.

The publisher also generates an AC policy from the
attributes defined by the trusted third party; the access policy then defines
which clients are authorized to access the content.
The publisher uses the AC policy to encrypt the symmetric key, 
which encrypts the content.
This encrypted symmetric key is the content name.
A client needs to retrieve the content name (possibly through some kind of domain 
name service) and extract the symmetric key using its attributes (only possible by 
an authorized client).
Despite its low overhead, the applicability of this scheme is questionable
due to the proposed naming scheme; the content name is generated by
encrypting the symmetric key with the AC policy.
Compromise of the symmetric key would necessitate re-keying and hence change the
content name, which undermines the spirit of immutable naming in ICN.
Also, client revocation remains a challenge. 

Da Silva~\kETAL~\cite{DasZor15} proposed an AC mechanism using attribute-based 
encryption for instantaneous access revocation.
The authors suggested the use of Ciphertext-policy ABE, in which the access
policy, generated by the provider, is embedded inside the encrypted content.
The content is encrypted with the required authorization attributes, which are stored in content 
routers.
Each content has an access policy, which is stored at a proxy.
Only the proxy can decrypt the access policy.

When the client registers with the application, it receives a key
(based on its attributes) and an ID.
For content retrieval, the client sends two interests: the first one retrieves
the encrypted content (from the publisher or a cache), and the second,
which includes the client ID and the content name, is sent to the proxy to decrypt
the access policy.
The proxy authenticates the client and decrypts the access policy on the client's
behalf; this decrypted policy is forwarded back to the client without being
cached in the network.
The client can decrypt the content if its attributes satisfy the
access policy retrieved from the proxy.
In order to perform immediate revocation, the publisher notifies
its proxy of each revoked client.
Because each client should be authenticated by the proxy for access policy decryption,
the proxy can deny access to the revoked clients.
The main drawback of this mechanism is its requirement for the third-party authentication by the 
proxy--a single point of failure that needs to be always online.

Raykova~\kETAL~\cite{RayKazLak15} proposed authentication-based AC for
pub/sub networks using distributed trust authorities, which play the roles of certificate 
and authorization authorities.
Before publishing a content, a publisher protects the payload using the
ciphertext-policy ABE.
Only a subscriber with the required attributes may decrypt the ciphertext.
In the pub/sub network, broker nodes match the published content to the subscriber's interest.
However, this matching process leaks some information such as the requested content 
name and the requester's subscription.

To limit this information exposure and preserve subscribers' privacy, the authors
suggested using a unique hashing function to hash interests and content tags.
These brokers may then use these hashed values instead of the raw interests and
content tags.
To limit the authorized brokers' access to these values, the hashed values
are also protected using ABE.
The overhead of interest hashing, ABE, and the corresponding per-hop hash matching procedure 
increase content retrieval latency significantly, thus undermining this approach.
%

%
%
\subsubsection{\it {\bfseries Identity-Based Access Control}}
\label{subsec04-01-04}
%
In identity-based cryptography, either entities' identities or the content names are used 
as the public keys. 
This allows providers or the network to authenticate a client using her identity.
Hamdane~\kETAL~\cite{HamSerFat13} proposed an identity-based cryptography 
AC system based on hierarchical tree-based content naming in which the 
entire sub-tree of a parent node inherits the AC policy of the parent.
In order to control the access to a sub-tree's content, the root of the sub-tree, is 
assigned an encryption/decryption key pair and a symmetric content encryption key.

The symmetric key is encrypted using the root's encryption key.
To give an entity read access on a content, the root decryption key is 
encrypted using the authorized entity's public key.
Upon successful authorization, the entity retrieves the encrypted symmetric key.
An entity with write access must also have access to the root's encryption key.
%
%
A lazy entity revocation can be performed in this scheme, which requires the root's 
encryption/decryption key pair to be updated.
This prevents a revoked client from accessing new content, 
however the client can access the contents published before revocation.
The old decryption key needs to be encrypted with the new key, so
that all newly added clients may access previously published content.
Considering that this procedure creates a chain of encrypted keys, each revocation
makes content access more expensive.
%
%

To overcome the above drawback, the authors proposed a credential and encryption-based
AC mechanism in~\cite{HamFat15}.
The proposed mechanism introduces an AC manager
(ACM), which possesses the root key for a namespace and defines and
enforces AC policies for the namespace.
Clients possess read and/or write capabilities so they can publish content and/or
request content.
To publish a content, a publisher queries the repository to check whether the target
namespace is subject to AC.
In the case that the name is protected, the publisher forwards its credentials, signed 
with its private key to the ACM, and requests an encryption and decryption key pair.
The ACM returns the encryption and decryption keys to an authorized publisher.

The publisher encrypts the content with a generated symmetric key, encrypts the 
symmetric key with the encryption key, and sends the encrypted content and the encrypted key 
to the repository to be cached.
When a client requests the content, the encrypted data will be delivered along
with the access policy.
The client then forwards its credentials to the ACM and retrieves the decryption
key, if its credentials satisfy the access policy.
However, the authors neglected the client revocation problem.
If a client that has access to several decryption keys is revoked, it can still 
keep using the keys.
To revoke it, all the corresponding publisher contents need to be re-keyed.
Also, the authors do not mention how the ACM verifies if a client is revoked or not and 
who performs the revocation. 
%

Aiash~\kETAL~\cite{AiaLoo15} proposed an identity-based AC mechanism for
NetInf.
This mechanism involves two steps: registration and
the authorization.
In the registration step, all clients and publishers share their public keys
(i.e., identities) with the name resolution service (NRS).
Upon a client's authentication, the NRS generates a sub-token (subscriber token)
and encrypts it with the client's public key.
To retrieve a content, a client retrieves both its (encrypted) token and a pointer to
the content object from the NRS.
The NRS replies with the identity of the publisher, and the client may use
its token to request the data from the publisher.

On receiving a client request, the publisher first queries the NRS to verify the
authenticity of the sub-token.
After token authentication, the publisher sends a challenge to the client to
verify its identity.
After authenticating the client, the publisher verifies the client's token against the content 
token, and if the client is authorized to access the content it returns the content.

This scheme's drawback is the communication overhead
introduced by both frequent queries to the NRS to verify tokens and the
challenge-response interaction between the client and the publisher.
Also, in this mechanism the authority of making content AC decisions lies with the NRS, instead of the publisher.
%

%
%

%
\subsection{Encryption Independent Approaches to Access Control}
\label{subsec04-02}
%
In this category, we discuss approaches where the AC mechanism is proposed as a generic framework and 
can use different available encryption mechanisms. 
We pay attention to the frameworks in these approaches without going into the details of the encryption mechanism 
used. 
%
For example, Kurihara~\kETAL~\cite{KurUzuWoo15} proposed an AC framework that 
can use any well-known cryptographic scheme.
%
%
This framework utilizes CCN's {\em manifest} feature, and can leverage 
AC mechanisms, such as group-based and broadcast-based AC.
The entities in the framework are content providers, clients, an encryption
and dissemination server, a key manager, and an access policy manager.
The key manager generates a symmetric key ({\it nonce key}) for content encryption
and sends it to the encryption and dissemination server, which performs content 
encryption and dissemination.

The nonce key is then encrypted by another encryption algorithm depending on the underlying 
AC structure, e.g., broadcast encryption, attribute-based encryption, or 
session-based encryption.
The decapsulation key, the key that decrypts the nonce key, is then encrypted by the access policy 
manager under the authorized client's public key and published into the network.
For content retrieval, an authorized client (authorization happens at the content provider using 
the client's credentials) downloads the encrypted content, uses the content manifest to locate 
the decapsulation key, and decrypts the content.
The authors suggested using lazy revocation, which would allow revoked
clients access to previously published content until it is re-encrypted and
re-disseminated. 
Overcoming this would require a significant overload--a downside for most proposed AC schemes.

Fotiou~\kETAL~\cite{FotMarPol12} proposed an AC enforcement method
for rendezvous-based ICN architectures.
The model proposes the use of an access control provider (ACP), which interacts
with publishers, rendezvous nodes (RNs), and subscribers to create AC 
policies and authenticate subscribers against the policy.
A publisher first provides its AC policy to the ACP, which assigns
a URI to the policy.
The publisher forwards the content, along with the policy URI, to the RNs.
A requesting subscriber will receive the URI of the AC  
as well as a nonce from the RN.
Simultaneously, the RN forwards the nonce and
the URI of the relevant AC policy to the ACP.
Upon receiving the client's credentials, the ACP verifies 
it against the policy and informs the RN whether the client is permitted access.
If permitted, the RN sends the content to the client. 

This approach has additional computation and communication overhead at RNs and/or routers 
which will increase response latency.
It requires the RN to store the AC policy URI for each content.
In addition, there is a need for a trusted ACP, which may become a single point of failure. 
%
%
Finally, the mechanism for subscriber revocation has not been discussed.

Singh~\cite{Sin12} proposed a trust-based approach for AC in
pub/sub networks.
In this scheme, a client has to establish trust with a broker, an
intermediate entity that authenticates clients and publishers.
During registration, a new client or publisher presents its credentials and attributes to
the broker, which results in the establishment of trust.  
The publisher defines an access policy and submits it to its broker.

A registered client requests content from its local broker.
If the local broker does not have the content, it returns the information
needed to locate the correct broker.
The broker possessing the content evaluates the trust and AC level of the client.
Despite the theoretically wide applicability of the proposed scheme, the authors
did not discuss client identification, and access level identification/verification, 
client revocation, communication overhead, and the broker network creation and management 
of publisher-broker network interactions.

Tan~\kETAL~\cite{TanZhoZou14} proposed a solution to copyright protection problem 
in the form of an AC mechanism.
They proposed to divide protected content into two portions: a large
cacheable portion, and a smaller portion which remains
at the publisher.
Each client retrieves the small portion from the publisher to reconstruct the content, 
thereby the publisher may enforce AC on its content.
In order to provide track-ability of authorized clients, the authors suggested
that the small portion be unique to each client; each client's copy stored at the publisher.
%
%
%

%
The request for this small portion allows publisher to track a client. 
According to the authors, this also allows identification of a malicious client that leaks its 
portion to an illegitimate user.
However, this verification may not be possible. 
If a malicious authorized client gives its content to an illegitimate user and the user 
downloads the rest of its content from the publisher, there is no way that the publisher 
can know, which user's small share was used. 
Another drawback of this mechanism is also the need for an always online provider.
%
%
%
%
%
%
\begin{table*}[!ht]
\caption{Summary and Classification of the Proposed Access Control Mechanisms}
\label{table:04_01}
\centering
\begin{tabular}{|l  c p{0.8cm} p{0.8cm} p{0.8cm} c c c c| c|}
 \hline
 \multirow{2}{*}{{\bf Mechanism}} & \multicolumn{1}{c}{{\bf Communication}} & \multicolumn{3}{c}{{\bf Computation Burden}} & \multicolumn{1}{c}{{\bf Additional}} & 
 \multicolumn{1}{c}{{\bf Client}} & \multicolumn{1}{c}{{\bf Cache}} & \multicolumn{1}{c|}{{\bf Access Control}} \\
 \cline{3-5}
 &{\bf Overhead}&{\bf Provider}&{\bf Network}&{\bf Client}&{\bf Infrastructure}&{\bf Revocation}&{\bf Utilization}&{\bf Enforcement} \\
 \hline 
 \hline 
 {\bf Encryption-Based} & & & & & & & & \\
  \hline 
 {\it Broadcast Encryption} & & & & & & & & \\
 Misra~\kETAL~\cite{MisTouMaj13,MisTouNat16}     & \cmark & \xmark & \xmark & \cmark & Not Required & Threshold Based       & Yes     & Client           \\
 \hline 
 {\it Session-Based} & & & & & & & & \\
 Renault~\kETAL~\cite{RenAhmAbi09, RenAhmAbi10}  & \cmark & \xmark & \xmark & \xmark & Required     & Not Considered        & No      & Network          \\
 Wang~\kETAL~\cite{WanXuFen14}                   & \cmark & \cmark & \xmark & \xmark & Not Required & Not Considered        & No      & Provider         \\
 \hline 
 {\it Proxy Re-Encryption} & & & & & & & & \\
 Wood~\kETAL~\cite{WooUzu14}                     & \cmark & \cmark & \xmark & \xmark & Not Required & Not Considered        & Yes     & Provider         \\
 Mangili~\kETAL~\cite{ManMarPar15}               & \cmark & \cmark & \xmark & \cmark & Not Required & Partial Re-encryption & Yes     & Client		 \\
 Zheng~\kETAL~\cite{ZheWanRav15}                 & \cmark & \cmark & \cmark & \xmark & Not Required & Not Considered        & Yes     & Network		 \\
 \hline 
 {\it Probabilistic Model} & & & & & & & & \\
 Chen~\kETAL~\cite{CheLeiXu14}                   & \cmark & \cmark & \cmark & \cmark & Not Required & Daily Re-encryption   & Limited & Provider/Network \\
 \hline 
 {\it Attribute-Based Encryption} & & & & & & & & \\
 Ion~\kETAL~\cite{IonZhaSch13}                   & \cmark & \xmark & \xmark & \xmark & Required     & Not Considered        & Yes     & Client   	 \\
 Li~\kETAL~\cite{LiVerHua14, LiWanHua14}         & \cmark & \cmark & \xmark & \cmark & Required     & Not Considered        & Yes     & Client           \\
 Da Silva~\kETAL~\cite{DasZor15}                 & \cmark & \xmark & \cmark & \xmark & Required     & Key Update per Revoc. & Yes     & Network          \\
 Raykova~\kETAL~\cite{RayKazLak15}               & \xmark & \cmark & \xmark & \cmark & Required     & Not Considered        & No      & Client		 \\
 \hline 
 {\it Identity-Based Encryption} & & & & & & & & \\
 Hamdane~\kETAL~\cite{HamSerFat13}               & \cmark & \xmark & \xmark & \xmark & Not Required & System Re-key         & Yes     & Provider         \\
 Hamdane~\kETAL~\cite{HamFat15}                  & \cmark & \xmark & \xmark & \cmark & Required     & Not Considered        & Yes     & Network          \\
 Aiash~\kETAL~\cite{AiaLoo15}                    & \cmark & \xmark & \xmark & \xmark & Required     & Not Considered        & No      & Provider	 \\
 \hline 
 \hline 
 {\bf Encryption-Independent} & & & & & & & & \\
 \hline 
 Kurihara~\kETAL~\cite{KurUzuWoo15}              & \cmark & \xmark & \cmark & \xmark & Required     & Lazy Revocation       & Yes     & Provider         \\
 Fotiou~\kETAL~\cite{FotMarPol12}                & \cmark & \xmark & \cmark & \xmark & Required     & Not Considered        & Yes     & Network          \\
 Singh~\cite{Sin12}                              & \cmark & \xmark & \cmark & \xmark & Required     & Not Considered        & Yes     & Network		 \\
 Tan~\kETAL~\cite{TanZhoZou14}                   & \cmark & \cmark & \xmark & \xmark & Not Required & Considered            & Yes     & Provider	 \\
 Ghali~\kETAL~\cite{GhaSchTsu15}                 & \xmark & \cmark & \cmark & \cmark & Not Required & Not Considered        & Limited & Provider/Network \\
 Li~\kETAL~\cite{LiZhaZhe15}                     & \cmark & \cmark & \cmark & \xmark & Not Required & Not Considered        & Yes     & Provider	 \\
 \hline 
\end{tabular}
\end{table*}

Ghali~\kETAL~\cite{GhaSchTsu15} tackled the AC problem using an
interest-based model, in contrast to popular encryption-based approaches.
The two major design aspects of this approach are $(1)$ name obfuscation, and
$(2)$ authorized disclosure.
The former prevents unauthorized clients from obtaining the content
name, the latter requires each entity responding to a content request to perform 
authentication/authorization on the publisher's behalf.
The authors proposed encryption-based and hash-based name obfuscation, in which
each authorized client (either individually or as part of a group) encrypts (with 
a symmetric key) or hashes a suffix of the content name with a key shared with the provider.

The interest for a content carries a nonce, a time-stamp, and a client identifier
in its payload, and is signed by the client using the client's private key (individual/group).
The provider, upon receiving an interest, verifies the client's signature and fetches 
the client's key to decrypt the encrypted portion of the content name.
The provider attaches the group's public key to the content (for signature verification) 
and forwards it to the client.
%
%
On receiving a content, the on-path routers, store the obfuscated content name and the  
public key to authenticate the subsequent requests for the same content from the same 
group of clients.
If the request cannot be authenticated it is dropped.  

This approach has several concerns. 
Obfuscated content names may result in several copies of a content being stored, 
undermining caching effectiveness. 
The use of hashing for name obfuscation would also require the provider
to pre-compute the hashed content names for each individual and group--not computation 
and storage efficient. 
A revoked client from a group can still request content until the provider 
revokes its membership and updates the group's keying material.

Li~\kETAL~\cite{LiZhaZhe15} designed a lightweight digital signature and AC  
scheme for NDN.
The access policies are enforced using provider generated tokens--metadata that indicate 
access levels.
%
%
Two private tokens, per authorized entity, enable content access and integrity verification.
Upon an entity's request for a token, the provider encrypts the token (generated by hashing 
a key vector) based on the requester's access level.

The provider combines a Merkle hash tree (generated using content blocks 
and tokens) and a new key vector to create hash-based signatures.
For signature verification, a client regenerates the Merkle hash tree, using the 
retrieved content and the new token, and combines it with the obtained signature 
to extract the original signing tokens.
The signature is valid if this token matches the token obtained from the content 
provider.

Although the proposed algorithm is faster than conventional RSA signing, the entities 
must synchronize with the provider for the correct version of the token.
The provider also must store, for each content, at least three tokens and their
corresponding key vectors at any time.
The tokens also need to be freshened at regular intervals for better security. 
Finally, client revocation, one of the most important concerns of AC  
in ICN, has not been discussed in this article.
%
%
%
\subsection{Summary and Future Directions in Access Control}
\label{subsec04-03}
%
%
Table~\ref{table:04_01} presents a summary of the proposed AC mechanisms
for ICN. 
It compares the existing mechanisms on the basis of their overhead: communication and computation, 
and the entities that bear the computation burden.
Client revocation method, ability of cache utilization, and the entities that enforce 
AC are other comparison features in Table~\ref{table:04_01}.

In this section, we reviewed the existing research in ICN AC enforcement and 
specifically focused on models including broadcast encryption-based~\cite{MisTouMaj13,MisTouNat16}, 
attribute-based~\cite{DasZor15,IonZhaSch13,LiVerHua14,LiWanHua14,RayKazLak15}, identity-based~\cite{HamFat15,HamSerFat13,AiaLoo15}
session-based~\cite{RenAhmAbi09,RenAhmAbi10,WanXuFen14}, proxy re-encryption-based~\cite{ManMarPar15,WooUzu14,ZheWanRav15}, 
and others~\cite{FotMarPol12,GhaSchTsu15,LiZhaZhe15,Sin12,TanZhoZou14} models.
Although almost all the proposed mechanisms introduce communication overhead, some of 
the proposed mechanisms~\cite{AiaLoo15,FotMarPol12} require extensive interactions between 
an AC manager and other network entities in order to enforce access constraints.
These interactions not only increase communication and computation overhead, but also require 
additional infrastructure. 

We believe that the availability of a content in caches is undermined significantly if content access 
requires authentication and/or authorization from an always-online server, which is difficult to guarantee.
To truly exploit ICN's intrinsic provisions for content availability, an 
AC mechanism should refrain from using an always-online entity. 
The work by Misra~\kETAL~\cite{MisTouNat16} is the first attempt in this direction.

Access right revocation is the other major concern of current proposals for
ICN AC management.
Attribute-based mechanisms~\cite{AiaLoo15,DasZor15,HamFat15,HamSerFat13,IonZhaSch13,LiVerHua14,LiWanHua14,RayKazLak15}, 
in general, either take the costly and inefficient approach of per-revocation re-keying, or allow clients to continue 
accessing cached content even after revocation.
Although we believe that the latter approach is more acceptable, as it imposes
less complexity, efficient access revocation is a key design factor
for scalable AC in ICNs.
Some of the proposed mechanisms~\cite{CheLeiXu14,DasZor15,HamFat15,RenAhmAbi09,ZheWanRav15,FotMarPol12,GhaSchTsu15,Sin12} 
require the network (routers) to enforce AC and authenticate clients.
The fact that the intermediate routers have to perform authentication procedure 
undermines the scalability of these mechanisms. 
There is scope for improvements on all these noted fronts.

We note that some of the proposed mechanisms target specific architectures, such as pub/sub based architectures~\cite{RayKazLak15,FotMarPol12,Sin12}, 
NetInf~\cite{RenAhmAbi09,RenAhmAbi10,AiaLoo15}, or CCN/NDN~\cite{KurUzuWoo15,LiZhaZhe15}.
However, the majority of the proposed mechanisms are generic and can apply to all ICN architectures.
There are some exceptions.
The work by Kurihara~\kETAL~\cite{KurUzuWoo15} is applicable to architectures with the manifest feature (e.g., CCN).
The mechanisms proposed by Li~\kETAL~\cite{LiVerHua14,LiWanHua14} and Ghali~\kETAL~\cite{GhaSchTsu15} modify the content 
name and hence are only applicable to architectures with flexible content naming scheme.
The proposal by Hamdane~\kETAL~\cite{HamSerFat13} is limited to architectures with a hierarchical naming scheme.
%

\section{Conclusions and Lessons Learned}
\label{sec05}
%
In this survey, we have comprehensively explored the existing work in the domain
of ICN security.
We divided the  content into three major sub-domains: security, privacy,
and access control enforcement.
We reviewed the existing work in each sub-domain, and highlighted the drawbacks
and benefits of each proposed solution.
Additionally, we provided potential future research directions to explore
to overcome the mentioned shortcomings.

In the security section, we explored attacks such as denial of service,
content poisoning, and cache pollution, and also presented the proposed models
for secure naming, routing, and applications.
The majority of the existing works in this sub-domain aim to prevent adversaries
from degrading the user QoS and QoE through malicious behavior, such as
interest flooding, cache pollution, and packet forgery.
However, the negative impacts of these solutions on legitimate clients have
not been studied in depth.
Among these attacks, DoS is the most widespread and the easiest to mount.
A simple rate limiting approach can mitigate the impact of the attack to some 
extent, however, it also can starve legitimate clients. 
Thwarting content poisoning attack, despite its detection simplicity, requires 
computational resources at the intermediate routers, which makes it more severe.

ICN privacy threats can affect content, caches, and the clients.
Timing and monitoring attacks specifically target cached content in the router
shared between a victim and an attacker threatening both the victim's and the cache's
privacy.
Proposed countermeasures such as applying random delay can protect the attack 
targets at the expense of latency.
Protocol attacks caused by ICN protocol design flaws target cache privacy,
while naming and signature privacy attacks target the name and signer privacy
respectively.
Among the privacy risks that we have explored, we believe requested content  
anonymity is of the utmost importance in ICNs.

The availability of content replicas at various locations outside the 
publisher's control creates need for more sophisticated 
access control mechanisms for ICN.
The majority of the access control mechanisms in the state of the art
rely on the existence of an online service to authorize each content request.
However, per-content online authorization dramatically increases the communication
overhead, and can also undermine content availability if the authorization
service goes down; regardless of the presence of the desired content in a nearby cache.
There is a need for an access control mechanism that guarantees the usability of 
the cached content, regardless of the content provider's availability.
This can be achieved through enforcing access control by network elements that cache the content. 
However, the computation and communication overheads at the routers of the authentication and 
authorization processes can become excessive. 

In what follows, we identify the lessons we have learned while
reviewing the state of the art in ICN security.

{\bf First}, the negative impacts of proposed security protocols on legitimate
clients can be significant and this impact's mitigation should be further investigated.
Approaches such as rate limiting on suspicious interfaces and name prefixes
may mitigate DoS attacks, however they come at the cost of quality of service
degradation for legitimate clients. 
By preventing content caching through either tunneling or request
flagging many privacy-focused schemes also inadvertently affect user QoE and QoS. 
For example, a privacy-sensitive client may unnecessarily mark all its content as private thus 
making caching ineffective. 
This will result in increased network load, and increased download latency for other users.

The architectures that use name based routing to route requests across the network (CCN, NDN, MobilityFirst) will fare 
better in the face of DoS/DDoS attacks on account of greater network-spread of interests and request aggregation; this is in contrast to 
architectures that route to specific set of nodes for efficiency (NetInf, PURSUIT) and hence adversely impact attack resilience. 
If end-to-end privacy by tunneling or other mechanisms are used, the network-wide routing approaches cannot benefit from in-network 
caching. 
At that point nothing separates the two architecture classes; the better the infrastructure the better the resilience. 

The {\bf second} lesson learned is that security concerns should be addressed at the intrinsic level. 
For example, content poisoning and cache pollution attacks are enabled due to 
lack of secure naming and caching schemes.
We believe that these attacks should be solved intrinsically by
employing strong cache verification mechanisms and self-certifying naming schemes, which
would inherently eliminate unpopular content from the cache and prevent forged
content from lingering in the network.
Similarly, a scalable naming scheme would not only eliminate many opportunities
for malicious behavior, but it also will improve the efficiency of content routing.
We note that despite these issues in-network caching is becoming a preferred approach, especially at the network-edge, propelled by the 
rapid developments in 5G technologies. 
Architectures that enable pervasive caching will thus receive more and more attention. 

{\bf Third}, in ICN, the privacy risks emanate from the data interest traveling in plaintext in the network. 
In the era of widespread consumer profiling, in which data consumption information are invaluable
to corporations, service providers, and censors, existing
ICN architectures have a wide attack surface for data collection. 
Although a handful of proposed mechanisms try to achieve communication anonymity,
they approaches have tended to port previous solutions from IP to the ICN paradigm.
We believe more needs to be done to develop a mechanism, which can preserve
privacy, while still leveraging the inherent ICN benefits. 
In this scenario, it is not very clear which class of architecture would perform the best for privacy; more research 
is needed to answer this question. 

{\bf Fourth}, the fundamental principles of ICN should be closely followed during
the design of new security mechanisms.
Here, we specifically refer to the necessity of efficient access control enforcement
mechanisms that are in agreement with ICN principles.
ICN, in principle, promotes content availability by allowing pervasive caching,
and hence requires more advanced, {\it service-independent} access control
mechanisms. 
In this survey, we have identified some initial attempts towards an independent
access control mechanism that can be enforced by any network caching entities 
efficiently. 
Again, in this context it is not clear if there is a specific architecture that stands out as best for access control; but 
we note that all architectures are nascent and still under a lot of flux. 
We suggest the research community must keep ICN principles in mind,
such that future access control schemes may protect content without undermining
features necessary for the future mobile devices and 5G-enabled Internet, such as in-network 
caching and use of multiple radio technologies concurrently for communication.
%
%


\bibliography{ICN_Security}
%
%
\vspace{-0.5in}
\begin{IEEEbiography}[{\includegraphics[width=1in,height=1.25in,clip,keepaspectratio]{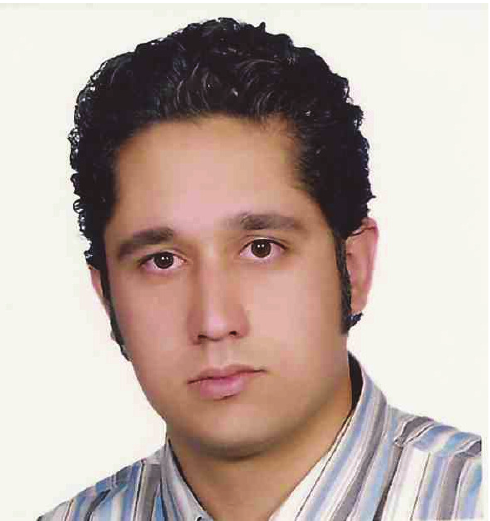}}]{Reza Tourani}
received his B.S. in Computer Engineering from IAUT, Tehran, Iran, in 2008, and M.S. in Computer Science from New Mexico State University, Las Cruces, NM, USA, in 2012. 
From 2013, he started his Ph.D. at New Mexico State University. His research interests include smart grid communication architecture and protocol, wireless protocols 
design and optimization, future Internet architecture, and privacy and security in wireless networks.
\end{IEEEbiography}
%
\vspace{-0.5in}
\begin{IEEEbiography}[{\includegraphics[width=1in,height=1.25in,clip,keepaspectratio]{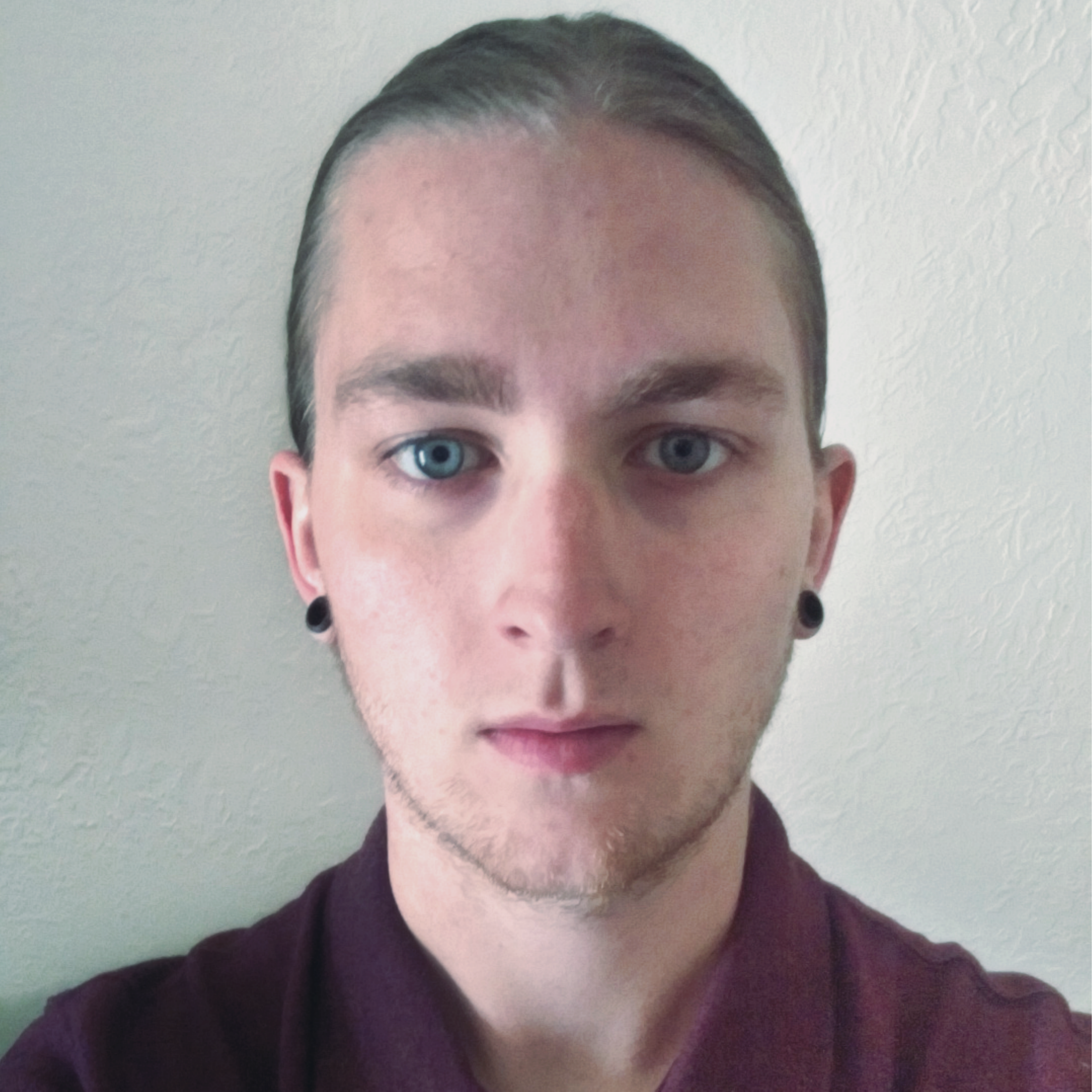}}]{Travis Mick}
completed his B.S. at New Mexico State University, Las Cruces, NM, USA, in 2014, and is now continuing toward an M.S., both in computer science.
His research focuses on information-centric networking, including security and privacy concerns, caching and forwarding strategies, and applications within the Internet of Things.
\end{IEEEbiography}
\vspace{-0.5in}
\begin{IEEEbiography}[{\includegraphics[width=1in,height=1.25in,clip,keepaspectratio]{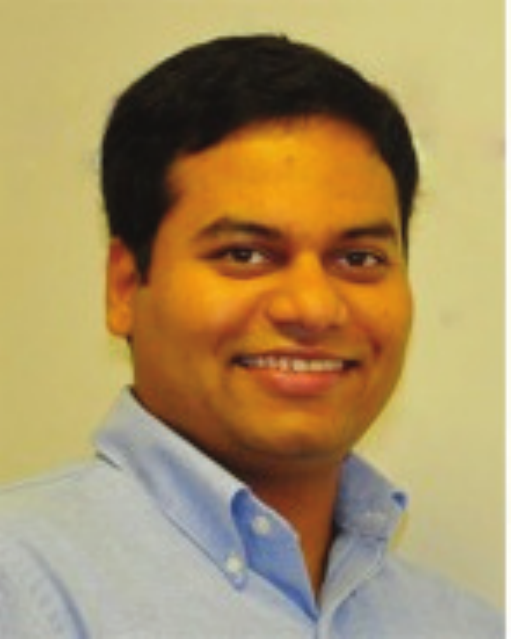}}]{Satyajayant Misra}
(SM'05, M'09) is an associate professor in computer science at New Mexico State University. He completed his M.Sc. in Physics and Information Systems from BITS, Pilani, 
India in 2003 and his Ph.D. in Computer Science from Arizona State University, Tempe, AZ, USA, in 2009. His research interests include wireless networks and the Internet, 
supercomputing, and smart grid architectures and protocols. He has served on several IEEE journal editorial boards and conference executive committees (Communications on Surveys 
and Tutorials, Wireless Communications Magazine, SECON 2010, INFOCOM 2012). He has authored more than 45 peer-reviewed IEEE/ACM journal articles and conference proceedings. 
More information can be obtained at www.cs.nmsu.edu/${\sim}$misra. 
\end{IEEEbiography}
\vspace{-0.5in}
\begin{IEEEbiography}[{\includegraphics[width=1in,height=1.25in,clip,keepaspectratio]{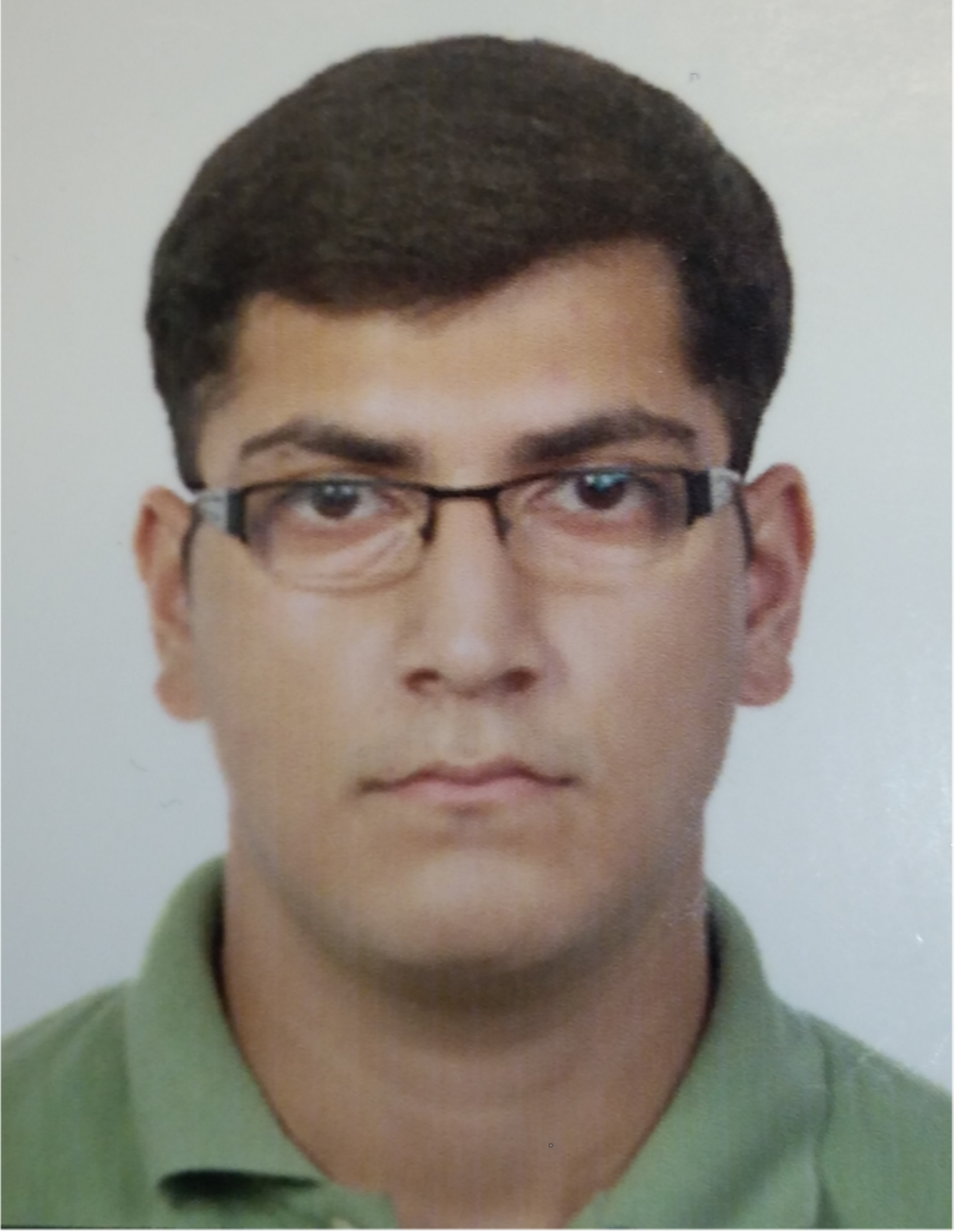}}]{Gaurav Panwar}
completed his B.Tech in electronics and communication engineering at Mahatma Gandhi Institute of Technology, Hyderabad, AP, India in 2013, and his M.S. in 
computer science at New Mexico State University in 2017. He is a Ph.D. student in computer science at New Mexico State University. His research is in wireless sensor networks, security, 
information-centric networking and smart grid technologies.
\end{IEEEbiography}
\end{document}